\newcommand{\zlabel}[1]{\label{#1} }
\newcommand{\fc}{\frac} 
\newcommand{\lt}{\left} 
\newcommand{\rt}{\right} 
\newcommand{\mr}{\mathbf{r}}
\newcommand{\pr}{\prime}

\newcommand{\mrp}{\mathbf{r}^\prime}

\newcommand{\al}{\alpha}
\newcommand{\be}{\beta}
\newcommand{\pa}[2]{\frac{\partial #1}{\partial #2}}

\newcommand{\bu}{\mathbf{u}}

\newcommand{\dive}{\text{\bf div}\;\!}
\newcommand{\grad}{\text{\bf grad}\;\!}

\newcommand{\mz}{\mathbf{z}}

\newcommand{\Ms}{\text{Ms}}
\newcommand{\Ma}{\text{Ma}}
\newcommand{\lab}[1]{ (\ref{#1}):\; }
\newcommand{\FIG}[2]{}
 
\newcommand{\rd}{)}

\newcommand{\sq}{[} 
\newcommand{\lsc}{\;\text{\Large ;}}
\newcommand{\lbar}{{\Large $|$}}
\newcommand{\rbar}{\;\;\text{\Large $|$}}
\documentclass[pra,preprint,preprintnumbers,amsmath,amssymb,floatfix]{revtex4}
\usepackage{graphicx}
\usepackage{dcolumn}
\usepackage{epic}%
\usepackage{eepic}%
\usepackage{amsthm}
\usepackage{color}
\usepackage{bm}
\usepackage{comment}
\usepackage{esvect}
\usepackage{textcomp}

\begin{document}

\preprint{APS/123-QED}

\title{The Fluid Dynamics of the One-Body Stationary States of Quantum Mechanics with Real Valued Wavefunctions}

\author{James P. Finley}
\email{james.finley@enmu.edu}
\affiliation{
Department of Physical Sciences,
Eastern New~Mexico University,
Station \#33, Portales, NM 88130}
\date{\today}

\begin{abstract}
It is demonstrated that the probability density function---given by the square of a quantum
mechanical wavefunction that is a real-valued eigenvector of a time-independent, one-body
Schr\"odinger equation---satisfies a compressible-flow generalization of the Bernoulli
equation, where the mass density is the probability density times the mass of the system; the
pressure and velocity fields are defined by functions depending on the probability density, and
the gradient and the Laplacian of the probability density. There are two possible directions of
the velocity on a streamline, called uphill and downhill flow, where the fluid particles move
in the direction of increasing and decreasing density, respectively.
The ``Bernoullian'' equation, mentioned above, is not a statement of the conservation of energy;
instead, it is a statement of local (spatial) conservation of specific energy, an intensive,
uniform scalar-field that is a constant of motion for each fluid element.
Relationships between the mathematical objects from the one-body quantum states considered,
called Q1 probability states, and the corresponding objects from the steady flow states, called
Q1 flow states, are discovered.  For example, the integrand of the expectation value of the
kinetic energy for a Q1 probability state is the sum of the kinetic-energy per volume and the
pressure of the corresponding Q1 flow state.
The velocity definition mentioned above implies a generalization of the steady-flow continuity equation where
mass is not locally conserved, and where the pressure is proportional to the mass
creation rate per volume. However, over all space, both mass and energy is conserved.
The gradient of the Bernoullian equation, mentioned above, is demonstrated to be equivalent to the steady flow Euler
equation for variable mass and irrotational flow. This implies that Q1 flow is a flow with
variable mass, that is steady, irrotational, inviscid, and compressible.
Speed of sound equations are obtained from a spherical, acoustic wave-pulse propagating along a
streamline of an ambient state of Q1 flow with a spherical flow geometry and a uniform speed
for both the fluid velocity and the wave pulse velocities (relative to the fluid velocity)
restricted to a streamline. One of the solutions of the derived quadratic equation is that the
wave-pulse velocity on a streamlime is equal in magnitude but opposite in direction of the
fluid velocity on the streamlime. Possible interpretations for a flow with this solution are
considered, including that it simply the solution for the case of steady flow with no wave
pulse. The other solution is given by explicit formulas which depends on scalar fields of the
ambient state, and this solution is the focus of attention from that point on.
It is proven that the extremums of the momentum per volume on a streamline occur at points
that are Mach 1 speed.
The developed formalism is applied to a particle in a one-dimensional box, the ground and first
excited-states of the one-dimensional harmonic oscillator, and the hydrogen 1s and 2s
states.
Some behavior is repeated in all the applications examined. For example, an antinode, a point
of local-maximum density, has zero velocity and zero Mach speed on the streamline, while a
node, a point of minimum density, has infinite velocity and Mach 2. In between the node and
antinode is an extremum of the momentum, and Mach 1. The antinode is also an unstable
equilibrium points, where, for downhill flow, the fluid particles are moving and accelerating
away from the unstable equilibrium point where they end of at a node, if one exist, with
infinite speed and zero mass.  Another type of behavior is identified, called a fall, where the
electron density is strictly decreasing away from the nucleus, as in the hydrogen 1s state.
So that different flows can be compared with ease, and flows can be characterized, a reference
flow, called ordinary Q1 flow, is defined.
(All the figures appear after the citations.)
\end{abstract}
  
\maketitle

\section{Introduction}

When one acquires knowledge of the fundamentals of quantum mechanics, it becomes clear that the
meaning of the word 'understand', as in the phrase `to understand quantum mechanics' is
difficult to ``define,'' and a definition is needed for the comprehension of the statement
``nobody understands quantum mechanics'' by Feynman \cite{Feynman:67}. However, for any
reasonable definition of the word ``understand,'' it is certainly true that, for all things
being equal, it is much more difficult to understand the information contained in a state of
quantum mechanics than one from classical mechanics. Therefore, it is not surprising that
entire monographs are available that go beyond the mathematical structures used in quantum
mechanics, including algebras and analysis \cite{Byron,Jordon}, and into topics such as quantum
logic \cite{Jauch,Hughes}.

For many applications of quantum mechanics, the values of physical properties, called
observables, are given by a finite or infinite series, in particular, the electronic energy,
and in these mathematical objects, very few terms can be assigned a ``physical meaning,'' as
indicated by Mullikan:
\begin{quote}
... the more accurate the calculations became, the more the concepts tended to vanish into thin air. \cite{Mullikan:64}
\end{quote}

However, for one-electron stationary-states, whose wavefunctions satisfy the time-independent,
electronic Schr\"odinger equation, there is one function, $-e|\psi|^2$, where $\psi$ is the
complex-valued wave-function and $-e$ is the electron charge, that is often given an
interpretation, as described by Raimes:
\begin{quote}
In many problems it is useful, if only as an intuitive aid, to picture the electron charge
as spread out in a cloud having charge density $-e|\psi(\mr)|^2$ at position $\mr$.~\cite{Raimes}
\end{quote}
This gives a simple meaning from classical electrostatics of the potential-energy functional
$\int v\rho\, d\mr$, where $\rho= |\psi|^2$ is the probability density of the quantum state,
and $v$ is the potential-energy of the corresponding classical system.  Also, for many electron
systems with wavefunctions that are approximated as a single Slater determinant, there is a
simple classical interpretation given for the Coulomb integral: It is the classical
Coulomb-repulsion between two charge clouds \cite{Raimes,Szabo}. For chemistry, many of the simple,
but very useful, models used to describe chemical stability and reactivity have components of
charge density, including Lewis dot structures. Crystal field theory also uses a simple
electrostatic repulsion model, based on the charge densities determined by occupied orbitals, to
describe the energy splitting of otherwise degenerate orbitals of metal complex ions.

In this paper, this idea of a state of a bound electron---and other one-body systems in a
stationary state---being described as a continuum, with charge density $-e\rho$, is specialized and
generalized to a steady-flow state of a fluid, such that $m\rho(\mr)$ is the mass density of
the state with (total) mass $m$. Also, the momentum per volume and the pressure are taken as
functions of the density $\rho$ and a subset of the partial derivatives of $\rho$: The momentum
per volume vector field is proportional to the gradient of $\rho$, as in Fick's law of
diffusion \cite{Gillespie}, and the pressure, a scalar field, is proportional to the Laplacian
of $\rho$, where in both cases, the proportionality constant depends on Plank's constant
$\hbar$. However, \emph{only} one-body quantum states with real-valued wavefunctions are considered.

The precise mathematical statements giving momentum, pressure, and other conditions given below
can be viewed as postulates. However, at this point, it is more convenient to categorize these
statements as definitions of a type of fluid. This is similar to the approach used by Currie
\cite{Currie} when he describes a Newtonian fluid, except that he calls the four statements
given for a Newtonian fluid, which can be viewed as definitions, postulates.

This paper is \emph{not} about challenging the fundamental axioms that give probabilities of
outcomes of the measurement of observables of a quantum state, or any other essential axiom of
quantum mechanics. Instead, the results presented below are about building upon and
contributing to what has been done.

Below (in Sec.~\ref{p0242}) it is proved that the probability density function $\rho$, such
that $\rho =\phi^2$, where $\phi$ is a real-valued eigenvector of a time-independent, one-body
Schr\"odinger equation for a system with mass $m$, satisfies a compressible-flow generalization
of the Bernoulli equation, where the mass density $\rho_m = m\rho$, pressure $p$, and vector
velocity $\bu$ fields are defined by (composite) functions depending on $\rho$, $\nabla\rho$
and $\nabla^2\rho$.  Also, there are two possible velocity directions $\bu_\pm$, where $\bu_- =
-\bu_+$, called uphill $\bu_+$ and downhill $\bu_-$ flow.  Furthermore, the ``Bernoullian''
equation, mentioned above, is a statement of the local spatial and temporal conservation of
(total) \emph{specific} energy $\bar{E}$, an intensive variable, e.g., the energy per mass,
that is a constant of motion for each fluid element. It is not a statement of the local
conservaton of the (total) energy, an extensive variable.

Relationships between the mathematical objects from the one-body quantum states considered,
called Q1 (or Q1.0) probability states, and the corresponding objects from the steady flow
states, called Q1 flow states, are developed (\ref{p0538}).  For example, the integrand of the
expectation value of the kinetic energy for a Q1 probability state is the sum of the
kinetic-energy per volume and the pressure of the corresponding Q1 flow state.  Also, if the
pressure of the Q1 flow state is \emph{not} the zero function, then it cannot be a nonnegative
function. Terminology are also developed (\ref{p0538}). For example, the kinetic energy per
mass $u^2/2$ for a classical fluid, ``corresponds'' to $mu^2/2$, for the quantum fluid, where
$mu^2/2$ is called the \emph{kinetic energy per amount}, $m$ is the (total) mass of the system,
and $mu^2/2$ has units of energy. (The scalar fields $u^2/2$ and $mu^2/2$ are also classified
as specific \emph{open} and \emph{closed} quantities, respectively, with respect to the
dimension of energy.)

The velocity, pressure and all terms terms within the Bernoullian equation, mentioned above,
are examined (\ref{p0247}), for the 1s state of the hydrogenic atom with atomic number $Z$.
Also, the average pressure in the subspace where the pressure is positive is calculated.

The velocity definition for the Q1 flow states is shown to imply a generalization of the
steady-flow continuity equation (\ref{p1530}), where mass is \emph{not} locally conserved, and
where the pressure is proportional to the mass creation rate per volume. For downhill flow
$\bu_-$, a positive and negative pressure implies a positive and negative mass creation-rate,
respectively. The total and positive mass creation rate, involving summing over the region of
space where the pressure is positive, is calculated for hydrogen 1s state with downhill flow.
In general, Q1 flow state do not conserve mass and energy \emph{locally}. However, over all
space, they, of course, do conserve mass and energy.

There is nothing novel presented in Sec.~\ref{p8250}.  It is the derivation of a form of the
Euler equation for variable mass and irrotational flow, obtained by using a standard derivation
\cite{Munson} (and a vector calculus approach like one the one used by Kelly \cite{Kelly}),
where the continuity equation is \emph{not} used. This equation (\ref{7288}) is derived because
it is not so common and it is needed in Sec.~\ref{p8252}, where it is proven that the gradient
of the Bernoullian equation from Sec.~\ref{p0242} is equivalent to the steady flow version of
(\ref{7288}). This implies that Q1 flow is a flow with variable mass, that is steady,
irrotational, inviscid, and compressible.

Speed of sound equations are obtained (\ref{1084c}) from a spherical, acoustic wave-pulse
propagating along a streamline $L$ of an ambient state of Q1 flow with a spherical flow
geometry, and a uniform speed for both the fluid $|u_\pm|$ and the wave pulse $-s_\pm$, relative
to the fluid velocity. After setting up the problem (\ref{1084c-1}), the momentum balance
(\ref{1084c-2}) and continuity equations (\ref{1084c-2}) are combined (\ref{1084c-4}), giving a
quadratic equation for the wave pulse (vector) component $-s_\pm$,
and then this equation is applied to the hydrogenic 1s flow state
(\ref{p7290}). Subsections~(\ref{7202}) and (\ref{p9022}) demonstrate that one of the solutions
to the quadratic equation is $s_\pm = u_\pm$, where $u_\pm$ is the fluid velocity (component)
on the streamline, and an explicit formula is derived for the other solution. A signed Mach
speed Ms is defined to be $\Ms = u_\pm/(-s_\pm)$, and $\Ma = \Ms$, where Ma is the Mach speed,
if $\Ms \ge 0$. Possible interpretations of the meaning of the $s_\pm = u_\pm$ solution are
given (\ref{p0822a}), including that it simply the solution for the case of steady flow with no
wave pulse.

The properties of Q1 flows are compared to classical flows that are mass conserving,
irrotational, steady, inviscid, compressible, and with no body force.  For convenience, these
classical flows are called the \emph{corresponding classical potential-free flows}.  The speed
of sound quadratic equation from (\ref{1084c-4}) is easily modified to yield the speed of sound
equation for the corresponding classical potential-free flows, and it is the same formulas as
the well known one that was derived for a static ambient state with one-dimensional flow
\cite{Munson,Pierce}: $\pm\sqrt{dp/d\rho_m}$.

It is proven that the extremums of the momentum density $\rho_mu_\pm$ on a streamline occur at points
$\mr\in L$ of space where $\rho(\mr) \ne 0$ and $\Ms(\mr) = \Ma(\mr) = 1$ (\ref{p9542}).

Applications (\ref{p3602}) of the formalism developed is applied to a fluid (or particle) in a
one-dimensional box (\ref{p7300}), the ground and first excited-state of the one-dimensional
harmonic oscillator (\ref{p7400}), and the hydrogen 2s state (\ref{p7500}). For the ground
state of a fluid in a one-dimensional box of length $[0,1]$ in atomic units, where the density
is maximum at $0.5$ and zero at $0$ and $1$, the point $r = 0.5$ of maximum $\rho$ is an
unstable equilibrium point. For $r > 0.5$, and a flow direction that moves away from the
maximum $\rho$ at $0.5$, the fluid particles are moving and accelerating to the right; for $r <
0.5 $ they move and accelerate to the left. A similar type of behavior is seen by all states
examined for similar regions where the density has a local maximum. Another behavior repeated
is a Mach speed of 0 and 2 at the maximum and minimum densities, and in between these point,
Mach 1 is obtained, where, as predicted, the momentum per volume is an extremum.  Another type
of behavior is identified, called a fall, where the electron density is strictly decreasing
away from the nucleus. Also, in the limit of a fluid particle reaching a node, the particle
have infinity speed and zero mass.
So that different flows can be compared with ease, and flows can be characterized, a reference
flow, called ordinary Q1 flow, is defined.

\section{A Bernoullian equation for quantum systems.} 

\subsection{The theoretic foundation \zlabel{p0242}}

{\bf Definition.}  A quantum 1 (Q1) wavefunction $\phi$ is a real-valued eigenfunction of a one-body
time-independent Schr\"odinger equation. A Q1 probability state is a state represented
by a quantum 1 wavefunction $\phi$, where all information about the state is determined by the
wavefunction $\phi$ (and the axioms of quantum mechanics).

Let $\phi$ be a Q1 wavefunction that is a solution of a one-body Schr\"odinger equation with external
potential $V$ and mass $m$:
\begin{equation} \zlabel{0001}
  -\fc{\hbar^2}{2m}\nabla^2\phi + V\phi = \bar{E}\phi,
\end{equation}
We next show that $\rho = \phi^2$ is a solution
of the following:
\begin{equation} \zlabel{0002}
  \fc12 mu^2 + p\rho^{-1} + V = \bar{E}; \qquad \rho(\mr)\ne 0,
\end{equation}
\begin{align}
\zlabel{0003a} \bu_\pm &= \pm\fc{\hbar}{2m}\fc{\nabla\rho}{\rho}, \\
 \zlabel{0003b}   p &= -\fc{\hbar^2}{4m}\nabla^2\rho,
\end{align}
and $u^2 = |\bu_\pm|^2$.  Equation~(\ref{0002}) is a compressible-flow generalization of the
Bernoulli equation \cite{Munson,Currie} with a mass density $\rho_m = m\rho$, and $\rho$ is the
probability density of the Q1 state represented by $\phi$. Henceforth, we call $\bu_\pm$ and
$p$ the velocity and pressure, respectively, regardless of their ``actual'' physical
meaning. Note that there are two possible velocities $\bu_+$ and $\bu_-$, giving two possible
directions along each streamlines. (An $N$--body generalization of (\ref{0002}), from the 
$N$--body generalization of (\ref{0001}), has been obtained elsewhere \cite{Finley1}.)

For the proof below, the external potential $V$ is required to be a function such that the
subspace $\{\mr\in\mathbb{R}^3|\rho(\mr) = 0\}$ has measure zero and the eigenfunctions of
(\ref{0001}) are three times continuously differentiable. Hence, the condition $\rho\ne 0$ for
(\ref{0002}) implies that $\rho$ satisfies the equation almost everywhere a.e, since
$\{\mr\in\mathbb{R}^3|\rho(\mr) = 0\}$ has measure zero.

The proof involves first obtaining the following:
\begin{equation} \zlabel{0000}
\longrightarrow  -\fc12\phi\nabla^2\phi = \fc18\rho^{-1}\nabla\rho\cdot\nabla\rho - \fc14\nabla^2\rho 
\end{equation}
To show that this is an identity, we obtain two other auxiliary identities. The first one,
given by
\begin{equation} \zlabel{0382}
\longrightarrow  -\fc12\phi\nabla^2\phi = \fc12\nabla\phi\cdot\nabla\phi - \fc14\nabla^2\rho,
\end{equation}
is obtained from the result of 
\begin{gather*}
  \fc12\nabla^2\rho = \fc12\nabla\cdot\nabla(\phi\phi) 
  = \nabla\cdot(\phi\nabla\phi) = \nabla\phi\cdot\nabla\phi + \phi\nabla\cdot(\nabla\phi)
  = \nabla\phi\cdot\nabla\phi + \phi\nabla^2\phi
\end{gather*}
For the second one,
\begin{equation} \zlabel{p3382}
  \longrightarrow \nabla\phi\cdot\nabla\phi = \fc14\rho^{-1}\nabla\rho\cdot\nabla\rho,
\end{equation}
we take the vector dot product of 
\[
\nabla\rho = \nabla(\phi^2) = 2\phi\nabla\phi
\]
with itself, giving an equation that is equivalent to (\ref{p3382}):
\[
  \nabla\rho\cdot\nabla\rho = 4\phi^2\nabla\phi\cdot\nabla\phi = 4\rho\nabla\phi\cdot\nabla\phi
\]
Substituting (\ref{p3382}) into (\ref{0382}) gives (\ref{0000}).
%
%
%
Multiply  (\ref{0001}) by $\phi$ and using (\ref{0000}) we have
\[
  \fc{\hbar^2}{8m}\rho^{-1}\nabla\rho\cdot\nabla\rho - \fc{\hbar^2}{4m}\nabla^2\rho + V\rho = \bar{E}\rho 
\]
Multiplying by $\rho^{-1}$ we obtain
\[
\fc{\hbar^2}{8m}\fc{\nabla\rho}{\rho}\cdot\fc{\nabla\rho}{\rho} - \fc{\hbar^2}{4m}(\nabla^2\rho) \rho^{-1} + V = \bar{E} 
\]
Substituting definitions (\ref{0003a}) and (\ref{0003b}) gives (\ref{0002}). Hence, if $\phi$
is a solution of (\ref{0001}), then $\rho = \phi^2$ satisfies (\ref{0002}).  It is easily seen
that the derivation is reversible giving the following: If $\rho$ is a solution of
(\ref{0002}), then a three times continuously differentiable function $\phi$ such that $\rho =
\phi^2$, satisfies (\ref{0001}). We indicate this state of affairs by stating that (\ref{0001})
and (\ref{0002}) are equivalent. Note that the function $|\phi| = +\sqrt{\rho}$ satisfies
(\ref{0001}) a.e, since the subspace $\{\mr\in\mathbb{R}^3|\rho(\mr) = 0\}$ has measure zero,
where $|\phi|$ is not necessarily differentiable.

For later use for systems with a (total) charge $q\ne 0$, we write (\ref{0002}) as
\begin{equation} \zlabel{0002b}
  \fc12 mu^2 + p\rho^{-1} + q\Phi = \bar{E}
\end{equation}
where $\Phi$ is defined by $V = q\Phi$, and the charge density is $q\rho$.

%

\subsection{Terminolgy, interpretations, and relationships of quantum particle states of
(\ref{0001}) and the corresponding classical fluid states of (\ref{0002}) \zlabel{p0538}}

The terminology and conventions used for states of a classical fluid and states of a quantum
(bound) particle, are different and similar in ways that can lead to confusion when
investigating relationships and the assignment of meaning to mathematical objects. Therefore,
it is necessary to introduce some terminology so that statements are clear and unambiguous.


{\bf Alias.} The kinetic energy per volume $\rho mu^2/2$ has the alias the
'kinetic-energy density,'  and this is also done for other per volume quantities, especially
ones involving energy.

{\bf Definitions.} A Q1 (or Q1.0) density $\rho$ is a real valued solution of the Bernoullian
(\ref{0002}), where the velocity and pressure are given by (\ref{0003a}) and (\ref{0003b}),
respectively. A Q1 (or Q1.0) flow state is a state represented by a Q1 density $\rho$, and all
information about the state is determined by the density $\rho$. A Q1 density is said to
normalized if
\[
\int_{\mathbb{R}^3} \rho\, d\mr = 1.
\]

Frequently we compare the properties of Q1 flows to classical flows that are mass conserving,
irrotational, steady, inviscid, compressible, and with no body force.
For convenience, these flows are called the corresponding classical potential-free flows.  For
example, in Sec.~\ref{1084c}, a speed of sound equation is derived for Q1 flows, and this
equation is easily modified to yield the speed of sound equation for the corresponding classical
potential-free flows.


{\bf Definitions.} Flow state with velocity fields $\bu_+$ and $\bu_-$ are said to have uphill
and downhill flows, respectively.

Note that it follows from definition (\ref{0003a}) that the direction of $\bu_+$ and $\bu_-$ on
a streamline are towards increasing and decreasing density $\rho$, respectively.  The velocity
field definition (\ref{0003a}) also indicates that Q1 flow states are irrotational. To obtain
the velocity potential $\omega$ from (\ref{0003a}), and avoid taking the natural logarithm of a
dimensioned quantity, both densities are replaced by the dimensionless density $\tilde{\rho} =
a_0^3\rho$, where $a_0$ is the Bohr radius, giving
\begin{gather}
\omega = \pm\fc{\hbar}{2m}\ln\tilde{\rho}
\end{gather}
If follows from (\ref{0003a}) that the potential of the momentum density $\rho_m\bu_\pm$ is
$\pm(\hbar/2m)\rho$.  As in the Bernoulli equation for irrotational flow \cite{Munson,Currie},
(\ref{0002}) holds for the entire flow field---there is only one constant $\bar{E}$, instead
one for each streamline.

{\bf Alias.} The pressure $p$ is also called the 'compression-energy density.'

Consider the following two equations: {\bf 1)} The equation obtained by multipling the
Schr\"odinger' (\ref{0001})
by $\phi$ and then followed by substituting $\rho = \phi^2$; and,
{\bf 2)} the Bernoullian' (\ref{0002}) multiplied by $\rho$. 
Subtracting of these two equations gives
\begin{equation} \zlabel{p7422}
-\fc{\hbar^2}{2m}\phi\nabla^2\phi = \fc12\rho_mu^2 + p
\end{equation}
Hence, the \emph{integrand} of the expectation value of the kinetic energy for a Q1 probability
state is the sum of the kinetic-energy and compression-energy densities of the corresponding Q1
flow state.

{\bf Definitions.} Recall that $m$ and $q$ are the \emph{total} change and mass of a system.  A
scalar-field is said to be open with respect to dimension of it has either energy per mass or
energy per charge units;  a closed scalar-field has units of energy.  The scalar fields $u^2/2$
and $mu^2/2$ (of a Q1 flow state) are the specific open and closed kinetic-energies,
respectively; these are also called the kinetic-energy per mass and per amount, respectively.
The scalar fields $\Phi$ and $q\Phi$ are the specific open and closed electrostatic
potential-energies, respectively; these are also called the potential-energy per charge and per
amount, respectively.  In general, an open scalar-field is closed by multiplying the open
scalar-field by either the total mass $m$ or the total charge $q$ of the system.  Hence,
$p\rho_m^{-1}$ and $m(p\rho_m^{-1}) = p\rho^{-1}$ are the specific open and closed
compression-energies, respectively. (An easy way to reconize this is to replace $p$ by $p_m$, since
the pressure unit contains a mass-unit factor: $p_m\rho_m^{-1}$ and $p_m\rho^{-1}$ are
certainly open and closed, respectively.)

For a one-electron atom with atomic number $Z$, the external potential $V$ can be written
\begin{equation} \zlabel{p5522}
V = (-e)\Phi, \qquad \Phi(r) = \fc{Z}{4\pi\epsilon_0} \fc{e}{r}
\end{equation}
where $(-e)$ and $e$ are the electron and proton charges, respectively. The scalar field $\Phi$
is the classical electrostatic-potential of a point source with charge $Ze$; $\Phi$ is the
specific open potential energy, or the potential energy per charge.  Hence, $(-e)\Phi =
V$---the external potential from the Schr\"odinger' (\ref{0001})---is the specific
closed potential energy for Q1 flow states of hydrogenic electronic systems.

Let the specific closed (total) energy $\bar{E}$ be defined by Eq.~(\ref{0002b}), where
$\bar{E}$ is also the energy eigenvalue of the Schr\"odinger equation (\ref{0001});
Eq.~(\ref{0002b}) is a statement of the local spatial and temporal conservation of (total)
\emph{specific} energy $\bar{E}$, an intensive variable with energy units, that is a constant
of motion for each fluid element. It is not a statement of the local conservaton of the (total)
energy, an extensive variable.


{\bf Definition.} The scalar field $\rho$ is called the open density; the scalar fields $\rho_m
\;\dot{=}\; m\rho$ and $\rho_q \;\dot{=}\; q\rho$ are called the (closed) mass and charge
densities, respectively. The symbol $\dot{=}$ is used for definitions.

Note that the kinetic energy density is equal to the specific open kinetic-enery times the
closed density $\rho_m$---the open-closed form---or the the specific closed kinetic-enery times
the open density $\rho$ ---the closed-open form:
\[
\lt(\fc12 u_\pm^2\rt)\rho_m = \lt(\fc12 m u_\pm^2\rt)\rho 
\]
These are also called the classical and quantum density forms, respectively.  Other energy
quantities are similar.

The various forms of (\ref{0002b}) can be written
\begin{align}
    \fc12 u^2 + p\rho_m^{-1} + \fc{q}{m}\Phi &= E, \qquad\quad \text{classical or open form} \\
    \fc12 u^2\rho_m + p + \fc{q}{m}\Phi\rho_m &= E\rho_m, \qquad \text{classical density form} \\
        \fc12 mu^2 + p\rho^{-1} + q\Phi &=  E_m, \qquad \text{quantum or closed form} \\
  \fc12mu^2\rho + p + q\Phi\rho &= E_m\rho, \qquad \text{quantum density form} 
\end{align}
where $E_m = mE = \bar{E}$.  If there is no chance of confusion, the words 'open' and 'closed'
in the above definitions can be suppressed.



For each cartesian coordiate $\al\in\{x,y.z\}$ we require that the wavefunction satisfy
\[
\lim_{\al \to \pm\infty} \phi(\mr) = \lim_{\al \to \pm\infty} \pa{\phi}{\al} = 0
\]
Hence
\[
\int_{-\infty}^{\infty} \pa{^2\rho}{\al^2}\, d\al = \lt.\pa{\rho}{\al}\rt|_{-\infty}^{\infty} =
2\lt.\phi\pa{\phi}{\al}\rt|_{-\infty}^{\infty} = 0 
\]
and therefore
\[
\int_{\mathbb{R}^3} \nabla^2\rho\, d\mr = 0
\]
This result combined with (\ref{0003b}) gives
\begin{equation} \zlabel{p2004}
\int_{\mathbb{R}^3}p\,d\mr = 0
\end{equation}
This equation implies that if $p(\mr) \ge 0$ a.e, then $p(\mr) = 0$ a.e.  Hence, if $p(\mr) \ne
0$ a.e, then we cannot have $p(\mr) \ge 0$ a.e. Hence, the pressure from quantum flows differ
in this regard from classical flows where the pressures are required to be thermodynamic
pressures, and such pressures satisfy $p(\mr) \ge0$.


If $V$ is a constant function, and there are no boundary condition imposed in $\mathbb{R}^3$,
then there is no \emph{real-valued} eigenfunction of (\ref{0001}). Since (\ref{0001}) and
(\ref{0002}) are equivalent, (\ref{0002}) also has no solutions if $V$ is constant.  Consider a
Q1 static flow ($\rho$ is uniform) where $V$ is not a constant scalar field. In that case,
(\ref{0003a}) and (\ref{0003b}) indicate that the velocity $\bu_\pm$ and pressure $p$ are the
zero vector- and scalar-field, respectively.  Hence, (\ref{0002}) has no
solutions. (Eq.~(\ref{0001}) only has the trivial solution, which is not an eigenfunction.)
Hence, there does not exists a static Q1 flow state.

The summed kinetic-energy (over all space) of the Q1 flow state represented by $\rho$ is
\begin{equation} \zlabel{p5022}
  \int_{\mathbb{R}^3} \lt(\fc12 m u_\pm^2\rt)\rho\,d\mr,
\end{equation}
where $m u_\pm^2/2$ is the specific, closed kinetic-energy of the Q1 flow state.
Analogous meanings also hold for the summed compression-, and (electrostatic)
potential-energy. Since the scalar field $\bar{E}$ is uniform and $\rho$ is normalized,
the summed total energy satisfies
\begin{equation} \zlabel{p9210}
\int_{\mathbb{R}^3} \bar{E}\rho\, d\mr = \bar{E}.
\end{equation}
Hence, the summed total energy of the Q1 flow state is equal to the specific, closed total-energy $\bar{E}$,
where $\bar{E}$ is also the energy of the corresponding Q1 probability state that satisfies (\ref{0001}).
(Note that the summed kinetic-energy, defined above, can also be called the total kinetic energy,
but we cannot do this when speaking about $\bar{E}$, since $\bar{E}$ is a kind of total energy.)

Let $S\in\mathbb{R}^3$.  The kinetic energy over the subspace $S$ (of the state represented by
$\rho$) is given by (\ref{p5022}) with $\mathbb{R}^3$ replaced by $S$. A similar definition is
used for the other energy-related scalar fields. For example, the compression energy over the
subspace $S$ is
\[
\int_{S} \lt(p\rho^{-1}\rt)\rho\, d\mr = \int_{S} p\, d\mr,
\]
and (\ref{p2004}) indicates that the summed compression energy (over all space) is zero.

The potential energy over $S$ satisfies 
\begin{gather*}
\int_{S} \phi^* V\phi\, d\mr =  \int_{S} V\rho\, d\mr
\end{gather*}
With $S = \mathbb{R}^3$, the above equation indicates that the expectation
value of the potential energy for a Q1 probability state is equal to the summed potential energy of
the corresponding Q1 flow state. Except for special cases, the kinetic energy satisfies
\[
\int_{S} \lt(-\fc{\hbar^2}{2m}\phi\nabla^2\phi\rt)\, d\mr \ne  \int_{S} \lt(\fc12 \rho_mu^2\rt) \, d\mr 
\]
By setting $S=\mathbb{R}^3$ for the integral on the left hand-side (lhs), and using
(\ref{p7422}) and (\ref{p2004}), we find that
\[
\int_{\mathbb{R}^3} \phi\hat{T}\phi\,d\mr = \fc12 \int_{\mathbb{R}^3}
\rho_mu^2\,d\mr, \qquad  \hat{T} = -\fc{\hbar^2}{2m} \nabla^2 
\]
Hence, the expectation value of the kinetic energy for a Q1 probability state is equal to the 
summed kinetic energy of the corresponding Q1 flow state.



\subsection{The ground state of hydrogenic atoms. \zlabel{p0247}}


Henceforth, a large semicolon, ending a displayed equations with a dedicated line, as in (\ref{p5902}) below,
is read ``where.''

Next we examine the velocity, pressure and the energy per amount terms within (\ref{0002b}) for
the 1s state of the hydrogenic atom with atomic number $Z$. The flow is spherical with each
unit vector $\hat{\mr}$ tangent to a streamline, where the direction of downhill and
uphill flow are $\hat{\mr}$ and $-\hat{\mr}$, respectively.  While examining the derivations it
is useful to be aware that, for the Hartee atomic units,  the derived velocity, pressure and energy are
\[
\fc{\hbar}{ma_0}, \quad \fc{\hbar^2}{ma_0^5}, \quad \text{and}\quad  \fc{\hbar^2}{ma_0^2}.
\]
respectively. Also, the Bohr radius $a_0$ is $0.5292\times 10^{-10}$ meter to four figures.

In spherical coordinates the open density, the probability density for hydrogenic 1s states
\cite{Bransden}, is
\begin{gather} \zlabel{p5902}
\rho(r) = \fc{Z^3}{a_0^3}\pi^{-1}e^{-2Zr/a_0}\lsc
\end{gather}
$a_0$ is the Bohr radius and $m$ is the electron mass. The velocity field is
\[
  \bu_\pm = \pm\fc{\hbar}{2m}\fc{\nabla\rho}{\rho} = \pm\fc{\hbar}{2m}\rho^{-1}\pa{\rho}{r}\hat{\mathbf{r}}
  = \pm\fc{\hbar}{2m}\rho^{-1}(-2Z/a_0)\rho\hat{\mr}
\]
In other words,
\begin{equation} \zlabel{p0404}
  \bu_\pm = \mp \fc{Z\hbar}{ma_0}\hat{\mr}, \qquad \fc12m u^2 = \fc12\fc{Z^2\hbar^2}{ma_0^2}
\end{equation}
Hence, the speed $|\bu_\pm|$ is constant. For $Z = 1$, $|\bu_\pm| = \hbar/ma_0$, and this
constant is both the derived Hartree unit of velocity and the speed of the electron in the first
Bohr orbit of hydrogen \cite{Bransden}.  The second equation from (\ref{p0404}), for use below,
is the specific closed kinetic energy, and $u$ can be chosen to be $|\bu_\pm|$, but there are
other choices, e.g., $u = \bu_\pm\cdot\hat{\mr}$.
(For higher accuracy, the electron mass $m$ can be replaced by the reduced mass $\mu$; $a_0$
then becomes the modified Bohr radius $a_\mu$.)

Let $\bar{z} = Z/a_0$.  Next we calculate the compression-energy density $p$ and the compression
energy per amount $p\rho^{-1}$, starting with the Laplacian of $\rho$:
\begin{equation*} 
\nabla^2 \rho = r^{-2}\pa{}{r}\lt(r^2\pa{\rho}{r}\rt) = -2\bar{z}r^{-2}\pa{}{r}(\rho r^2)
= -2\bar{z}r^{-2}(-2\bar{z}r^2\rho + 2r\rho)
= 4\bar{z}^2\rho  - 4\bar{z}r^{-1}\rho 
\end{equation*}
The pressure $p$, defined by (\ref{0003b}), is 
\begin{gather} \zlabel{p0102}
  p(r) = -\fc{1}{4}\fc{\hbar^2}{m}\nabla^2 \rho =  \lt(-\fc{\hbar^2}{m}\bar{z}^2 + \fc{\hbar^2}{a_0m}Zr^{-1}\rt)\rho(r) 
\end{gather} 
Using $\bar{z} = Z/a_0$, this equation can also be written
\[
  p(r) = \lt(\fc{\hbar^2}{a_0m}Zr^{-1} - \fc{\hbar^2}{a_0m}\fc{Z^2}{a_0}\rt)\rho(r) 
  = \fc{Z\hbar^2}{a_0m}\lt(\fc{1}{r} - \fc{Z}{a_0}\rt)\rho(r) 
\]
Substitute (\ref{p5902}) for $\rho$ gives
\begin{gather*}
  p(r) = \fc{Z^4\hbar^2}{a_0^4m}\lt(\fc{1}{r} - \fc{Z}{a_0}\rt)\pi^{-1}e^{-2Zr/a_0}
\end{gather*}

Finally we express the pressure in a manner so that it is clearly dependent on $r/a_0$:
\begin{equation} \zlabel{p5220}
  p(r) = Z^4\fc{\hbar^2}{ma_0^5}\lt(\fc{a_0}{r} - Z \rt)\pi^{-1}e^{-2r/a_0},
\end{equation}
where $\hbar^2/ma_0^5$ is the derived Hartee unit of pressure.  For the hydrogen atom, the
pressure is zero at $r = a_0$; it is positive and negative for $r < a_0$ and $r > a_0$,
respectively. The pressure is plotted for the hydrogen atom in Fig.~\ref{pressure-1s} for a
calculation with $a_0$ in the Angstrom unit and $\hbar^2/ma_0^5$ in the Pascal unit.
(All the figures appear after the citations.)
\FIG{pressure-1s}{The pressure of hydrogen atom from Eq.~(\ref{p5220}) \zlabel{pressure-1s}}


The positive $\mathbf{P}_+$ and negative $\mathbf{P}_-$ pressure subspaces of $\mathbf{R}^3$
for a Q1 flow are $\mathbf{P}_+ = \{\mr\in\mathbf{R}^3|p(\mr)>0\}$ and $\mathbf{P}_+ = \{\mr
\in\mathbf{R}^3|p(\mr)< 0\}$, respectively.  The compression energy $X$ over the positive
pressure subspace $\mathbf{P}_+$ is
\begin{equation} \zlabel{p8210}
X = \int_{\mathbf{P}_+} \lt(p\rho^{-1}\rt)\rho\, d\mr = \int_{\mathbf{P}_+} p\, d\mr
\end{equation}
Since the summed compression energy (over all space) is zero, as pointed out in the previous
subsection, and since the set $\{\mr\in\mathbf{R}^3|p(r) = 0\}$ does not contribute to $X$, the compression
energy over the negative pressure subspace $\mathbf{P}_-$ is $-X$.

For the hydrogenic 1s state, the pressure $p$ is a function of the radial sperical-coordinate
only and $\mathbf{P}_+ = \{\mr\in\mathbf{R}^3|r < a_0\}$. Hence
\[
X = 4\pi\int_0^{a_0} r^2p(r) \, dr = -\pi\fc{\hbar^2}{m}\int_0^{a_0} r^2\nabla^2\rho \, dr
\]
and we substitutied (\ref{0003b}). Expressing the Laplacian in spherical coordinates, we discover that
\[
r^2\nabla^2\rho = \pa{}{r}\lt(r^2\pa{\rho}{r}\rt)
\]
Substituting this equation into the one above it and then using (\ref{p5902}) we get
\begin{gather} \notag
  X = -\pi\fc{\hbar^2}{m}a_0^2 \pa{\rho(a_0)}{r} = -\pi\fc{\hbar^2}{m}a_0^2 \lt(-2\fc{Z^4}{a_0^4}\pi^{-1}e^{-2Z}\rt), \text{ i.e.,}
  \\ \zlabel{p2808}
 X = 2Z^4\fc{\hbar^2}{ma_0^2}e^{-2Z}   
\end{gather}
and $\hbar^2/(ma_0^2)$ is the derived energy unit from the Hartree atomic units. 
For $Z=1$, $X\approx 0.27$~a.u.

The average value $\bar{p}_+$ of the pressure in $\mathbf{P}_+$ is
\[
\bar{p}_+ = \fc{1}{\text{Vol}(\mathbf{P}_+)}\int_{\mathbf{P}_+} p\, d\mr 
= \fc34 \pi^{-1} a_0^{-3} \lt(2Z^4\fc{\hbar^2}{ma_0^2}e^{-2Z}\rt) 
  =   \fc32e^{-2Z}\pi^{-1}Z^4\fc{\hbar^2}{ma_0^5} 
\]
where we used (\ref{p8210}), (\ref{p2808}) and $\text{Vol}(\mathbf{P}_+)$ is the volume of
$\mathbf{P}_+$, a ball of radius $a_0$. For $Z=1$, we have
\[
\bar{p}_+ = \fc32e^{-2}\pi^{-1}\fc{\hbar^2}{ma_0^5} \approx 0.065 \fc{\hbar^2}{ma_0^5},
\]
about nineteen hundred gigapascals.
The average value $\bar{p}_-$ of the prssure in $\mathbf{P}_-$ is zero, because the volume $\text{Vol}(\mathbf{P}_-)$ is infiite.

Returning to the Eq.~(\ref{p0102}), the specific compression energy $p\rho^{-1}$ is
\[
p\rho^{-1} =  -\fc{\hbar^2}{m}\bar{z}^2  + \fc{\hbar^2}{a_0m}Zr^{-1}.
\]
Using the definition \cite{Bransden}
\[
\fc{\hbar^2}{a_0m} = \fc{e^2}{4\pi\varepsilon_0}\lsc
\]
$e$ is the proton charge, we obtain the desired form:
\begin{gather} \zlabel{p3814}
 p\rho^{-1} =  -\fc{\hbar^2}{m}\bar{z}^2 + \fc{e^2}{4\pi\varepsilon_0}Zr^{-1}
\end{gather}
Adding the second equation from (\ref{p0404}) to equations (\ref{p3814}) and (\ref{p5522}), we
get
\[
\fc12m u^2 +  p\rho^{-1} + (-e)\Phi_p(r) =  
\fc12\fc{Z^2\hbar^2}{a_0^2m} - \fc{\hbar^2}{m}\bar{z}^2 + \fc{e^2}{4\pi\varepsilon_0}Zr^{-1} -\fc{Z}{4\pi\epsilon_0} \fc{e^2}{r} 
\]
In other words,
\[
\fc12m u^2 +  p\rho^{-1} + (-e)\Phi_p(r) =  -\fc12\fc{Z^2\hbar^2}{ma_0^2}
\]
Comparing this result with (\ref{0002b}) for $q = -e$, we obtain
\[
\bar{E} = -\fc{1}{2}  Z^2\fc{\hbar^2}{ma_0^2} = -\fc{e^2}{(4\pi\varepsilon_0)a_0}\fc{Z^2}{2}
\]
The second form of the energy is obtained from $1/(a_0m) = e^2/(4\pi\varepsilon_0\hbar^2)$. As
expected, $\bar{E}$ is the eigenvalue of the Schr\"odinger equation (\ref{0001}) for the 1s
ground state with atomic number $Z$ \cite{Bransden}.

For later use we compute ($p\rho^{-1} + (-e)\Phi$), a constant function,  using (\ref{p3814}) and (\ref{p5522}):
\begin{equation} \zlabel{p0880}
p\rho^{-1} + (-e)\Phi = -\fc{Z^2\hbar^2}{a_0^2m}
\end{equation} 




\section{The Continuity equation for quantum systems \zlabel{p1530}}

The differential and integral forms of the time-dependent
continuity equations are \cite{Munson,Currie}
\begin{gather} \zlabel{Cont1} 
  \partial_t\rho_m + \nabla\cdot\rho_m\bu = \mathbf{0}
  \\ \zlabel{Cont2} 
  \int_V\partial_t\rho_m \; dV + \int_S \rho_m \bu\cdot\hat{\mathbf{n}} \; dS
  = \mathbf{0}
\end{gather}
where $\partial_t = \partial/\partial t$ and $\nabla\cdot\rho_m\bu$ is the mass
creation-rate density.  For the quantum flows under consideration, a generalizations of
the differential form of the continuity equation (\ref{Cont1}) for steady flows is obtained by
multiplying the velocity definition (\ref{0003a})
by the mass density $\rho_m$ and then taking the divergence of the result:
\begin{equation} \zlabel{cont1}
  \nabla\cdot\rho_m\bu_\pm = \pm\fc{\hbar}{2}\nabla^2\rho 
\end{equation}
Hence, for this steady-flow generalization of the continuity equation (\ref{Cont1}), mass is not locally conserved.
Using $\nabla^2\rho = \nabla\cdot\nabla\rho$ with the divergence theorem, we obtain a
generalization of (\ref{Cont2}) for steady flows:
\begin{equation} \zlabel{cont2}
 \int_S \rho_m\bu_\pm\cdot\hat{\mathbf{n}}\, dS = \pm\fc{\hbar}{2} \int_S \nabla\rho\cdot\hat{\mathbf{n}}\, dS
\end{equation}
Generalizations of both (\ref{Cont1}) and (\ref{cont1}), and both (\ref{Cont2}) and
(\ref{cont2}) are, respectively,
\begin{gather} \zlabel{cont1-time}
  \partial_t\rho_m + \nabla\cdot\rho_m\bu_\pm = \pm\fc{\hbar}{2}\nabla^2\rho  
  \\ \zlabel{cont2-time}
  \int_V\partial_t\rho_m \; dV + \int_S \rho_m \bu_\pm\cdot\hat{\mathbf{n}} \; dS
  = \pm\fc{\hbar}{2}\int_S \nabla\rho\cdot\hat{\mathbf{n}} \; dS 
\end{gather}

Reversing the order of (\ref{cont1}), followed by multiplying the equations
by $\pm\hbar/(2m)$, and using $\rho_m/m = \rho$, we have
\[
  \fc{\hbar^2}{4m}\nabla^2\rho = \pm\fc{\hbar}{2}\nabla\cdot(\rho\bu_\pm) 
\]
Comparing this result with the pressure definition (\ref{0003b}) 
gives
\begin{equation} \zlabel{press-2}
  p = \mp\fc{\hbar}{2}\nabla\cdot\rho\bu_\pm
\end{equation}
Hence, the pressure is proportional to the mass creation rate per volume
$\nabla\cdot\rho_m\bu$.  Note that both velocity vectors $\bu_\pm$ give the same pressure
scalar field, but their formulas differ.  The previous equation can also be written
\begin{equation} \zlabel{press-2b} 
  \nabla\cdot\rho_m\bu_\pm = \mp\fc{2m}{\hbar}p 
\end{equation}
Hence, for a point $\mr\in\mathbb{R}^3$ on a streamline with downhill flow $\bu_-$, a positive pressure
($\mr \in \mathbf{P}_+$) implies a positive creation rate $[\nabla\cdot\rho_m\bu_\pm](\mr) > 0$;
a negative pressure ($\mr \in \mathbf{P}_-$) implies a negative creation rate
$[\nabla\cdot\rho_m\bu_\pm](\mr) < 0$.  The analogous ``switched around'' conclusion for uphill
flow is easily determined by noting the following: $[\nabla\cdot\rho_+\bu_+](\mr) > 0$ if and
only if $[\nabla\cdot\rho_-\bu_-](\mr) < 0$ and vice versa.
Using (\ref{press-2b}), the total (and positive) mass-creation rate $\dot{\text{M}}$ for a Q1 downhill flow state is
\[
\dot{\text{M}} = \int_{\mathbf{P}_+} \nabla\cdot\rho_m \bu_- \, d\mr = \fc{2m}{\hbar} \int_{\mathbf{P}_+}p\, d\mr
= \fc{2m}{\hbar} X\lsc
\]
$X$ is the compression energy, defined by (\ref{p8210}).  To compute $\dot{\text{M}}$ for a
hydrogenic 1s state we use (\ref{p2808}), giving
\[
  \dot{\text{M}}_{1s} = 4Z^4\fc{\hbar}{a_0^2}e^{-2Z}
\]
The constant $\hbar/a_0^2$, which looks like a derived Hartee unit, is
\[
\fc{\hbar}{a_0^2} = 3.767\times 10^{-14}~\text{kg s$^{-1}$}  
\]
to four figures. Using $4e^{-2} = 0.5413$ for the hydrogen atom ground state we have
\[
\dot{\text{M}}_{1s} = 2.039 \times 10^{-14}\text{kg s$^{-1}$}
\]
The total electron mass-creation rate of one mole of noninteracting hydrogen atoms is about 12 billion
kilograms per second.  The \emph{electronic effective refresh rate} for a one electron flow is the number
of electron masses $m_e$ created per second. For the hydrogen 1s state, it is
\[
\fc{\dot{\text{M}}}{m_e} =  2.238\times 10^{16}~\text{e}^-~\text{Hz},
\]
about twenty thousand trillion electron masses per second.


Suppose the Q1 model presented here can be extended to many-body systems, and that it holds
for all fundamental particles of atomic nuclei with mass. Then, according to the model, all ordinary
matter is being simultaneously created and destroyed by an unknown nonlocal process, with separate
mass creation and annihilation zones.

Since equation (\ref{cont1}) is determined by the velocity definition (\ref{0003a}), the mass
creation (and annihilation) rate is determined by the velocity.  It might be interesting, if
possible, to develop a model where time dependent mass creation rates of different particles
determines the velocity and other field quantities needed to describe motion, including both
translation and rotation, of all objects with mass.  The universe can then be modeled as a
three-dimensional visual display, analog or digital, where, in the digital model, volumetric
pixels of mass are created and destroyed as time proceeds. Relativistic effects might appear as
a ``natural'' generalization. A mass increase caused by a process of an object moving to a
higher speed, may, somehow, slow the mass creation and annihilation rates, which could be a
fundamental for the relative speed that time passes.

\section{The variable mass Euler equation for irrotation flows \zlabel{p8250}}

In the vast majority of derivation of the Euler equation, the differential form of the
momentum-balance equation is used that does not hold for systems that do not conserve mass.
Since our systems under consideration have variable mass, we need a form of the Euler equation
that does not use the continuity equation. It is trivial to derive variable mass
momentum-balance equations, one simply uses a standard derivation \cite{Currie}, but refrain from 
utilizing the continuity equation, leading to some extra terms in the working equation.
Here we derive the variable-mass Euler equation for irrotation flows.

We start with the momentum
balance equation for a fluid subject to a Coulombic body-force with force per charge
$(-\nabla\Phi)$:
\begin{equation} \zlabel{3920}
\fc{d}{dt}\int_V \rho_m\bu \; dV = \int_S \pmb{\sigma}\mathbf{n}\, dS + \int_V q\rho(-\nabla\Phi)\, dV
\end{equation}
where $\pmb{\sigma}$ is the stress tensor and $\mathbf{n}$ is the normal unit vector to the
surface $S$, the border of the subspace $V$.  First we work on the lhs.  Using the Reynolds'
transport theorem, and the definition,
\[
\int_{V}g\fc{d(dV)}{dt}\; \dot{=} \; \int_{V}g\nabla\cdot \mathbf{u} \; dV
\]
where $g$ is an arbitrary function, we obtain
\begin{gather*}
  \fc{d}{dt}\int_{V(t)}\rho_m\bu \; dV =
  \int_V\lt(\rho_m\fc{d\bu}{dt} + \bu\fc{d\rho_m}{dt}\rt)\; dV +  \int_V \rho_m\bu\fc{d(dV)}{dt}
  \\
  = \int_V \lt(
  \rho_m\pa{\bu}{t} + \rho_m (\grad \bu)\bu + \lt(\pa{\rho_m}{t} + \nabla\rho_m\cdot\bu\rt)\bu + (\nabla\cdot\bu)\rho_m\bu
   \rt)\; dV
\end{gather*}
Hence
\begin{equation} \zlabel{2302}
 \fc{d}{dt}\int_V \rho_m\bu\; dV =
\int_V \lt(\rho_m\pa{\bu}{t} + \rho_m (\grad\bu)\bu  + \mathbb{M}\bu \rt)\, dV
\end{equation}
where
\begin{equation} \zlabel{4922}
\mathbb{M}  = \pa{\rho_m}{t} + \nabla\cdot(\rho_m\bu)  = \dot{\rho}_m + (\nabla\cdot\bu)\rho_m 
\end{equation}
and the continuity equation for systems that conserve mass is $\mathbb{M} = \mathbf{0}$.  Note
that the definition above permits the use of a product rule of differentiation on the factors
of the integrand: $\rho_m\times\bu\times dV$, where $dV$ is ``considered'' a factor, even
though the symbol is excluded in some notations for integrals.

For the surface integral of Eq.~(\ref{3920}), we apply the  divergence theorem:
\begin{equation} \zlabel{8372}
  \int_S \pmb{\sigma}\mathbf{n} \, dS  = \int_V \dive {\pmb{\sigma}} \, dV
\end{equation}

Substituting Eq.~(\ref{2302}) and (\ref{8372}) into (\ref{3920}) gives the differential
momentum-balance equation with variable mass:
\begin{equation} \zlabel{5973} 
\rho_m\pa{\bu}{t} + \rho_m (\grad\bu)\bu + \mathbb{M}\bu = \dive \pmb{\sigma} + q\rho(-\nabla\Phi),
\end{equation}
where we removed the integrations and obtained a true statement, since the equation with the
integrations holds for all subspaces $V$.

Next we consider only inviscid flows. By definition, these satisfy $\pmb{\sigma} =
-p\mathbf{I}$, where $p$ is the pressure.  Using this equality and a vector identity, the first
term on the right-hand side (\ref{5973}) for inviscid fluids  becomes
\[
\dive \pmb{\sigma} = -\dive(p\mathbf{I}) = -\nabla p
\]
Next we require $\bu$ to be irrotational, i.e., $\nabla\times \bu = \mathbf{0}$. This permits
the use of the following equality:
\begin{equation} 
(\grad\bu)\bu = \fc12\nabla u^2, \quad \text{if} \quad \nabla\times \bu = \mathbf{0} 
\end{equation}
Substituting the above two equations into (\ref{5973}) we obtained the desired equation:
\[
\rho_m\pa{\bu}{t} + \fc12\rho_m\nabla u^2 + \lt(\pa{\rho_m}{t} + \nabla\cdot(\rho_m\bu)\rt)\bu = -\nabla p - q\rho\nabla\Phi
\]
and we also used (\ref{4922}).  This equation can also be written can be written
\begin{gather}
 \zlabel{7288} \pa{}{t}(\rho_m\bu) + \fc12\rho_m\nabla u^2 + \nabla\cdot(\rho_m\bu)\bu + \nabla p + q\rho\nabla\Phi = 0
\end{gather}
This equation is the variable-mass Euler equation for the special case of irrotational flow.
In other words, Eq.~(\ref{7288}) is applicable to flows that are irrotational, compressible, invsicid, and
variable-mass. We also require the body force to be Coulombic, but, obviously,
$q\rho\nabla\Phi$ can be replaced by $\rho_m F$, where $F$ is a force per mass. In the special
case where the flow is steady, incompressible, and mass is conserved, the division of
(\ref{7288}) by $\rho_m$, followed by integration, yields the Bernoulli equation.

\section{The relationship between the Bernoullian and the Euler equations for Q1 flows. \zlabel{p8252}}

Let $\mathbb{A}$ and $\mathbb{B}$ be two equations. It is said that $\mathbb{A}$ implies
$\mathbb{B}$, if $\phi$ is a solution of $\mathbb{A}$ implies that $\phi$ is a solution of
$\mathbb{B}$, where $\phi$ is not the zero function.  If $\mathbb{A}$ implies $\mathbb{B}$ and
$\mathbb{B}$ implies $\mathbb{A}$, then equations $\mathbb{A}$ and $\mathbb{B}$ are
equivalent.

Consider the gradient  of the Beroullian (\ref{0002b})
\begin{gather} \zlabel{4838}
  \fc12 m\nabla u^2 + \nabla\lt(\fc{p}{\rho}\rt) + q\nabla\Phi = \mathbf{0},
\end{gather}
In this section we show that the time dependent generalization of this equation, given by
\begin{gather} \zlabel{p4880}
\rho^{-1}\pa{}{t}(\rho_m\bu) + \fc12 m \nabla u^2 + \nabla\lt(\fc{p}{\rho}\rt) + q\nabla\Phi = 0
\end{gather}
is equivalent to the variable-mass Euler' (\ref{7288}) with the condition that the
velocity and pressure are given by (\ref{0003a}) and (\ref{0003b}).

{\it The gradient of the Beroullian (\ref{p4880}) implies variable-mass Euler' (\ref{7288}).}
Substituting the function $p$, defined by Eq.~(\ref{0003b}), into (\ref{p4880})
and multiplying the result by $\rho$ we obtain
\begin{equation} \zlabel{3802}
 \pa{}{t}(\rho_m\bu) +
\fc12\rho_m \nabla u^2 - \fc{1}{4}\fc{\hbar^2}{m}\rho\nabla\lt(\fc{\nabla^2\rho}{\rho}\rt) + q\rho\nabla \Phi =\mathbf{0}
\end{equation}
For later use, note that
\begin{equation} \zlabel{ident-1}
\fc{\hbar}{2m}\rho \nabla \fc{1}{\rho}
= -\fc{\hbar}{2m}\rho \rho^{-2} \nabla \rho
= -\fc{\hbar}{2m}\fc{\nabla\rho}{\rho} = \bu_-
\end{equation}
Next we expand out a factor from the second term from (\ref{3802}) and then multiply it by a constant:
\begin{align*}
\rho\nabla\lt(\fc{\nabla^2\rho}{\rho}\rt) &=
\nabla(\nabla^2\rho) + \rho(\nabla^2\rho)\nabla\fc{1}{\rho} \\
-\fc{1}{4}\fc{\hbar^2}{m}\rho\nabla\lt(\fc{\nabla^2\rho}{\rho}\rt) &=
\fc{1}{2}\fc{\hbar}{m}\nabla\lt(-\fc{\hbar}{2}\nabla^2\rho\rt)
+ \lt(-\fc{\hbar}{2}\nabla^2\rho\rt)\lt(\fc{\hbar}{2m}\rho\rt)\nabla\fc{1}{\rho}
\end{align*}
Using (\ref{ident-1}) and the continuity Eq.~(\ref{cont1}) for the $\bu_-$ velocity field, we obtain
\[
-\fc{1}{4}\fc{\hbar^2}{m}\rho\nabla\lt(\fc{\nabla^2\rho}{\rho}\rt) =
\fc{1}{2}\fc{\hbar}{m}\nabla(\nabla\cdot\rho_m\bu_-) + \nabla\cdot(\rho_m\bu_-)\bu_-
\]
which can be written
\[
-\fc{1}{4}\fc{\hbar^2}{m}\rho\nabla\lt(\fc{\nabla^2\rho}{\rho}\rt) =
\nabla\lt(\fc{\hbar}{2}\nabla\cdot\rho\bu_-\rt) + \nabla\cdot(\rho_m\bu)\bu
\]
where $\bu =\bu_\pm$
Using (\ref{press-2}) we get
\[
-\fc{1}{4}\fc{\hbar^2}{m}\rho\nabla\lt(\fc{\nabla^2\rho}{\rho}\rt) = \nabla p + \nabla\cdot(\rho_m\bu)\bu
\]
Substituting this into (\ref{3802}), we obtain the Euler equation (\ref{7288}) for a steady
flow. {\it Hence, (\ref{p4880}) implies (\ref{7288}).  Also, (\ref{4838}) implies the steady flow form of (\ref{7288}).}


{\it The  variable-mass Euler' (\ref{7288}) implies the gradient of the Beroullian (\ref{p4880}).}
Let $\bu = \bu_-$. Using (\ref{press-2b}) and (\ref{0003a}) we have
\begin{equation} \zlabel{p7292}
\nabla\cdot(\rho_m\bu)\bu = \fc{2m}{\hbar}p\lt(-\fc{\hbar}{2m}\fc{\nabla\rho}{\rho}\rt) = -\fc{p}{\rho}\nabla\rho
\end{equation}
Since $\bu_+\bu_+ = \bu_-\bu_-$, this same result also holds for velocity $\bu_+$. Let
$\bu = \bu_\pm$. The substitution of (\ref{p7292}) into (\ref{7288}) gives
\[
\pa{}{t}(\rho_m\bu) + \fc12\rho_m\nabla u^2 - \fc{p}{\rho}\nabla\rho + \nabla p + q\rho\nabla\Phi = 0
\]

Multiplying by $\rho$ and rearrangement gives
\begin{equation} \zlabel{p5282}
\rho\pa{}{t}(\rho_m\bu) + \fc12\rho_m\rho\nabla u^2 + \rho\nabla p - p\nabla\rho + q\rho^2\nabla\Phi = 0
\end{equation}
Next we set $v = \rho^{-1}$ and derive an identity:
\[
\rho\nabla p - p\nabla\rho = v^{-1}\nabla p - p\nabla v^{-1} = v^{-1} \nabla p + pv^{-2}\nabla v = v^{-2}(v\nabla p + p\nabla v)
\]
Hence, the identity:
\[
  \rho\nabla p - p\nabla\rho = \rho^2 \nabla\lt(\fc{p}{\rho}\rt)
\]
Substituting this result into (\ref{p5282}) and dividing by $\rho^2$ gives the the Bernoullian
(\ref{p4880}).  {\it Hence, (\ref{7288}) implies (\ref{p4880})}.  {\it Also, the steady
  form of (\ref{7288}) implies (\ref{4838}).}

Since $\bar{E}$ is determined by $m$, $q$, $\rho$, and $\Phi$, (\ref{4838}) determines
$\bar{E}$ and its integral, the Bernoullian (\ref{0002b}).  Therefore, the Bernoullian
(\ref{0002b}) and its integral (\ref{4838}) are equivalent.  {\it Hence, the Beroullian
  (\ref{0002b}) is equivalent to the variable-mass Euler' (\ref{7288}) for steady flows with
  the condition that the velocity and pressure are given by (\ref{0003a}) and (\ref{0003b}),
  respectively. A Q1 flow is a flow with variable mass, that is steady, irrotational, inviscid,
  and compressible.}

A Newtonian fluid is defined by four equations: 1) continuity, 2) Navier Stokes, 3) energy and
4) an equation of state for the pressure \cite{Currie}. For Q1 flows, there is a generalized continuity
equation and a fluid that satisfies the Euler equation is a special case of one that satisfies
the Navier Stokes \cite{Currie}. However, there is no equation of state of the pressure, as in
the pressure as a function of the density that is used for  an isentropic flows an isentropic
mediums \cite{Munson, Pierce}.  Instead, Q1 flow has the pressure as a functioan of the
\emph{Laplacian} of the density $\rho$ (\ref{0003b}), where the actual form of the function $p$
depends on the ``formula'' for the density $\rho$. An energy equation has been derived for Q1
flow \cite{Finley2}, but it does not provide any useful information. Henceforth, we ignore any
thermodynamic properties a Q1 state may have, and see if this approach leads to any short
comings. For example, we do not define an isentropic process for Q1 fluid particles.


For later use, we write (\ref{0002b}) and (\ref{p4880}) in a more condensed form
\begin{gather}
   \zlabel{2492}  \fc12 m u^2 + P\rho^{-1} = \bar{E},
  \\ \zlabel{2392R}
  \rho^{-1}\pa{}{t}(\rho_m\bu) + \fc12 m\nabla u^2 + \nabla(P\rho^{-1}) = \mathbf{0};
\end{gather}
where the effective pressure $P$ is defined $P = p + q\rho \Phi$, which for all $\mr\in\mathbf{R}^3$
such that  $\rho(\mr) \ne 0$, can be written
\begin{equation}
 \zlabel{2392-def} P\rho^{-1} = p\rho^{-1} + q\Phi,
\end{equation}
and $P\rho^{-1}$ is the specific effective compression-energy or the (total) specific potential-energy.
For the hydrogen 1s flow, $P\rho^{-1}$ can be obtained by combining (\ref{2392-def}) and
(\ref{p0880})
with $q = -e$:
\begin{equation} \zlabel{p0422}
P\rho^{-1} = -\fc{Z^2\hbar^2}{ma_0^2}
\end{equation}
Hence, the hydrogen ground-state has both a constant specific potential-energy $P\rho^{-1}$ and,
from (\ref{p0404}), a constant specific kinetic-energy $mu^2/2$.






\section{The speed of sound \zlabel{1084c}}

\subsection{Setting up the problem \zlabel{1084c-1}}


In this section we derive the speed of sound for a spherical, acoustic wave-pulse propagating
along a streamline $L$ of an ambient state of Q1 flow with a spherical flow geometry and a
uniform fluid speed $|u|$. The restriction of the velocity field $\bu$ to $L$ is $\bu|_L =
u\hat{\mz}$, where $\hat{\mz}$ is a unit tangent vector of the streamline in an arbitrary
direction, and $u$ is the (one-dimensional) velocity of the fluid in the streamline, the
streamline velocity.  To reduce clutter, henceforth, we suppress $|_L$ on $\bu|_L$, and write
$\bu = u\hat{\mz}$ for the velocities on the point set $L$, called the $\bu$-velocity
streamline.

Consider three coordinate frames. 1) {\it The static coordinate frame.} A coordinate frame such
that the ambient state satisfies the variable-mass Euler equation, Eq.~(\ref{7288}),
\emph{without} the time derivative. In this frame, the velocity of fluid elements on the
$\bu$-velocity streamline is, of course, $\bu$. 2)~{\it The fluid velocity frame.}  An inertial
coordinate frame that moves along the $\bu$-velocity streamline with velocity $\bu$ relative to
the static frame.  In this frame, the velocity of fluid elements in the $\bu$-velocity
streamline is, of course, $0$.
\begin{quote}
  Let the velocity of the wave pulse in the streamline $L$ in the fluid-velocity frame be
  $-\mathbf{s}|_L = -s\hat{\mz}$. Hence, the velocity of the wave pulse in the static
  coordinate frame is $u\hat{\mz} - s\hat{\mz}$, where $|s|$ is the speed of
  sound, and we do not require $s\ge0$.
\end{quote}
3) {\it The wave-pulse frame.}  A coordinate frame that moves with the wave pulse.  The
\emph{fluid} velocity in the wave-pulse frame is $\mathbf{s} = s\hat{\mz}$, right in front of
the wave pulse, and $\mathbf{s} = (s + d\bar{u})\hat{\mz}$, right behind the wave pulse. All
three coordinate frames have a Cartesian coordinate axis that contains the $\hat{\mz}$ unit
vector. The difference between the velocity right behind the wave pulse, and the velocity right
in front of the wave pulse, denoted $d\bar{u}$, is the same from all three frames. It not
necessary to know or to determine $d\bar{u}$, because it is eliminated below by combining the
momentum-balance and continuity equations.



In order to derive a solvable equation, we only consider the special case where the speed of
sound $|s|$ is constant. With this additional constraint, the wave-pulse coordinate frame, like
the the other two frames, is inertial.  For the static frame, let the parameter $z$ be the
position along the $\bu$-velocity streamline; hence, each vector-field and scalar-field
restriction to the $\bu$-velocity streamline is a function of $z$.  At time $t$, let $z_1$ and
$z_2$ be the positions right in front and right behind the wave pulse, respectively.  Except for
the velocity field $\bu$, the values of the pertinent fields for the velocity- and wave-pulse
frames are equal to, and can be obtained from, the static frame, including velocity
\emph{differences} between two point on the $\bu$-streamline.
The later convenience, unit vector $\hat{\mathbf{w}}$ is defined by $\hat{\mathbf{w}} =
-\hat{\mz}$.
\FIG{wavepulse}{A representation of a spherical wave pulse and some of the functions needed
  for an ambient state with a Q1 flow and a corresponding classical potential-free
  flow.\zlabel{wavepulse}}

Figure~\ref{wavepulse} gives a summary of the state of affairs for a wave pulse where the
ambient state is either a Q1 flow or the corresponding classical potential-free flow.  The
speed of sound equations for the corresponding classical potential-free flows can be obtained
from the Q1 flow equations by replacing $d(P\rho_m^{-1})$ with $\rho_m^{-1}dp$ and the constant
$\hbar$ with zero.  For a classical ambient state, if the restrictions of $\rho$ and $p$ to the
streamlime on the interval $(z_1,z_2)$ are strictly increasing and the ambient state is static,
then the direction of the wave pulse is $\hat{\mathbf{w}} = -\hat{\mz}$.

Next we derive and combine both the continuity and momentum-balance equations for an arbitrary
inertial coordinate frame with a fluid velocity $\bar{\bu}_\pm = \bar{u}_\pm\hat{\mz}$, where
final equations also holds in the special where the wave pulse is absent.  The combined general
equation is applied to the wave pulse frame by setting $\bar{u}_\pm = s_\pm$, in other words,
$\bar{\bu}_\pm|_L = \mathbf{s}_\pm|_L$, yielding equations containing $s_\pm$, where $|s_\pm|$
is the speed sound. Note the the steady flow case is covered by setting $\bar{u}_\pm = u_\pm$ and
by removing the time dependent terms.

In the derivations that follow, for a wave pulse propagating on an ambient state, we assume
that the time-dependent linear-momentum balance equation (\ref{2392R}), a generalization of
(\ref{4838}), holds; we also assume that the time-dependent continuity equation (\ref{cont2-time}), a
generalization of (\ref{cont2}), holds. 

\subsection{The Momentum Balance equation \zlabel{1084c-2}}

In order to improve efficiency, imperative sentences are used for some
derivations that start and end with a large vertical line.

\lbar Use the definition $\partial_t = \partial/\partial t$. After dividing the momentum
balance equation (\ref{2392R}) by $m$, take the dot product of the resulting equation with
$d\mz = \hat{\mz}\,dz$, and then rearrange:
\begin{gather} \notag
  \rho^{-1}\partial_t(\rho\bar{\bu}\cdot d\mz) + \nabla(P\rho_m^{-1})\cdot d\mz + \fc12(\nabla\bar{u}^2)\cdot d\mz  = 0
  \\ \zlabel{p0850}
  \rho^{-1}\partial_t[\rho(z)\bar{u}(z)] \, dz + d(P\rho_m^{-1}) + \bar{u}(z)\,d\bar{u} = 0
\end{gather}
Use
\[
  \fc{\partial_t[\rho(z)\bar{u}(z)]}{\rho\bar{u}(z)} =
  \fc{\partial_t[\bar{u}(z)]}{\bar{u}(z)}\, dz + \fc{\partial_t\rho(z)}{\rho(z)}\, dz
\]
and solve (\ref{p0850}) for $-d\bar{u}(z)$:
\begin{equation} \zlabel{p7342} 
  -d\bar{u} = \fc{d(P\rho_m^{-1})}{\bar{u}} + \fc{\partial_t\bar{u}}{\bar{u}}\, dz + \fc{\partial_t\rho}{\rho}\, dz,
\end{equation}
Set $\bar{u}=\bar{u}_\pm$ and multiply the result by the factor
$-\rho_m\bar{u}_\pm/d\rho_m$ to put it into a more convenient form:
%
%
\begin{equation} \zlabel{p7321} 
  \rho_m\bar{u}_\pm \fc{d\bar{u}_\pm}{d\rho_m} = -\rho_m\fc{d(P\rho_m^{-1})}{d\rho_m} - \rho_m \partial_t\bar{u}_\pm \, \fc{dz}{d\rho_m}
  - \bar{u}_\pm\partial_t\rho_m\, \fc{dz}{d\rho_m} \rbar 
\end{equation}

For later use (\ref{p7342}) is rearranged:
\begin{equation} \zlabel{p7352} 
  \bar{u} = -\fc{d(P\rho_m^{-1})}{d\bar{u}} - \fc{\partial_t\bar{u}}{d\bar{u}}\, dz - \bar{u}\rho^{-1}\fc{\partial_t\rho}{d\bar{u}}\, dz
\end{equation}

\subsection{The continuity equation for spherical flow \zlabel{1084c-3}}

An \emph{orthonormal control region or volume} is a connected and closed region of space
$\mathbb{R}^3$ that is bordered by a set of disjoint surfaces $\{S_i\}$, where $i\in I$ and $I$
is an index set, such that two elements of $\{S_i\}$, $S_f$ and $S_b$, have unit normal-vectors
$\hat{\mathbf{n}}$ that are anti-parallel and parallel to the streamline unit-vectors,
respectively; also, the other $S_i$ remaining members of $\{S_i\}$ have normal vectors that are
perpendicular to any subset of a streamline contained in $\{S_i\}$.  The elements $S_f$ and
$S_b$ are said to be the front- and back-doors of the orthogonal control volume. A streamline
enters the control volume through the front door and exits the backdoor.

Let $f$ be a scalar field; let $\mathbf{F}$ be a vector field. A steady flow is one dimensional if and
only if the restriction of the scalar fields $f$ and $\mathbf{F}\cdot\hat{\mz}$ to a door is a constant function,
where $\hat{\mz} = \bu/|\bu|$ and $\bu$ is the velocity vector field. Spherical flow is one
dimensional flow.

\lbar Consider the orthonormal control volume represented in Figure~\ref{wavepulse}, where, for
spherical flow, each streamline is a signed ray, and all rays begin at the same starting point.
Let $z = 0$ be the center of the spherical flow, the starting point. The area $A(z_1)$ of the
frontdoor $S(z_1)$ is $az_1^2$, where $a \le 4\pi$. Since, later the limit $z_2 \to z_1$
is taken, make a first-order approximation for the area $A(z_2)$ of backdoor $S(z_2)$:
\begin{equation} \zlabel{0100}
  A(z_2) = A(z_1) + dA, \quad A = az_1^2, \quad dA = 2az_1\, dz
\end{equation}
The volume $V(z)$ of the fluid between the origin and front door $S(z_1)$ is
$(a/3)z_1^3$. Make a first-order approximation for the volume of the system
between the origin and the backdoor $S(x_2)$:
\begin{equation} \zlabel{0980}
V(z_2) = V(z_1) + dV,\quad V(z_1) = \fc{a}{3}z_1^3  \quad dV = az_1^2\, dz \rbar
\end{equation}
Hence, $dV = V(z_2) - V(z_1)$ is the approximate volume of the control volume.

\lbar Next compute the individual terms from (\ref{cont2-time}), which, for later convenience, is written
\begin{gather} \zlabel{cont3-time} 
 \fc{1}{A(z)}\int_V\partial_t\rho_m \; dV + \fc{1}{A(z)}\int_S\rho_m\bar{\bu}_\pm\cdot\hat{\mathbf{n}} \; dA
 - \fc{\al_{\pm}}{A(z)}\int_S \nabla\rho\cdot\hat{\mathbf{n}} \; dA = 0\lsc \\
 \zlabel{alph} \al_\pm = \pm\fc{\hbar}{2} 
\end{gather}
Let $\bu = \bu_\pm$. For the integral from the second term on the lhs, separate it into two
parts:
\begin{gather} 
  \zlabel{0102}
  \int_S \rho_m \bar{\bu}\cdot\hat{\mathbf{n}} \; dA
   = \int_{S(z_2)} \rho_m \bar{\bu}\cdot\hat{\mathbf{n}} \; dA + \int_{S(z_1)} \rho_m \bar{\bu}\cdot\hat{\mathbf{n}} \; dA 
\end{gather}
Treat the front door term of the rhs of this equation in two steps:
\begin{gather} \notag
  \int_{S(z_1)} \rho_m \bar{\bu}\cdot\hat{\mathbf{n}} \; dA = [\rho_m\bar{\bu}](n_1)\cdot\hat{\mathbf{n}}(n_1)\, A(z_1) 
  = -[\rho_m\bar{\bu}](z_1)\cdot\hat{\mathbf{z}}(z_1)\, A(z_1)
  \\ \zlabel{0104}
\int_{S(z_1)} \rho_m \bar{\bu}\cdot\hat{\mathbf{n}} \; dA = -\rho_m(z_1)\bar{u}(z_1)A(z_1)
\end{gather}
In a two-step evaluation of the backdoor term on the rhs of (\ref{0102}), neglect the
second-order terms and the product of differentials:
\begin{gather*} 
  \int_{S(z_2)} \rho_m \bar{\bu}\cdot\hat{\mathbf{n}} \; dA = [\rho_m\bar{\bu}](n_2)\cdot\hat{\mathbf{n}}(n_2)\, A(z_2)
  = [\rho_m\bar{\bu}](z_2)\cdot\hat{\mathbf{z}}(z_2)\, A(z_2) \\
  = (\rho_m(z_1)+ d\rho_m)(\bar{u}(z_1) + d\bar{u})(A(z_1) + dA) \\
  = [\rho_m(z_1)\,d\bar{u} + \bar{u}(z_1)\,d\rho_m + \rho_m(z_1)\bar{u}(z_1)](A(z_1) + dA)
\end{gather*}
\begin{gather} \zlabel{0108} 
  \int_{S(z_2)} \rho_m \bar{\bu}\cdot\hat{\mathbf{n}} \; dA
  = \hspace{50ex} \\ \notag \hspace{8ex}
  \rho_m(z_1)A(z_1)\,d\bar{u} + \bar{u}(z_1)A(z_1)\,d\rho_m + \rho_m(z_1)\bar{u}(z_1)A(z_1) + \rho_m(z_1)\bar{u}(z_1)\,dA
\end{gather}

Substite (\ref{0104}) and (\ref{0108}) into (\ref{0102}), set $z_1 = z$, divide
by $A(z)$, and use $dA/A(z) =2\, dz/z$ from (\ref{0100}):
\begin{equation} \zlabel{0110} 
  \fc{1}{A(z)}\int_S \rho_m \bar{\bu}\cdot\hat{\mathbf{n}} \; dA  =
  \rho_m(z)\,d\bar{u} + \bar{u}(z)\,d\rho_m + 2\rho_m(z)\bar{u}(z)z^{-1}\,dz
\end{equation}

For the last term on the lhs of (\ref{cont3-time}), first set $\rho(\mr) =1 $ in (\ref{0110}):
\[
   \fc{1}{A(z)}\int_S \bar{\bu}\cdot\hat{\mathbf{n}} \; dA  = \,d\bar{u} + 2\bar{u}(z)z^{-1}\,dz, \quad \rho = 1 \\
\]
To obtain the desired result, substitute the values
\[
 \bar{\bu} = \nabla\rho, \qquad \bar{u} = [\nabla\rho\cdot\hat{\mz}](z), \qquad d\bar{u} = d(\nabla\rho\cdot\hat{\mz}) 
\]
into the previous equation and multiply the equation by $-\al_\pm$:
\begin{equation} \zlabel{0120}
-\fc{1}{A(z)}\al_\pm\int_S \nabla\rho\cdot\hat{\mathbf{n}} \; dA  = -\al_\pm d(\nabla\rho\cdot\hat{\mz})
 - 2\al_\pm [\nabla\rho\cdot\hat{\mz}](z)z^{-1}\,dz
\end{equation}

For the first term on the lhs of (\ref{cont3-time}), treat the integral using
the mean value theorem, approximating the average value $[\partial_t\rho_m](z^*)$ of the
integrand over the volume $dV$ by $[\partial_t\rho_m](z_1)$, also use (\ref{0980}) for $dV$ and set $z_1 = z$:
\begin{equation} \zlabel{0140}
\fc{1}{A(z_1)}\int_V\partial_t\rho_m \; dV = \partial_t\rho_m(z_1)\, \fc{dV}{A(z_1)} = \partial_t\rho_m(z_1)\fc{az_1^2\, dz}{az_1^2}
  = \partial_t\rho_m(z)\,dz
\end{equation}
Obtain the satisfaction of (\ref{cont3-time}) by adding the rhs of (\ref{0110}), (\ref{0120}),
and (\ref{0140}), and setting the sum to zero:
\begin{gather} \notag
  \rho_m(z)\,d\bar{u} + \bar{u}(z)\,d\rho_m - \al_\pm d(\nabla\rho\cdot\hat{\mz}) + \beta\, dz = 0;
  \\ \zlabel{beta}
 \beta(z) = 2\rho_m(z)\bar{u}(z)z^{-1} + \partial_t\rho_m(z) - 2\al_\pm [\nabla\rho\cdot\hat{\mz}](z)z^{-1} 
\end{gather}
Obtain the desired result by solving for $-d\bar{u}$:
\begin{equation*}
-\,d\bar{u} = \rho_m^{-1}(z)\bar{u}(z)\,d\rho_m - \al_\pm \rho_m^{-1}(z)d(\nabla\rho\cdot\hat{\mz}) + \rho_m^{-1}(z)\beta\, dz 
\end{equation*}
Set $\bar{u}_\pm= \bar{u}$ and multiply it by the factor $\rho_m \bar{u}_\pm/d\rho_m$ to put it into a more convenient form:
\begin{equation} \zlabel{p4928}
  -\,\rho_m\bar{u}_\pm \fc{d\bar{u}_\pm}{d\rho_m} = \bar{u}_\pm^2 - \al_\pm \bar{u}_\pm\fc{d(\nabla\rho\cdot\hat{\mz})}{d\rho_m}
  + \beta\bar{u}_\pm\,\fc{dz}{d\rho_m} \rbar
\end{equation}
For uniform flow, (\ref{p4928}) still holds with
\begin{equation} \zlabel{beta-unif}
\beta = \partial_t\rho_m(z), \quad \text{uniform flow}
\end{equation}
This follows because $dA = 0$ for uniform flow, and since $dA/A(z) = 2\, z^{-1}\,dz$, the terms
with $z^{-1}$ factor do not appear for uniform flow.

%

\subsection{The speed of sound quadratic equation \zlabel{1084c-4}}

\lbar Next eliminate $d\bar{u}$ from (\ref{p4928}) and (\ref{p7321}) by adding the two equations in two steps:
\begin{gather} \notag
  \bar{u}_\pm^2 - \al_\pm \bar{u}_\pm\fc{d(\nabla\rho\cdot\hat{\mz})}{d\rho_m} +\beta\bar{u}_\pm\,\fc{dz}{d\rho_m}
  - \rho_m\fc{d(P\rho_m^{-1})}{d\rho_m} - \rho_m \partial_t\bar{u}_\pm \, \fc{dz}{d\rho_m} - \bar{u}_\pm\partial_t\rho_m\, \fc{dz}{d\rho_m}  = \mathbf{0}
  \\ \zlabel{u-quad}
  \bar{u}_\pm^2 - \al_\pm \fc{d(\nabla\rho\cdot\hat{\mz})}{d\rho_m}\bar{u}_\pm - \rho_m\fc{d(P\rho_m^{-1})}{d\rho_m} 
  +  \gamma\fc{dz}{d\rho_m}  = \mathbf{0}\lsc \\ \zlabel{gamma}
    \gamma = \beta\bar{u}_\pm - \rho_m \partial_t\bar{u}_\pm - \bar{u}_\pm\partial_t\rho_m \rbar
\end{gather}
%

Consider the equation obtain by setting $\bar{u}_\pm = s_\pm$ in (\ref{u-quad}). This equation
represents the case of a wave pulse contained in, and moving with, the control volume.  The
function $\rho_m$ is an independent variable at our disposal with the constraint that
$\rho_m(z_1)$ is taken from the ambient state.  Let $\rho_m$ be the limit of a sequence
$\phi_1,\phi_2,\cdots$ of continuous function that converge \emph{point wise} such that
$\rho_m$ has a jump discontinuity at $z_1$ and $\rho_m(z_2) = \lim_{z\to z_1^+} \phi(z)$.
Hence, $\lim_{n\to\infty} d\phi_n(z_1)/dz = \infty$, giving $\lim_{n\to\infty} dz/d\phi_n(z_1)
= 0$, and (\ref{u-quad}) becomes
\begin{equation} \zlabel{p0024}
  s_\pm^2 - \lt(\pm\fc{\hbar}{2}\rt)\fc{d(\partial\rho)}{d\rho_m}s_\pm - \rho_m\fc{d(P\rho_m^{-1})}{d\rho_m}  = \mathbf{0},
\end{equation}
where the following notation is used for the directional derivative $\nabla f\cdot\hat{\mz}$ in
the direction of the unit vector $\hat{z}$:
\begin{equation} \zlabel{4320}
\partial f \:\dot{=} \nabla f\cdot\hat{\mz},
\end{equation}
where, in this case, $f = \rho$.  The speed of sound $|s|$ can be obtained from the quadratic
formula applied to (\ref{p0024}).
(Substituting (\ref{beta}) for $\beta$ into (\ref{gamma}), I got
$\gamma =
2\rho_m\bar{u}^2z^{-1} - 2\al_\pm {[\nabla\rho\cdot\hat{\mz}]}z^{-1}\bar{u} - \rho_m
\partial_t\bar{u}$, and this simplifies to $\gamma = \partial_t\bar{u}$ for uniform flow.)

Equation (\ref{u-quad}) also holds for uniform flow with the $\beta$ formula (\ref{beta})
replaced by (\ref{beta-unif}), and since $\beta$ does not appear in (\ref{p0024}), this
equation also holds for uniform flow. Furthermore, (\ref{p0024}) probably holds for a variety of
flow geometries, with a similar limit taken that removed the $\gamma$ term from (\ref{u-quad}).




\subsection{The hydrogenic 1s state \zlabel{p7290}}




Since, for hydrogenic atoms in the 1s state, the specific effective-compression-energy
$P\rho^{-1}$ (\ref{p0422}) is constant,
\[
\fc{d(P\rho_m^{-1})}{d\rho_m} =\fc{d(P\rho^{-1})}{d\rho} = \mathbf{0}
\]
and (\ref{p0024}) reduces to
\begin{equation} \zlabel{p0225}
  s_\pm^2 - \lt(\pm\fc{\hbar}{2}\rt)\fc{d(\partial\rho)}{d\rho_m}s_\pm = \mathbf{0}
\end{equation}
A solution is $s_\pm = 0$ for both velocity directions. This is Mach $\infty$. Hence, if
$P\rho^{-1}$ as a function of $\rho$ is a constant, then the fluid has Mach speed $\infty$.
Such a flow is ``silent'' since it cannot propagate a wave pulse. A disturbance is only
carried downstream with the fluid velocity.

\lbar For the second solution, divide (\ref{p0225}) by $s$:
\[
s_\pm  = \pm\fc{\hbar}{2}\fc{d(\partial\rho)}{d\rho_m}
\]
To evaluate the expression use $\rho$ from (\ref{p5902}) and notation (\ref{4320}) with the choice $\hat{\mz} = \hat{\mr}$,
where $\hat{\mr}$ is the radial spherical-coordinate:
\[
s_\pm = \pm\fc{\hbar}{2}\fc{d(\nabla \rho\cdot\hat{\mr})}{d\rho_m}
= \pm\fc{\hbar}{2m}\fc{d}{d\rho}(\partial_r\rho) = \pm\fc{\hbar}{2m}\fc{d}{d\rho}\lt(-\fc{2Z}{a_0}\rho\rt)
= \mp\fc{Z\hbar}{ma_0} = u_\pm\rbar
\]
Interpretations of this result are given at the end of the next subsection and in (\ref{p0822a}).

\subsection{Another speed of sound quadratic equation. \zlabel{7202}}


In order to apply (\ref{p0024}), it is assumed that all the functions from this equation can be
taken from the ambient medium without the wave pulse. Let $X$ be a scalar field that can be
taken from the ambient medium.  For many of the derivation and calculation that follow, it is
useful to be aware that, we have
\begin{equation} \zlabel{p8372}
  \fc{dX}{d\rho} = \fc{\nabla X\cdot d\mz}{\nabla \rho\cdot d\mz}
  = \fc{\nabla X\cdot d\hat{\mz}}{\nabla \rho\cdot \hat{d\mz}}
  = \fc{(\partial X)}{\partial \rho}
\end{equation}
and notation (\ref{4320}) is used.  Note that the brackets $(\cdots)$ matter: $(\partial
X)/\partial \rho$ is the ratio of directional derivatives and the \emph{not} the same as the
partial derivative $\partial X/\partial \rho$.

In order to assign another meaning to $dX/d\rho$, besides being a ratio of differentials of the
variable $\mr\in \mathbb{R}$, consider a subspace $S\subset \mathbb{R}^3$ such that both $\rho$
and $X$ are strictly monotone. In that case, the map $\rho \rightarrow X$ exists with domain $W
= \text{Range}\lt(\rho|_S\rt)$. Hence, the derivative $dX/d\rho$ with domain
$\text{Range}\lt(\rho|_S\rt)$ exists, and it satisfies (\ref{p8372}). This also holds for other
scalar fields besides $\rho$, in particular the streamline velocity $u_\pm$.

In the special case of $X = \partial\rho$, from (\ref{p8372}), we obtain
\begin{equation} \zlabel{4870}
\fc{d(\partial\rho)}{d\rho} = \fc{(\partial^2\rho)}{\partial \rho}
\;\dot{=}\;\fc{(\partial(\partial\rho))}{\partial \rho}
\end{equation}
where the last statement defines $(\partial^2\rho)$. Examples of the use of this notation is
obtained from speed of sound equation (\ref{p0024}) and the definition (\ref{0003a}) of
$\bu_\pm$:
\begin{gather}
\zlabel{p0025}
  s_\pm^2 - \lt(\pm\fc{\hbar}{2}\rt)\fc{(\partial^2\rho)}{\partial\rho_m}s_\pm - \rho_m\fc{d(P\rho_m^{-1})}{d\rho_m}  = \mathbf{0} \\
\zlabel{4880}
u_\pm = \bu_\pm\cdot\hat{\mz} = \pm\fc{\hbar}{2m}\rho^{-1}\nabla\rho\cdot\hat{\mz} = \pm\fc{\hbar}{2m}\rho^{-1}\partial\rho
\end{gather}
%
For use below, we obtain an equality by taking the open-density derivative of this equation where $u = u_+$:
\begin{gather*}
  \fc{du}{d\rho} = \fc{\hbar}{2m}\fc{d}{d\rho}\lt(\rho^{-1}\partial\rho\rt)
  = \fc{\hbar}{2m}\fc{d(\rho^{-1})}{d\rho}\partial\rho + \fc{\hbar}{2m}\rho^{-1}\fc{d(\partial\rho)}{d\rho} =
  -\lt(\fc{\hbar}{2m}\rho^{-1}\partial\rho\rt)\rho^{-1} + \rho^{-1}\fc{\hbar}{2m}\fc{d(\partial\rho)}{d\rho} 
\end{gather*}
The use of (\ref{4880}) and (\ref{4870}) gives the result
\begin{equation} \zlabel{4025}
  \fc{du}{d\rho} = -u\rho^{-1} + \rho^{-1}\fc{\hbar}{2m}\fc{(\partial^2\rho)}{\partial\rho},
  \quad u = u_+
\end{equation}

In order to find another speed of sound equation, we derive a relationship between the two
density derivatives in (\ref{p0025}).  Solving (\ref{2492}) for $P\rho^{-1}$, followed by
taking the density derivative and the use of (\ref{4025}), with $u = u_+$, gives
\begin{gather*}
  \fc{d(P\rho^{-1})}{d\rho} = -\fc{d}{d\rho}\lt(\fc12mu^2\rt) = -um\fc{du}{d\rho}
    = u^2m\rho^{-1} -um\rho^{-1}\fc{\hbar}{2m}\fc{(\partial^2\rho)}{\partial \rho}
\end{gather*}
Hence,
\[
\rho\fc{d(P\rho^{-1})}{d\rho} = u_+^2m - \fc{\hbar}{2}\fc{(\partial^2\rho)}{\partial\rho}u_+
\]
Substituting $u_+ = -u_-$ gives another equation. The combination of the two equations can be written
\begin{equation}  \zlabel{7352} 
  -\rho_m\fc{d(P\rho_m^{-1})}{d\rho_m} = \lt(\pm\fc{\hbar}{2}\rt)\fc{(\partial^2\rho)}{\partial\rho_m}u_\pm - u_\pm^2
\end{equation}
For later use, it can also be written
\begin{equation}  \zlabel{p8352}
  \pm\fc{\hbar}{2}\fc{(\partial^2\rho)}{\partial\rho_m} = -u_\pm^{-1}\rho\fc{d(P\rho^{-1})}{d\rho_m} + u_\pm  
\end{equation}
Substituting (\ref{7352}) 
into (\ref{p0025})
we obtain the desired equality:
\begin{equation} \zlabel{8350n}
  s_\pm^2 - \lt(\pm\fc{\hbar}{2}\rt)\fc{(\partial^2\rho)}{\partial\rho_m}s_\pm
  + \lt(\pm\fc{\hbar}{2}\rt)\fc{(\partial^2\rho)}{\partial\rho_m}u_\pm - u_\pm^2
   = \mathbf{0}
\end{equation}
For later use, we write down the quadratic formula for this equation:
\begin{gather} \zlabel{8352n} 
  2s_\pm = -b_\pm \pm \sqrt{b_\pm^2 - 4C};
  \\ \zlabel{p8358n}  b_\pm = -\lt(\pm\fc{\hbar}{2}\rt)\fc{(\partial^2\rho)}{\partial\rho_m},
\qquad C = -b_\pm u_\pm - u_\pm^2 
\end{gather}
By inspection, it is immediate obvious that one solution of (\ref{8350n}) is $s_\pm = u_\pm$.

For the case of a corresponding classical potential-free flow, (\ref{p0850}) and
(\ref{cont2-time}) hold with $d(P\rho_m^{-1})$ replaced by $\rho_m^{-1}dp$ and $\hbar$ replaced
by zero, giving, from (\ref{p0025}), the well known result that was derived for a static ambient
state with one-dimensional flow \cite{Munson,Pierce}:
\begin{equation} \zlabel{p8370n}
s = \pm\sqrt{\fc{dp}{d\rho_m}}
\end{equation}
Hence, (\ref{p0025}) is a generalization of (\ref{p8370n}). However, (\ref{8350n}) does not
have a classical corresponding equation with gravity neglected, because it is derived from
(\ref{p0025}) using the velocity definition (\ref{0003a}),
implying variable mass. Hence, for a wave pulse, $s = u$ is not necessarily a solution of
(\ref{p8370n}).  In the special case where $s = u$ is a solution of (\ref{p8370n}), the other
solution must be $s = -u$, and this is Mach 1. These two solutions $s = \pm u$ also satisfy
(\ref{8350n}) in the so-called classical limit, where $\hbar$ is replaced by zero, and for a
static fluid this gives $s = 0$.

One possible interpretations for a Q1 flow with the solution $s_\pm = u_\pm$ from (\ref{8350n})
is that this is case where there is no wave pulse, and this state can obtained by setting
$\bar{u} =u$ in (\ref{u-quad}) in the derivation. This possibility and a Mach 1 possibility, is
discussed in Section ~(\ref{p0822a}).

\subsection{The other solution of the quadratic formual for the speed of sound \zlabel{p9022}}


As shown above, $s_\pm = u_\pm$ is one of the solutions of the quadratic equation
(\ref{8350n}), and possible interpretations of this solution is discussed in
subsection~\ref{p7290}.  \lbar For the quadratic formula (\ref{8352n}), let the solution set be
$\{u_\pm,s_\pm\}$, and let $u_\pm$ satisfy the quadratic formula with the positive sign $+$ in
front of the square root.
Add the two solutions of the quadratic formula:
\begin{align}  \zlabel{p3789n} 
&\hspace{1ex} 2u_\pm = -b_\pm + \sqrt{b_\pm^2 + 4C}
  \\ \zlabel{p3788n}
  &\underline{+ \quad 2s_\pm = -b_\pm - \sqrt{b_+^2 + 4C}}
  \\ 
  &\hspace{4ex} s_\pm = -b_\pm - u_\pm \rbar
\end{align}
By inspection, this result also covers the case where $u_\pm$ satisfies the quadratic formula
with the negative sign ($-$) in front of the square root. Substituting (\ref{p8358n}) into the previous equation
gives the desired result:
\begin{equation} \zlabel{p2445n}
s_\pm = \pm\fc{\hbar}{2}\fc{(\partial^2\rho)}{\partial\rho_m} - u_\pm
\end{equation}
Also, using $-u_+ = u_-$ we discover that
\[
-s_+ = -\lt(+\fc{\hbar}{2}\fc{(\partial^2\rho)}{\partial\rho_m} + u_-\rt) = s_-
\]
Substituting (\ref{p8352}) 
into (\ref{p2445n}) gives us another formula for the speed of sound $|s|$:
\begin{equation} \zlabel{4944n}
s_\pm = -u_\pm^{-1}\rho\fc{d(P\rho^{-1})}{d\rho_m}\;|
\end{equation}

Recall that the fluid velocity on the $L$ streamline relative in the static frame is
$u_\pm\hat{\mz}$; recall that the velocity of the wave pulse in the fluid velocity frame is
$-s_\pm\hat{\mz}$.  In other words, the one-dimensional velocity of the wave pulse on the $L$
streamline is $-s_\pm\hat{\mz}$ \emph{relative} to the velocity of the fluid on the $L$
streamline. With this in mind, and utilizing (\ref{4944n}),
let the signed Mach speed $\Ms$, a generalization of the the Mach speed $\Ma$, be
defined by
\begin{equation} \zlabel{p0370}
\Ms = \fc{u_\pm}{-s_\pm\hspace{1ex}} = u^2\lt(\rho\fc{d(P\rho^{-1})}{d\rho_m}\rt)^{-1},
\end{equation}
and note that $\Ms$ is nonnegative, if and only of $d(P\rho^{-1}/d\rho_m)$ is nonnegative.
This formula on the rhs does \emph{not} hold for the flow with the solution ($s_\pm = u_\pm$)
of the quadratic equation (\ref{8350n}). However, if we set $\Ms = u_\pm/(-s_\pm)$ for this
case, then $\Ms = -1$. Also, if $\Ms \ge 0$, we then set $\Ms = \Ma$.

According to Equation~(\ref{4944n}) for a specified direction of a velocity fluid $u_\pm$, the direction of
the wave pulse in the velocity frame is determined by the sign of $d(P\rho^{-1})/d\rho_m$,
This issue is discussed in Section~(\ref{p0822c}).

Let $f$ be a scalar field.  Consider uniform flow with the velocity vector direction along the
$\hat{\mz}$ axis. The Laplacian $\nabla^2f$ and second directional derivative $(\partial^2 f)$
of $f$ are then equal. For example, $f = \rho$ gives
\begin{equation} \zlabel{4939}
(\partial^2\rho) \:\dot{=}\;\nabla(\nabla\rho\cdot\hat{\mz})\cdot\hat{\mz} = \partial^2\rho/\partial z^2  = \nabla^2\rho
\end{equation}
If $p(z) = 0$, then (\ref{0003b}) and the above equality gives $[\partial^2\rho](z) = 0$.
Combining this result with (\ref{p2445n}) and (\ref{p0370}) gives the following: Uniform flow
and $p(z) = 0$ implies $\Ms = \Ma = 1$.

\section{Effect of Variation in flow cross-sectional area \zlabel{p9542}}


The following equation is applicable on a streamline of the corresponding classical,
one-dimensional potential-free flows \cite{Munson}:
\begin{gather}
  \zlabel{p5040} 
  \fc{du}{u} = -\fc{dA}{A}\fc{1}{\lt(1 - \text{Ma}^2\rt)}\lsc
\end{gather}
$A$ is the variable cross sectional area.  This equation indicates that an area increase along
the streamline results in a decrease in fluid speed, for subsonic flow ($\text{Ma} < 1$), and
an increase in fluid speed, for supersonic flow ($\text{Ma} > 1$). A rearrangement of this
equation for $du\ne 0$,
\begin{equation} \zlabel{p7210}
  \fc{dA}{du} = -\fc{A}{u}\lt(1 - \text{Ma}^2\rt), 
\end{equation}
indicates that Mach 1 is an extremum for the cross sectional area $A$ in any region on the
streamline where $A$ is a function of the the velocity $u$. Both Eqs.~(\ref{p5040}) and
(\ref{p7210}) are important for characterizing duct flow \cite{Munson}. For example, if a duct
is sufficiently long, a converging-diverging duct will obtain sonic speed (Ma = 1) at the
minimum cross sectional area, called the throat, for either subsonic or supersonic flow.

The above two equations are, of course, applicable for spherical flow.

In this section we derive equations that are analogous to (\ref{p5040}) and (\ref{p7210}) for
Q1 flows: We follow the procedure explained by Munson, Young, and Okiishi \cite{Munson} in
their derivation of (\ref{p5040}) and (\ref{p7210}), deviating as little as possible, but
using, of course, the analogous equations for Q1 flows. At key points of the derivation, the
derived equations for Q1 flow are compared to the corresponding classical, one-dimensional
potential-free flow equations.

Let $u^2\;\dot{=}\, u_-^2 = u_+^2$.  Let the restriction of the velocity field $\bu$ to the
streamline $L$ be $u\hat{\mz}$, where $\hat{\mz}$ is the unit tangent vector of the streamline
in the direction of the velocity field.  Taking the dot product of the momentum balance
equation (\ref{2392R})
with $d\mz = \hat{\mz}\, dz$, for steady flow, yields $mu_\pm\,du_\pm + d(P\rho^{-1}) =
\mathbf{0}$, or $u_\pm\,du_\pm + d(P\rho_m^{-1}) = \mathbf{0}$, which, for $u\ne 0$, can be
written
\begin{align} \zlabel{p4920}
  \fc{d(P\rho_m^{-1})}{u^2} &= -\fc{du_\pm}{u_\pm}, \quad \text{quantum 1 flow} \\ \notag
  \fc{\rho_m^{-1}dp}{u^2} &= - \fc{du}{u}, \quad \text{classical flow,}
\end{align}
and the second one is the corresponding equation for a potential-free classical flow: A
special case of the Q1 with $d(P\rho_m^{-1}) = \rho_m^{-1}dp$.
The continuity equation (\ref{cont2-time}) for an othonormal control region with doors $S_f$
and $S_b$ is
\[
\int_{S_b+S_f} \lt(\rho_m\bu_\pm\cdot\hat{\mathbf{n}} - \al_\pm\nabla\rho\cdot\hat{\mathbf{n}}\rt) \; dA = 0\lsc
\]
$\al_\pm$ is defined by $(\ref{alph})$.  Since $\hat{\mathbf{n}}(\mr_1) = -\hat{\mz}(\mr_1)$
and $\hat{\mathbf{n}}(\mr_2) = \hat{\mz}(\mr_2)$, for  $\mr_1 \in S_f$ and $\mr_2 \in S_b$, we have
\[
\int_{S_b} (\rho_mu_\pm - \al_\pm\partial\rho) \; dA - 
\int_{S_f} (\rho_mu_\pm - \al_\pm\partial\rho) \; dA  = 0\lsc
\]
$\partial\rho$ is the directional derivative in the direction of $\hat{\mz}$. We require
the flow to be one-dimensional, implying that
\[
[\rho_mu_\pm](\mr) = [\rho_mu_\pm](\mrp), \qquad \partial\rho(\mr) = \partial\rho(\mrp);\quad  \mr,\mr^\pr\in S_i,
\]
and these equations are satisfied in two separate cases: $i=f,b$, giving
\[
[\rho_mu_\pm A](\mr_2)  - [\al_\pm\partial\rho A](\mr_2) 
- \lt\{[\rho_mu_\pm A](\mr_1)  - [\al_\pm\partial\rho A](\mr_1) \rt\}
=  \mathbf{0}
\]
Since we are interested in the limit of $\mr_2\to\mr_1$, we approximate this equation to first order
\[
d[\rho_mu_\pm A  - \al_\pm\partial\rho A] = \mathbf{0}
\]
\lbar Expand this equation out. Divide by $\rho_mu_\pm A$. Rearrange.
\begin{align} 
\notag \fc{du_\pm}{u_\pm} + \fc{dA}{A} + \fc{d\rho_m}{\rho_m}  &= \al_\pm\fc{d(A\partial\rho)}{\rho_m Au_\pm}  \\
\zlabel{p4940}
-\fc{du_\pm}{u_\pm}  - \fc{d\rho_m}{\rho_m} &= \fc{dA}{A} + \al_\mp\fc{d(A\partial\rho)}{\rho_m Au_\pm}, \quad \text{Q1 flow} \rbar
\\ \notag
-\fc{du}{u} - \fc{d\rho_m}{\rho_m} &=  \fc{dA}{A}, \quad \text{classical flow,}
\end{align}
The second one, again, is for the corresponding classical flow: A special case with $\al_\pm = 0$.

\lbar Add equations (\ref{p4920})
and (\ref{p4940}),
and then continue with some algebraic manipulations so that the signed Mach Ms~(\ref{p0370})
can be substituted:
\begin{align*}
  \fc{d(P\rho_m^{-1})}{u^2} - \fc{d\rho_m}{\rho_m} &= \fc{dA}{A}  + \al_{\mp}\fc{d(A\partial\rho)}{\rho_m Au_\pm} \\
    \fc{d(P\rho_m^{-1})}{u^2}\lt(1 - \fc{u^2}{[d(P\rho_m^{-1})/d\rho]\rho}\rt)
    &= \fc{dA}{A} + \al_{\mp} \fc{d(A\partial\rho)}{\rho_m Au_\pm}
\end{align*}
Make the substitution 
\begin{gather}
 \zlabel{p4980}  \fc{d(P\rho_m^{-1})}{u^2}\lt(1 - \Ms\rt) = \fc{dA}{A} + \al_\mp\fc{d(A\partial\rho)}{\rho_m Au_\pm} \rbar
  \\ \notag
 \fc{\rho_m^{-1}dp}{u^2}\lt(1 - \text{Ma}^2\rt) =  \fc{dA}{A}, \quad \text{classical flow,} 
\end{gather}
The second one, again, is for the corresponding classical flow. The significant
difference between the two arises from the difference in the formulas for Ma and $\Ms$.

\lbar Substitute (\ref{p4920}) 
into (\ref{p4980}):
\begin{gather}
\zlabel{p5045} 
  \fc{du_\pm}{u_\pm} = \lt(-\fc{dA}{A} + \al_\pm\fc{d(A\partial\rho)}{\rho_m Au_\pm}\rt)\fc{1}{\lt(1 - \Ms\rt)}\rbar
\end{gather}
The corresponding classical flow equation is (\ref{p5040}), and
(\ref{p5040}) implies (\ref{p7210}).

It is our objective to obtain equations that are analogous to (\ref{p5040}) and (\ref{p7210})
for Q1 flows. In order to accomplish this objective, below we expand out and then
simplify the two terms involving the cross sectional $A$ in (\ref{p5045}).
However, before we do that, we first we obtain a useful equality by taking the dot
product of $d\mz = \hat{\mz}\, dz$ with the velocity equation (\ref{0003a}) in the form
$\bu_\pm\rho_m = \pm \al \nabla\rho$, giving, for the streamline, the momentum per volume
equation:
\begin{equation} \zlabel{p2284}
\rho_mu_\pm = \al_\pm\partial\rho,
\end{equation}
and $u_\pm\rho_m$ is also called the 'streamline momentum-density.'  Substitute this equality
into the expansion of the factor from (\ref{p5045}) involving the cross sectional area $A$, we
obtain
\begin{gather*}
  -\fc{dA}{A} + \al_\pm\fc{d(A\partial\rho)}{\rho_mAu_\pm}
    = -\fc{dA}{A} + \fc{\al_\pm\partial\rho}{\rho_mu_\pm} \,\fc{dA}{A} + \al_\pm\fc{d(\partial\rho)}{\rho_mu_\pm}
  = \al_\pm\fc{d(\partial\rho)}{\rho_mu_\pm}
\end{gather*}
Hence, the area terms cancel in (\ref{p5045}), giving
\begin{equation} \zlabel{p7250}
\fc{du_\pm}{u_\pm} =  \al_\pm\fc{d(\partial\rho)}{\rho_mu_\pm}\fc{1}{\lt(1 - \Ms\rt)} 
\end{equation} 
Hence, for the Q1 flows, we do not obtain the same type of relationship involving the
cross sectional area, briefly mentioned in the introduction to this section, and described in
detail elsewhere \cite{Munson}. Instead, next we obtain a different relationship involving the 
streamline momentum-density $\rho_m u_\pm$.

Substituting (\ref{p2284}) into (\ref{p7250}), we get
\begin{equation} \zlabel{p1120}
\fc{du_\pm}{u_\pm} =  \fc{d(\rho_mu_\pm)}{\rho_mu_\pm}\fc{1}{\lt(1 - \Ms\rt)}
\end{equation}
If $du_\pm \ne 0$, this equation can also be written
\begin{equation} \zlabel{p2222}
  \fc{d(\rho_mu_\pm)}{du_\pm} = \rho_m(1 - \Ms)
\end{equation}
Hence, the extremums of the momentum density $\rho_mu_\pm$ on a streamline occur at points
$\mr\in L$ of space where $\rho(\mr) \ne 0$ and $\Ms(\mr) = 1$.  Also, since $\Ms \ge 0$ implies
$\text{Ma} = \Ms$, $\Ms(\mr) = 1$ implies Mach 1.

Consider a point $\mr^\pr \in \mathbb{R}$ such that $\rho(\mr^\pr) = 0$. In that case, for the
flows considered in the next section, we find that the limit $u_\pm \to 0$ as $\mr\to \mr^\pr $
does not exists and that $|u_\pm| \to \infty$ as $\mr\to \mr^\pr $. Hence, the derivative
$d(\rho_mu_\pm)/du_\pm$ does not exists at points where $\rho(\mr^\pr) = 0$. Hence, extremums of
the momentum density $\rho_mu_\pm$ do not necessarily occur at such points.

{\it Definition.}  A connected surface such that the momentum density is minimum and maximum on
the streamlines are called nodes and antinodes, respectively. With the assumption mentioned
above, the nodes and antinodes are defined by $\rho(\mr) = 0$ and $\text{Ma}(\mr) = 1$, respectively.

In the special case where $u_\pm \ne 0$ and $u_\pm$ is constant, (\ref{p1120})
reduces to
\begin{gather*}
   \fc{d\rho_m}{\rho_m}\fc{1}{\lt(1 - \Ms\rt)} = \mathbf{0}
\end{gather*}
Hence, if $d\rho_m \ne 0$, $\text{Ma} = \infty$, implying that $|s_\pm| = 0$. One example of
this type of silent flow is presented in Sec.~\ref{p7290} for the hydrogenic 1s
states,

(Next we show that $u_\pm \ne 0$ implies $d\rho_m \ne 0$, so the the statement '$d\rho_m \ne
0$' above is an unnecessary condition. Let the restriction of $\rho_m$ to the streamline $L$ be
a function of the streamline coordinate $z$. With this requirement, $d\rho_m = d\rho_m(z;dz)$,
and $d\rho_m = \partial\rho_m\, dz$. The requirement $u_\pm \ne 0$ and (\ref{p2284}) implies
that $\partial\rho_m \ne 0$; $\partial \rho_m \ne 0$ and $d\rho_m = \partial\rho_m\, dz$ implies
that $d\rho_m \ne 0$.)



\section{Applications \zlabel{p3602}}

The formula for the speed of sound for classical compressible flows (\ref{p8370n}), $s =
+\sqrt{dp/d\rho_m}$, is derived with the restriction that the one-dimensional velocity $u_\pm$
and wave pulse $s_\pm$ are constant along the streamline (and the flow is either spherical or
uniform). However, this same formula is, apparently, also very successful in describing
compressible flows without these restrictions \cite{Munson,Shapiro}.  With this in mind, it is assumed
that (\ref{p2445n}), and other derived speed of sound formulas, including (\ref{4944n}), hold,
or are useful approximations, for Q1 flows.

In this section a set of Q1 flows are examined using the formulas derived in the previous sections
for many physical properties, including the speed of sound.
Flow that are neither uniform nor spherical are not considered, but I suspect that
(\ref{p2445n}) could be derived with a variety of flow geometry with the restriction of
constant $u_\pm$ and $s_\pm$. (The speed of sound equations derived actually also hold if the
sum $u_\pm + s_\pm$ is constant, but $u_\pm$ and $s_\pm$ are not constant, since the
wave pulse frame is still inertial.)


All the calculations in this section use atomic units.

\subsection{A fluid (or particle) in a one-dimensional box and ordinary Q1 flow\zlabel{p7300}}




For a particle, or fluid, in a one-dimensional box of length $\ell$, let $\ell$ be the Bohr
radius $a_0$. In one-dimensional cases, the directional derivatives reduce to ordinary
derivatives, but the general notation is still used below, with the definition
$\partial\rho\;\dot{=}\; d\rho/dr$, and $r\in \mathbf{R}$ is the position variable for this
subsection.

\rbar Start with the well known open density $\rho$ \cite{Raimes,Levine} with domain $[0,\ell]$
and quantum number ($n = 1,2,\cdots$) for the Q1 probability states and compute the uphill flow
velocity-component using (\ref{4880}), i.e., $u_\pm = \pm\partial\rho/(2\rho)$, and the
momentum density $\rho u_\pm$:
\begin{gather} \notag
  \rho(r) = 2\sin^2(n\pi r)
  \\ \zlabel{p0430}
  \partial\rho(r) = 4n\pi\sin(n\pi r)\cos(n\pi r)
  \\ \zlabel{p0530} u_\pm(r) = \pm n\pi\cot(n\pi r) 
  \\ \zlabel{p0532} \rho u_\pm(r) 
  = \pm 2n\pi\sin(n\pi r)\cos(n\pi r)
\end{gather}
Compute the first term on the rhs of the wave pulse formula $-s_+$ (\ref{p2445n}),
given by $-s_+ = -(1/2)\partial^2\rho/\partial\rho_m + u_+$:
\begin{gather}
  \zlabel{p7302} \partial^2\rho(r) = 4n^2\pi^2[\cos^2(n\pi r) - \sin^2(n\pi r)] \\
   \notag \fc{1}{2}\fc{(\partial^2\rho)}{\partial\rho_m}(r) = \fc12n\pi\lt(\fc{\cos(n\pi r)}{\sin(n\pi r)} - \fc{\sin(n\pi r)}{\cos(n\pi r)}\rt)\\
  \label{p5310}-\fc{1}{2}\fc{(\partial^2\rho)}{\partial\rho_m}(r) = -\fc12n\pi\cot(n\pi r) + \fc12n\pi\tan(n\pi r) 
\end{gather}
Note (\ref{p2445n}): Obtain the wave pulse component $-s_+$ by adding (\ref{p5310}) and
(\ref{p0530}):
\begin{gather}
  -s_+(r) = \fc12n\pi[\cot(n\pi r) + \tan(n\pi r)] 
\end{gather}
Calculate the signed Mach speed $\Ms$ by first obtaining its reciprocal:
\begin{gather}
  \notag  (\ref{p0370}): \; (\Ms)^{-1}(r) = -[s_+ u_+^{-1}](r) = \fc12[1 + \tan^2(n\pi r)] = \fc12 \sec^2(n\pi r)
  \\ \zlabel{p5240}
  \Ms(r) = 2\cos^2(n\pi r) \rbar
\end{gather}
Hence, the signed $\Ms$ is nonnegative, giving $\text{Ma} = \Ms$, and the minimum and maximum
values of $\Ms$ are 0 and 2 for all flow states, respectively.  Also, $\Ms(r) = 1$ at
$nr\in\{1/4,3/4\}$, and this agrees with the extremum of momentum density, as predicted by
Eq.~(\ref{p2222}), that occurs at points $r \in [0,1]$ such that $\partial[\rho u_+](r) =
0$, where, from (\ref{p0532}), the derivative is
\begin{gather*}
  \partial[\rho u_+](r) 
  = 2n\pi[\cos^2(n\pi r) - \sin^2(n\pi r)]  
\end{gather*}
\lbar Substitute ($\cos^2(n\pi r) = 1 - \sin^2(n\pi r)$) into (\ref{p7302}) and use (\ref{0003b}), i.e.,
($p = -\partial^2\rho/4$), to calculate the pressure:
\begin{equation} \zlabel{p5241}
    p(r) = -n^2\pi^2[1 - 2\sin^2(n\pi r)]
\end{equation}
Use $\rho^{-1}(r) = \csc^2(n\pi r)/2$ and $\csc^2 x = 1 + \cot^2 x$ to calculate the specific compression energy~$p\rho^{-1}$:
\begin{gather} \notag
    [p\rho^{-1}](r) = -\fc12n^2\pi^2[\csc^2(n\pi r) - 2] \\
  [p\rho^{-1}](r) = -\fc12n^2\pi^2\cot^2(n\pi r) + \fc12n^2\pi^2 
\end{gather}
Calculate the specific kinetic-energy $mu^2/2$ using (\ref{p0530}) and the well known total
energy $\bar{E}_n$ using the formula for the specific total-energy (\ref{2492}), given by $\bar{E}_n = m u^2/2 +
p\rho^{-1}$:
\begin{gather}
  \fc12 [mu_\pm^2](r) = \fc12n^2\pi^2\cot^2(n\pi r) \\
    \bar{E}_n = \fc12n^2\pi^2\; \rbar
\end{gather}


Recall from subsection~\ref{p9022}, after Eq.~(\ref{4939}), that it was proven that $p(r) = 0$
implies $\Ms(r) = \Ma(r) = 1$ for uniform flow. Equations.~(\ref{p5240}) and (\ref{p5241}) give
$p(r) = 0$ and $\Ma(r) = 1$ for $r\in\{1/4,3/4\}$, if $n=1$, and
$r\in\{1/8,3/8,5/8,7/8\} = \{.125,.375,.625,.875\}$, if
$n=2$.  These Mach 1 points are represented by vertical lines in the following plots.

Figure~\ref{box1-energy} presents a plot of the specific- compression $p\rho^{-1}$ and the
total $\bar{E}_1$ -energies with the density $\rho$ (or the momentum-density potential
$\rho/2$) and the downhill velocity $u_-$ of the ground-state flow in a one-dimensional box of
length $a_0$ in atomic units. The specific kinetic $mu^2/2$ energy can be approximated from the
figure by noting that $mu^2/2 = \bar{E}_1 - p\rho^{-1}$.  The point $r = 0.5$ is an unstable
equilibrium point. The maximum density is also at $r = 0.5$ where the velocity is zero.
For $r > 0.5$,
and downhill flow, the fluid particles are moving and accelerating to the right, in other
words, they move downhill with respect to the ``density hill,'' so to speak; for $r < 0.5 $
they move and accelerate to the left.  In the limit of a fluid particle reaching a node, i.e.,
$r\rightarrow 0$ or $r\rightarrow 1$, the particle will have infinity speed and zero mass.
From (\ref{press-2b}), given by $\partial(\rho_mu_\pm) = \mp 2 p$ for one-dimension and atomic
units, it follows that for downhill flow, the streamline segments $(0.25,0.75)$, where the
pressure is positive, is the creation zone, or the source, and $(0,0.25)$ and $(0.75,1)$, where
the pressure is negative, are the annihilation zones, or the sink.
\FIG{box1-energy}{The specifc- compression $p\rho^{-1}$ and total $\bar{E}_1$ energies with the density $\rho$
  and downhill velocity $u_-$ of the ground-state of a fluid in a one-dimensional box of length $a_0$
  in atomic units. \zlabel{box1-energy}}

Figure~\ref{box2-energy} presents a plot of the same variables for the first excited state for
$r\in[0,0.5]$. The same general behavior as the ground state is observed, but with an unstable
equilibrium point of $0.25$ and nodes at $0$ and $0.5$. For downhill flow, the source is
$(0.125,0.375)$ and the sinks are $(0, 0.125)$ and $(0.375,0.5)$.  The plot from $r\in[0.5,1]$
can be obtained from Figure~\ref{box2-energy} using the periodic property $f(x + 0.5) = f(x)$
for $x \le 0.5$.  Hence, the other unstable equilibrium point is $0.75$ and the nodes are $0$,
$0.5$ and $1$.
\FIG{box2-energy}{Same is Fig~(\ref{box1-energy}) but for the first excited state and
  for $r\in[0,0.5]$ \zlabel{box2-energy}}

\lbar  Let $\bar{p}_n = p/(n\pi)$. Relate the momentum density $u_-\rho$ and the pressure by using
$(2\sin x\cos x = \sin 2x)$ in (\ref{p0532}) and ($1 - 2\sin^2 x = \cos 2x$) in (\ref{p5241}):
\begin{gather}
     \zlabel{p0533} [\rho u_-]_n(r) = -2n\pi\sin(n\pi r)\cos(n\pi r) = -n\pi\sin(2n\pi r) \\
     \zlabel{p5242} \bar{p}_n(r) = -n\pi[1 - 2\sin^2(n\pi r)]  = -n\pi\cos(2n\pi r) 
\end{gather}
Hence
\begin{equation} \zlabel{p2352}
    \bar{p}_n\lt(r - \fc{1}{4n}\rt) = -n\pi\cos(2n\pi r - \pi/2) = -n\pi\sin(2n\pi r) = [u_-\rho]_n(r) \rbar
\end{equation}


{\bf Definition.} An equation of state for a Q1 flow, when it exists, for the pressure $p$ and
the effective pressure $P$, are equations that define the maps $\rho(\mr) \rightarrow p(\mr)$
and $\rho(\mr) \rightarrow P(\mr)$, respectively, where the maps hold for $\text{Range}(\rho)$.

Note that the above definition gives a stronger condition for an equation of state to exist
than one where the pressure $p$ is determined by $\rho$ and a given nonempty set of its partial
derivative, or the pressure is a functional of the density $\rho$.  Also, the definition is
similar to how a thermodynamic equations of state from static fluids are used for
substances undergoing laminar flow, or when a homogeneous substances becomes only a continuous one
with independent variables represented by scalar fields \cite{Kestin1}.

By the definition above, (\ref{p0422}), which can be written
\[
P(n) = -\fc{Z^2\hbar^2}{ma_0^2}\eta, \qquad \eta \in \text{Range}(\rho)
\]
is an effective pressure equation of state for the ground-states of hydrogenic atoms, and $P$ is a
linear function of $\rho(r)\in \text{Range}(\rho)$. Combining (\ref{p5241}) with $\rho =
2\sin^2(n\pi r)$ gives the pressure equation of state for all flow states in a one-dimensional box:
\[
 p(\eta) = -n^2\pi^2(1 - \eta),
\]
and $p$ is an affine function of $\eta = \rho(r)$.

A function $f:\mathbf{R}\longrightarrow \mathbf{R}$ is said to be symmetric with respect to the
point $c\in \mathbb{R}$ within $l\in \mathbb{R}$ if $f(c + \delta) = f(c - \delta)$ for
$\delta\in[0,l]$, and antisymmetric if $f(c + \delta) = -f(c - \delta)$. For the $n= 1$ flow
state, the ``$\pm$ functions'' $u_\pm$, $\rho u_\pm$, and $s_\pm$ are antisymmetric with respect
to the (maximum $\rho$) point $1/2$ within $1/2$, and $\Ms$ and $\rho$ are symmetric under the
same conditions.  For the $n= 2$ flow state, the ``$\pm$ functions'' are antisymmetric with
respect to the (maximum $\rho$) points $1/4$ and $3/4$, within $1/4$ for both points, and $\Ms$
and $\rho$ are symmetric under the same conditions.

Figure~\ref{box1} presents a plot of the Mach speed and the pressure $p$ with the
density $\rho$ and the downhill momentum density $\rho u_-$ of the ground-state flow. 
 \FIG{box1}{The signed Mach speed Ms and the pressure $p$ with the density $\rho$ and the
    downhill momentum density $\rho u_-$ of the ground-state of a fluid in a one-dimensional box of
    length $a_0$ in atomic units\zlabel{box1}}
In atomic units, $\rho(r) + \text{Ma} = 2$, so a point $r$ where $\rho(r)>1$, $\Ma(r)<1$, and
vice versa; hence, when $\rho(r) > \Ma(r)$ the flow is subsonic, and when $\rho(r) < \Ma(r)$
the flow is supersonic.  (Note if $\rho(r)$ is dimensioned, e.g., per cubic meter, it cannot be
added to the dimensionless Ma.)  Also, when $p > 0$ the flow is subsonic, and when $p < 0$ the
flow is supersonic, and this is consistent with the result from the end of
subsection~\ref{p9022}, where it is demonstrated that $p(r) = 0$ implies $\Ms(r) = \Ma(r) = 1$
for uniform flow.
Furthermore, from (\ref{p2352}) for $n = 1$ we have $p(r - 1/4)/\pi = u_-\rho(r)$.

As mentioned above, for the ground state, $\Ms \ge 0$, giving $\Ms = \text{Ma}$, and Mach 1
occurs at the points $r = 1/4$ and $3/4$, where $\rho u_-$ is either a maximum or minimum. The
points $r = 1/4$ and $3/4$ divide the flow into subsonic and supersonic regions; they are
analogous to choke points for the corresponding classical potential-free flows with variable
cross section \cite{Munson,Shapiro}.

Fig.~\ref{box2} presents a plot of the downhill wave pulse $-s_-$ and velocity $u_-$ with,
again, the density $\rho$ and the downhill momentum density $\rho u_-$ of the ground-state for
the rhs region, $r\in[1/2,1]$.  As expected, at the Mach 1 point $r = 0.75$, $-s_- = u_-$, where
$\rho u_-$ is maximized. The behavior of the functions for $r\in[0,1/2]$ are obtained from the
plots given by noting that the $\pm$ functions are antisymmetrical with respect to the maximum
$\rho$ point of $1/2$ within $1/2$, and $p$ is symmetric.  Therefore, at the
Mach 1 point $r = 1/4$, $\rho u_-$ is minimized.
\FIG{box2}{The downhill wave pulse $-s_-$ and velocity $u_-$ with the density $\rho$ and
  the downhill momentum density $\rho u_-$ of the ground-state of a fluid in a dimensional box of
  length $a_0$ in atomic units for $r\in[1/2,1]$. \zlabel{box2}}

So that different flows can be compared with ease, and flows can be characterized, a reference
flow, called ordinary Q1 flow, is defined below. The properties defining ordinary Q1 flow are
the ones viewed to be important that are exhibited repeatedly in the applications; they might
hold for all Q1 flows. These properties are considered to be fluid (no pun intended), so
they can be modified without hesitation.

\vspace{2ex}

{\bf Definitions.}  A node $\mr_{\text{node}}\in\mathbb{R}^3$ is a point on the streamline such
that $\rho(\mr_{\text{node}}) = 0$. An antinode $\mr_{\text{antinode}}\in\mathbb{R}^3$ is a
point on the streamline such that $\nabla\rho(\mr_{\text{antinode}})\cdot\hat{\mz} = 0$. Let
$\mr_\infty\notin\mathbb{R}^3$ be a point at infinity and let
$\mr_{\text{nuc}}\in\mathbb{R}^3$ be the location of point charge, as in the nucleus of an
atom, where the nucleus is treated as a point charge.

A (matter) closed flow segment $S$ of a streamline $L(\hat{\mz})$ is a subset of $L$ such that
no matter can enter or leave $S$ by convection. A primitive flow segment is a (matter) closed
flow subset of a continuous streamline $L(\hat{\mz})$ that has the following properties: 1) The
restriction $\rho|_S$ is continuously differentiable, i.e, $\nabla\rho(\mr)\cdot\hat{\mz}$
exist and is continuous for all $\mr \in S$.  2) The flow segment $S$ contains \emph{either}
{\bf i)} the endpoints $\mr_{\text{nuc}}$ and
$\mr_0\in\{\mr_{\text{node}},\mr_{\text{anti-node}},\mr_\infty\}$ \emph{or} {\bf ii)} the end
points $\mr_{\text{anti-node}}$ and $\mr_0\in\{\mr_{\text{node}},\mr_\infty\}$, and, for both
cases, there are no other nodes or anti-nodes in (the closure) of $S$.  (For the fluid in a
one-dimensional box, each endpoint, 0 and $\ell$, is taken as a node $\mr_{\text{node}}$.)

Let the parameter $z$ be the position along the streamline $L(\hat{\mz})$.  A continuous
streamline $L(\hat{\mz})$ can be specified by a parametric function $\mr$, a path that depends
on the independent variable $z$, with an interval $J$ as the domain, i.e., $\mr: J \rightarrow
\mathbb{R}^3$. Since the path $\mr$ is signed, i.e., it has a direction, the function $\mr$,
with its signature, determines the unit tangent vector $\hat{\mz}$ of the streamline.

A primitive flow segment $S$ can be specified by the restriction $\mr|_I$, where, for example
$I = [z_{\text{anti-node}},z_{\text{node}}]$ and $\mr(z_{\text{anti-node}}) =
\mr_{\text{anti-node}}$. Hence, if $L$ is specified by the function $\mr$, a particular $S$ is
completely specified by the interval $[z_1,z_2]\subset J$, and the meaning of $S =
\{(\mr,[z_1,z_2])\}$ is clear, where $z_1$ and $z_2$ are end points of an inteval. We also
write $S = [\mr_1,\mr_2]$ where the path function is $\mr$, $\mr(z_i) = \mr_i$, $i = 1,2$, and
$[\mr_1,\mr_2]$ is an ``line or arc interval,'' with vectors $\mr_1$ and $\mr_2$ as end points.

The intervals $I$ are required to be half open, as in $\sq z_{\text{anti-node}},z_\infty\rd$,
if $I$ has an end point at infinity $z_\infty$, otherwise $I$ is required to be closed.  For
one-dimensional models, uniform and spherical flows, the streamline is defned by $L =
(J,\hat{\mz})$, where $J$ is an interval and $\hat{\mz}$ is the unit tangent vector, a constant
vector.

Since Q1 flows always have two flow directions, we ofen use unsigned streamlines $L$ and
primitive flow segments.  Therfore, $L = \text{Range}(\mr)$, and a tangent vector $\hat{z}$ is
not part of the definition. Also, for one-dimensional models, uniform and spherical flows, the
streamline and the primitive flow segments are intervals, e.g., $L = J$, where, for spherical
flow, an inteval $[r_1,r_2]$ is defined by the radial spherical-coordinate of the end
points, and, because of symmetry, the two angles require to completely specify a streamline
can be suppressed.

The ground-state flow in a one-dimensional box presented in Fig~\ref{box1} has the following
two primitive flow segments: $[0,1/2]$ and $[1/2,1]$, where $z_{\text{node}} = 0,1$,
$z_{\text{anti-node}} = 1/2$, and the streamline $L$ satisfies $L = [0,1/2] \cup[1/2,1]$, and
$\{\sq 0,1/2\rd ,[1/2,1]\}$ is a partitioning of $L$.  For flow in the $[1/2,1]$ segment,
$u_\pm = 0$ at $r = 1/2$.  Therefore, all the fluid particles in the segment $[1/2,1]$ do not
flow into the other segment $[0,1/2]$, and, similarly, the fluid particles from the flow
segment $[0,1/2]$ does not flow into the segment $[1/2,1]$. Hence, the segments are (matter)
closed. In general, since the velocity is zero at antinodes and the density is zero at nodes,
primitive flow segments are closed, and they do not contain a proper subset that is closed.


A Q1 flow restricted to a streamline $L$ is an {\it ordinary Q1 streamlime flow} if $\Ms|_L$ is
non-negative and the following conditions hold for each primitive flow segment $S$ of $L$ that
does not contain a point-charge location $\mr_{\text{nuc}}$ as an end point.

\vspace{1ex}


\noindent
{\bf 1)} $S$ contains one and only one unstable equilibrium point at $\mr_{\text{anti-node}}$.
    
\noindent
{\bf 2)} There exist a map $\rho(\mr)\rightarrow \,\Ma(\mr)$, $\mr\in S$,
such that the restriction of the Mach speed $\Ma$ to $\text{Range}(\rho|_S)$ is a strictly
decreasing function of~$\rho(\mr)\in \text{Range}(\rho|_S)$.

\noindent
{\bf 3)}  For $\mr$ restricted to $L$ and $s\in\{s_+,s_-\}$, $u\in\{u_+,u_-\}$, the following
conditions are satisfied for all nodes $\mr_{\text{anti-node}}$, antinotes $\mr_{\text{node}}$,
and points at infinity $\mr_\infty$ that are members of the closure of $L$:

 i) As $\mr \to \mr_{\text{anti-node}}$, $u\to 0$,\hspace{2ex} $\Ma \to 0$.
\\ \mbox{\hspace{2ex}} ii) As $\mr \to \mr_{\text{node}}$,\hspace{2.5ex} $|u|\to\infty$,  $\Ma \to 2$
\\ \mbox{\hspace{2ex}}iv) As $\mr \to \mr^\pr\in\{\mr_{\text{node}}, \mr_{\text{anti-node}}\}$, $|\rho u| \to 0$ and $|s|\to\infty$.
\\ \mbox{\hspace{2ex}}v) As $\mr \to \mr_\infty,\;$  $\Ma \to \infty$, \quad $|s|,|\rho u|  \to 0$

\noindent
The point $u_\infty$, defined by $|u| \to u_\infty$ as $\mr \to \mr_\infty$, is called the
speed limit of the streamline.  An ordinary Q1 flow is a Q1 flow that is an ordinary Q1
streamlime flow for each streamline.
\vspace{2ex}

{\bf Definition.} Let $P$ be the set of all primitive flow segments of streamline $L$. The
primitive flow segments $I$ and $I^\pr$ are related if they contain the same antinode end-point
$\mr_{\text{anti-node}}$, and this state of affairs is denoted by
$(\mr_0,\mr_{\text{anti-node}},\mr_0^\pr)$, where $I = [\mr_0,\mr_{\text{anti-node}}]$ and
$I^\pr = [\mr_{\text{anti-node}},\mr_0^\pr]$, and $(\mr_0,\mr_{\text{anti-node}},\mr_0^\pr)$ is
called a flow pair or a flow hill. For a particle in a
one dimensional box, the number of flow hills is $(n-1)$, where $n$ is the number of nodes
including the end points for the fluid in a box.

By examining the pertinent functions, $\rho$, $u$, $\rho u$, $s$, and Ms, it is easily
verified, as it is illustrated in figures~\ref{box1-energy}, \ref{box1} and \ref{box2}, that
the ground state fluid in a box is an ordinary Q1 flow containing one flow hill $(0,1/2,1)$. 

It is also easily verified that the first excited state is an ordinary Q1 flow containing the
two flow hills $(0,\,.25,\,.5)$, $(.5,\,.75,\,1)$.  Figure (\ref{box3}) and (\ref{box4}) are
plots of the same variables as in (\ref{box1}) and (\ref{box2}), respectively, but for the
first excited state. Fig.~(\ref{box4}) is for the $[3/4,1]$ primitive flow segment, and the
behavior of the functions in $r\in[0,3/4]$ can be obtained by noting that the $\pm$-functions
and $\rho$ are antisymmetric and symmetrical, respectively, with respect to the points $1/4$,
$2/4$ and $3/4$, all within $1/4$, as displayed in Fig.~(\ref{box3}) for $\rho u_-$ and $\rho$.
\FIG{box3}{Same as (\ref{box1}), but for the first excited state
  and the two flow hills $(0,\,.25,\,.5)$ and $(.5,\,.75,\,1)$. \zlabel{box3}}

\subsection{The harmonic oscillator \zlabel{p7400}}

In this subsection we treat the ground state and first excited state of the harmonic
oscillator, with quantum number $n = 0,1$.  The $Q1$ wavefunctions for the ground- and
first excited-states of the harmonic oscillator are \cite{Raimes,Levine}
\begin{equation} \zlabel{p3940}
  \phi_0(x) =\lt(\fc{2\al}{\pi}\rt)^{1/4}e^{-\al x^2}, \quad  \text{\quad} \phi_1(x) = 2\lt(\fc{2\al^3}{\pi}\rt)^{1/4}xe^{-\al x^2}
\end{equation} 
where $\al = m\omega/(2\hbar)$, $\omega = \sqrt{k/m}$, and the specific potential-energy is
$V(x) = m\omega^2x^2$/2.  We choose $m,k = 1$, giving $\omega = 1$ and $\alpha = 1/2$ in atomic
units.
\FIG{box4}{Same as (\ref{box2}), but for the first excited state and the primitive flow
  segment $[3/4,1]$. \zlabel{box4}}

\subsubsection{The harmonic oscillator ground state}

\lbar Start with the open density formula $\rho = \phi_0^2$  and (\ref{p3940}) to compute the
uphill flow velocity-component using (\ref{4880}), i.e., $u_\pm = \pm\partial\rho/(2\rho)$, the
momentum density $\rho u_\pm$, and the specific kinetic energy $mu^2/2$:
\begin{gather}
  \notag  \rho(x) = \phi^2(x) = \pi^{-1/2} e^{-x^2}
  \\ \zlabel{p5832}\partial\rho(x) =  -2x\rho(x)
  \\ \zlabel{p5834} u_+(x) = \fc12[\rho^{-1}\partial\rho](x) = -x
  \\ \zlabel{p5844} \rho u_+ = -x\pi^{-1/2} e^{-x^2} \\
  \fc12mu^2(x) = \fc12 x^2
\end{gather}
Using (\ref{p5832}), compute the first term on the rhs of the wave pulse formula (\ref{p2445n}),
given by $-s_+ = -(1/2)\partial^2\rho/\partial\rho_m + u_+$:
\begin{gather} \zlabel{p4252}
  \partial^2\rho(x) = -2\pi^{-1/2}\partial(x e^{-x^2}) = (-2 + 4 x^2)\rho
  \\ \zlabel{p4282}
  -\fc12\fc{(\partial^2\rho)}{(\partial\rho)}(x) = -\fc12\lt(\fc{-2 + 4 x^2}{-2 x}\rt) = -\fc12x^{-1} + x 
\end{gather}
Compute the pressure $p$ using (\ref{p4252}) and (\ref{0003b}), i.e., ($p =
-\partial^2\rho/4$), the specifc compression energy $p\rho^{-1}$, the specific
effective-compression energy $P\rho^{-1} = p\rho^{-1} + V$, with $V(x) = x^2/2$, and the well
known specific total-energy $\bar{E}_0$ from (\ref{2492}):
\begin{gather}
  \zlabel{p5272} p(x) =  \lt(\fc12 - x^2\rt)\rho \\
  [p\rho^{-1}](x) =  \fc12 - x^2 \\
  [P\rho^{-1}](x) =  \fc12 - \fc12 x^2 \\
  \bar{E}_0 =  \fc12 mu^2 + P\rho^{-1} = \fc12 \rbar
\end{gather}

Fig.~\ref{hosc1-energy} presents the specific- effective-compression $P\rho^{-1}$, potential
$V(x) = x^2$/2. and total $\bar{E}_0$ \mbox{-energies} with the density $\rho$ and downhill
velocity $u_-$ of the ground-state harmonic oscillator with $m,k = 1$ in atomic units.  The
same type of behavior is observed as in the particle in a box states considered,
Figs.~\ref{box1-energy} and \ref{box2-energy}, with $x = 0$ being an unstable equilibrium
point. For $p(x) = 0$ we obtain $[P\rho^{-1}](x) = V(x)$, and this is Mach 1. From
(\ref{p5272}), $p(x) = 0$ is satisfied at $x = \pm \sqrt{1/2}$. In the figure $x = \pm
\sqrt{1/2}$ are represented by vertical lines where $[P\rho^{-1}](x)$ and $V(x)$ cross. A
special case of this same type of crossing appears in Figs.~\ref{box1-energy} and
\ref{box2-energy}, where $V(x) = 0$, giving $P\rho^{-1}(x) = p\rho^{-1}(x) = 0$ at the crossing
points.
\FIG{hosc1-energy}{The effective compression $P\rho^{-1}$, potential $V(x) = x^2/2$. and total
  $\bar{E}_0$ \mbox{-energies} represented as closed and specific quantities with the density
  $\rho$ and downhill velocity $u_-$ of the ground-state harmonic oscillator
  $(-\infty,0,+\infty)$ in atomic units. \zlabel{hosc1-energy}}

Note the the total specific potential-energy satisfies $P\rho^{-1} = -V + E_0$ , where $V(x) =
x^2/2$. It goes without saying that the behavior of a ground-state fluid-particle from the
harmonic-oscilator flow with the specific potential-energy $P\rho^{-1}$, and with an unstable
equilibrium point at $x=0$, is totally different, and in many ways opposite, of a point-mass
particle of the corresponding classical harmonic-oscillator with the potential-energy function
$V$, and with a stable equilibrium point $x=0$.

\lbar  Note (\ref{p2445n}) in the form  $-s_\pm = -(1/2)(\partial^2\rho)/\partial\rho_m + u_\pm$: Obtain the
wave pulse component $-s_+$ by adding (\ref{p5834}) and (\ref{p4282}), and then obtain the
signed Mach speed Ms:
\begin{gather*}
  -s_+(x) = -\fc12x^{-1} \\
  \Ms(x) = u_+(-s_+^{-1}) = 2x^2\rbar 
\end{gather*}
Hence, $\Ms \ge 0$, giving $\Ms = \Ma$; also, $\Ma = 1$ at $x = \pm \sqrt{1/2}$, and these
values agrees with the extremums of the momentum density that satisfies $\partial(\rho u_+) = 0
$, where from (\ref{p5844}) we have
\begin{gather*}
  \partial[\rho u_+](x) = -\pi^{-1/2}\partial(xe^{-x^2})  = \pi^{-1/2}(-1 + 2x^2)  e^{-x^2}
\end{gather*}

Figure~\ref{hosc1} presents the Mach speed Ma and the pressure $p$ with the density $\rho$ and
the downhill momentum density $\rho u_-$ of the ground-state harmonic oscillator for the subset
$(-2,0,2)$ of the only flow hill $(-\infty,0,\infty\rd$.  Figure~\ref{hosc2} presents the
downhill wave pulse $-s_-$ and velocity $u_-$ with the density $\rho$ and the downhill momentum
density $\rho u_-$ for the same flow, and for the primitive flow segment subset
$[0,2]\subset[0,\infty\rd$.
The behavior of the functions for $x\in[2,0]$ can be obtained by noting that the
$\pm$-functions and density $\rho$ are antisymmetric and symmetrical, respectively, with respect to the
antinode point $0$ within $\infty$.

By examining the pertinent functions, it is readily verified, and it is also illustrated in
figures~\ref{hosc1-energy}, \ref{hosc1} and \ref{hosc2}, that the ground-state
harmonic-oscillator flow is an ordinary Q1 flow with speed limit $\infty$ for both primitive
flow segments, i.e., $|u|\to \infty$ as $r \to \pm\infty$.

One significant difference with this flow compared to the fluid in a box flows is that
ground-state harmonic-oscillator flow contains points at infinity $\mr_\infty = \pm\infty$,
while the flows for the fluid in a box contain node points $\mr_{\text{node}}$. The two
different types of points, $\mr_{\text{node}}$ and $\mr_\infty$, yield the following
differences for the functions Ma and $|s|$:
\[
\text{As $\mr \to \mr_{\text{node}}$, $\Ma \to 2$ and $|s|\to\infty$,
and as $\mr \to \mr_\infty$, $\Ma \to \infty$ and $|s|\to 0$.}
\]
\FIG{hosc1}{The Mach speed Ma and the pressure $p$ with the density $\rho$ and the downhill
  momentum density $\rho u_-$ of the ground-state harmonic oscillator $(-\infty,0,+\infty)$ in
  atomic units. \zlabel{hosc1}}
\FIG{hosc2}{The downhill wave pulse $-s_-$ and velocity $u_-$ with the density $\rho$ and
  the downhill momentum density $\rho u_-$ of the ground-state harmonic oscillator in
  atomic units for the flow segment subset $[0,2]\subset[0,\infty)$. \zlabel{hosc2}}

\subsubsection{The harmonic oscillator first excited state} 

\lbar Starting with the open density formula $\rho = \phi_1^2$ and (\ref{p3940}), compute the
uphill flow velocity-component using (\ref{4880}), i.e., $u_\pm = \pm\partial\rho/(2\rho)$, the
momentum density $\rho u_\pm$, and the specific kinetic energy $mu^2/2$:
\begin{gather} \notag
  \rho(x) = \be x^2e^{-x^2}, \qquad \beta =  2\lt(\fc{1}{\pi}\rt)^{1/2} \\
  \zlabel{p5835} \partial\rho(x) = 2x\be e^{-x^2} + (-2x) x^2\be e^{-x^2} = 2(x -x^3)\be e^{-x^2} \\
  \notag u_+(x) = \fc12[\rho^{-1}\partial\rho](x) = \fc12 x^{-2}(2x -2x^3) = x^{-2}(x -x^3) \\
  \zlabel{p5845}  u_+(x) = x^{-1} - x \\
  [\rho u_+](x) =  \be (x -x^3)e^{-x^2} ]\\
\fc12 [mu^2](x) = \fc12 (x^{-1} - x)^2  \rbar
\end{gather}


Since $\rho$ has maximums at $x = \pm 1$ and a minimum at $x=0$, the flow has two flow hills
$(-\infty,-1,0)$ and $(0,1,\infty)$, and $x = 0$ is a node.  \lbar Using (\ref{p5835}), compute
the first term on the rhs of the wave pulse formula $-s_+$ (\ref{p2445n}), given by $-s_+ =
-(1/2)\partial^2\rho/\partial\rho_m + u_+$:
\begin{gather} \notag
  \be^{-1}\partial^2\rho(x) = 2\partial[(x -x^3)e^{-x^2}] \\ \notag
  \be^{-1}\partial^2\rho(x) = 2(1 -3x^2)e^{-x^2} - 4x(x -x^3)e^{-x^2} \\ 
  \zlabel{p4253} \partial^2\rho(x) = (2 - 10x^2 + 4x^4)\be e^{-x^2} \\ \notag
  \fc{(\partial^2\rho)}{(\partial\rho)}(x) = \fc{2 - 10x^2 + 4x^4}{2(x -x^3)}
  \\ \zlabel{p5235}
 -\fc12\fc{(\partial^2\rho)}{(\partial\rho)}(x) = -\fc12\lt(\fc{1 - 5x^2 + 2x^4}{x -x^3}\rt)
\end{gather}
Compute the pressure $p$ using (\ref{p4253}) and (\ref{0003a}), the specific compression energy
$p\rho^{-1}$, the specific effective-compression energy $P\rho^{-1} = p\rho^{-1} + V$, with
$V(x) = x^2/2$, and the well known specific total-energy $\bar{E}_1$ from (\ref{2492}):
\begin{gather} \zlabel{p0902}
  p(x) = -\fc12(1 - 5x^2 + 2x^4)\be e^{-x^2} 
  = -\fc12 x^{-2}(1 - 5x^2 + 2x^4)\rho \\ \notag
  [P\rho^{-1}](x) = -\fc12 x^{-2}(1 - 5x^2 + 2x^4) + \fc12 x^2  \\ \notag
  [P\rho^{-1}](x) = -\fc12x^{-2}  + \fc52  - x^2    + \fc12 x^2 = \fc52 - \fc12x^{-2} - \fc12 x^2 \\ \notag
  \fc12 [mu^2](x) = \fc12 (x^{-1} - x)^2 = \fc12 (x^{-2} + x^2 -2) = \fc12x^{-2} + \fc12x^2 - 1 \\ \notag
  \bar{E}_1 = \fc32
\end{gather}

Substitute (\ref{p5235}) into (\ref{p2445n}), in the form $-s_\pm =
-(1/2)(\partial^2\rho)/\partial\rho_m + u_\pm$, to obtain the wave pulse component $-s_+$
with $u_+$ given by (\ref{p5845}):
\begin{equation} \zlabel{p4320}
    -s_+(x) = -\fc12\lt(\fc{1 - 5x^2 + 2x^4}{x -x^3}\rt) + u_+(x) \rbar
\end{equation}
This equation and $(\ref{p5845})$ gives a formula for the signed Mach Ms, using $\Ms = u_+/(-s_+)$.

Consider $(0,1,\infty)$, the flow hill of the rhs of the streamline. From (\ref{p0902}), it
follows that for $x\in\mathbb{R}$, $p(x) = 0$ if and only if $(1 - 5x^2 + 2x^4) = 0$.  By
examining a plot of $(1 - 5x^2 + 2x^4)$, or the function $p(x)$, as in Fig~\ref{hosc3}, it is
easily verified that
$p(x) = 0$ for $x\in[0,\infty)$ is satisfied at approximately $x= 0.47$ and $1.51$.  Since,
$p(r) = 0$ implies $\Ms(r) = \Ma(r) = 1$ for uniform flow, $x= 0.47$ and $x= 1.51$ are
approximately Mach 1 points. By examining a plot of the function Ms restricted to the subsets
of $\sq 0,\infty\rd$, as in Fig.~(\ref{hosc3}), where Ms has a minimum at $x=1$, and where
$\Ms(1) = 0$, it is almost certain that Ms is non-negative. Hence, with great certainty,
$\Ms\ge 0$ and $\Ms = \Ma$ for Ms restricted to $[0,\infty]$.  Since $\Ms$ is a symmetric with
respect to the point $0$ within $\infty$, $\Ma(x) = 1$ is satisfied at approximately $\pm 0.47$
and $\pm 1.51$, and the great certainty $\Ms\ge 0$ and $\Ms = \Ma$ for the entire domain
$(-\infty,\infty)$. These Mach 1 points are represented by vertical lines in the following
plots.
  
Figure~(\ref{hosc2-energy}) presents the same functions as in Fig.~\ref{hosc1-energy}, for the
flow-hill subset $[0,2]\subset\sq 0,1,\infty \rd$ of the first excited state. Some of the same type
of general behavior, as seen in the ground state, is observed, including an unstable
equilibrium point at the antinode ($x = 1$).  However, unlike the ground state, the functions
restricted to $[0,\infty]$ are neither symmetric nor antisymmetric with respect to the antinode
(at $x=1$). The flow hill $\sq 0,1,\infty \rd$ is also distinct from the others considered, since
it contains \emph{both} a node (at $x=0$) and a point at infinity, and for ordinary Q1 streamline
flow, the limit values at the node and points at infinity differ for $|s|$ and Ma.  The
behavior of the functions for the flow hill $(-\infty,-1,0)$ can be obtained by noting that the
$\pm$-functions and the density $\rho$ are antisymmetric and symmetrical, respectively, with
respect to the node point at $x = 0$ within~$\infty$.
\FIG{hosc2-energy}{Same as (\ref{hosc1-energy}), but for the first excited state and the
  flow-hill subset $[0,2]\subset[0,1,\infty)$. \zlabel{hosc2-energy}}

Figure~(\ref{hosc3}) and (\ref{hosc4}) present the same functions as Fig.~(\ref{hosc1}) and
(\ref{hosc2}), respectively, but for the first excited state and the flow-hill subset
$[0,2.5]\subset\sq 0,1,\infty \rd$. Note that, unlike the previous cases, the maximum pressure
does not occur at the anti-node, as indicated in Fig.~\ref{hosc3}.

By examining the pertinent functions, it is readily verified, and it is also illustrated in the
three figures, that the first excited-state harmonic-oscillator flow is an ordinary Q1
flow. Also, as in the ground state, the speed limit is $\infty$ for both primitive
flow segments.
\FIG{hosc3}{Same as (\ref{hosc1}), but for the first excited state and the
  flow-hill subset $[0,2.5]\subset[0,1,\infty)$. \zlabel{hosc3}}
\FIG{hosc4}{Same as (\ref{hosc2}), but for the first excited state and the flow-hill subset
  $[0,2.5]\subset[0,1,\infty)$. \zlabel{hosc4}}
%

\subsection{The hydrogen 2s state \zlabel{p7500}} 







In this subsection we treat the 2s state of the hydrogen atom. In spherical coordinates, each
streamline depends on the radial coordinate $r$ only. So the results hold for any streamline given
by a ray $r\in[0,\infty)$ for and any polar and azimuth angles.

\lbar Let $h = 1/2$.  To calculate the pressure $p$, note Eq.~(\ref{4927}) from
Appendix~\ref{p4387}.  Start with the Q1 wavefunction $\phi$ of the hydrogen 2s state
\cite{Raimes,Levine,Bransden} and calculate the radial directional-derivative in spherical
coordinates: $\partial\phi\;\dot{=}\;\nabla\phi\cdot\hat{\mr} = \partial\phi/\partial r$.
%
\begin{gather}
  \zlabel{p0282} \phi(r) =  \eta(2 - r)e^{-hr}, \quad \eta = \fc{1}{4\sqrt{2\pi}} 
\\ \zlabel{p0285} \partial\phi = \eta[-1  - h(2 - r)]e^{-hr} = \eta(hr - 2)e^{-hr}
\end{gather}
Use this result for $\partial\phi$ and also [$h - h(hr - 2) = h - h^2r + 2h = 3h - h^2r$] to
calcluate the second directional derivative $\partial^2\phi$.
\begin{equation} \zlabel{p0290} 
  \partial^2\phi = \eta[h - h(hr - 2)]e^{-hr} = \eta(3h - h^2r)e^{-hr}
\end{equation}
Use (\ref{p0285}), (\ref{p0290}) and
\[
 2r^{-1}(hr - 2) + (3h - h^2r) = 1 - 4r^{-1} + 3h - h^2r
\]
to calculate the Laplacian ($\nabla^2\phi = 2r^{-1}\partial\phi + \partial^2\phi$) of the
wavefunction times $(-\phi/2)$ in spherical coordinates, the first term on the rhs of
(\ref{4927}), given by $p = -\phi\nabla^2\phi/2 - \nabla\phi\cdot\nabla\phi/2$.
\begin{gather} \notag
 \nabla^2\phi = \eta[2r^{-1}(hr - 2) + (3h - h^2r)]e^{-hr}\\ \notag
 \nabla^2\phi = \eta(1 +3h - 4r^{-1} - h^2r)e^{-hr} 
\\  \zlabel{p9252} -\fc12\phi\nabla^2\phi = \fc12\eta^2(r - 2)(1 +3h - 4r^{-1} - h^2r)e^{-r}
\end{gather}
Calulate the second term on the rhs of (\ref{4927}) using (\ref{p0285}).
\[
 -\fc12\nabla\phi\cdot\nabla\phi =-\fc12(\partial\phi)^2  =  -\fc12\eta^2(hr - 2)^2e^{-r} \\
 \]
Obtain the pressure $p$, by adding this result to the rhs (\ref{p9252}) as indicated in
(\ref{4927}).
\begin{equation}
  p(r) = \fc12\eta^2(r - 2)(1 +3h - 4r^{-1} - h^2r)e^{-r} - \fc12\eta^2(hr - 2)^2e^{-r}
\end{equation}

Compute the streamline uphill velocity $u_+= \bu_+\cdot\hat{\mr}$ by using
(\ref{8925}), given by $u_+ = \partial u_+/u_+$, (\ref{p0285}), and $\phi = \eta(2 -
r)e^{-hr}$, and then compute the specific kinetic-energy~$mu^2/2$.
\begin{gather}
  \zlabel{p9279}
  u_+ = \fc{\partial\phi}{\phi} = (hr - 2)(2 - r)^{-1} = -\fc12(r - 4)(r - 2)^{-1}
  \\ \notag \fc12 mu^2 = \fc18(r - 4)^2(r - 2)^{-2}
\end{gather}
Use this formula for $mu^2/2$, the well known eigenvalue of the Schr\"odiger equation for
the 2s state, $E_2 = -1/8$, and (\ref{2492}), which can be written $P\rho^{-1} = \bar{E} -
\fc12 u^2$, to compute the specific compression-energy~$P\rho^{-1}$.
\begin{gather*}
     P\rho^{-1} = -\fc18[1 +(r - 4)^2(r - 2)^{-2}] \rbar
\end{gather*}

The 2s state has the primitive flow segment $[0,2]$ and the flow hill $\sq2,4,\infty\rd$.  The
segment $[0,2]$ contains the nucleus $r_{\text{nuc}} = 0$ and the node $r_{\text{node}} = 2$,
but it does not contain an antinode. As discussed below, there two Mach points at 2.59 and
5.41, and these are represented by vertical lines in the following plots.

\FIG{2s-energy}{The specific- effective-compression $P\rho^{-1}$, potential $V(r) =
  -r^{-1}$. and total $\bar{E}_{2}$ \mbox{-energies} with the density $\rho$ and downhill
  velocity $u_-$ of the 2s state of the hydrogen atom and the flow-hill subset
  $[2,7]\subset[2,4,\infty)$. The constant $\eta$ is defined in Eq.~(\ref{p0282}). \zlabel{2s-energy}}
Fig.~\ref{2s-energy} presents the same type of specific energy plot as (\ref{hosc1-energy}) and
(\ref{hosc2-energy}) but for the hydrogen 2s state with the flow-hill subset
$[2,7]\subset\sq2,4,\infty\rd$. It follows from the shape of the specific potential-energy
$P\rho^{-1}$ that, overall, the absolute value of the acceleration is much greater on the lhs
side of the unstable equilibrium point at 4.0 than on the rhs. This flow-hill subset resembles
the corresponding subset $[0,2]\subset\sq 0,1,\infty\rd$ for the first excited state of the
harmonic oscillator flow from Fig.~(\ref{hosc2-energy}), but the two differ significantly on the rhs
of the unstable equilibrium point.  Also, unlike the harmonic oscillator flows, the speed limit of $1/2$
is finite.

\FIG{2sl-energy}{Same as (\ref{2s-energy}), but for the electron fall
  $[0,2]$. \zlabel{2sl-energy}}
Fig.~\ref{2sl-energy} presents the same scalar fields as in Fig.~\ref{2s-energy}, but for the
primitive flow segment $[0,2]$, where $\mr =0$ is the location of the nucleus.  The density
restricted to the segment $[0,2]$ is a strictly decreasing function.  Primitive flow segment
that do not contain an antinode, like this one, are called fluid falls.  The density $\rho$,
velocity component $u_\pm$, and specific- effective-compression $P\rho^{-1}$ have finite values
at $\mr_{\text{nuc}} = 0$. The specific potenitals $V = -1/r$ and $P\rho^{-1}$ seem to be
approximate mirror images of each other, with respect to the vertical line at about 1.

\lbar Use $\phi(r) =  \eta(2 - r)e^{-hr}$ and (\ref{p0285}) to calculate $\partial\rho(r)$ and $\partial^2\rho(r)$.
\begin{gather*}
  \partial\rho(r) = [\partial\phi^2](r) = 2[\phi\partial\phi](r) = 2\eta(2 - r)e^{-hr}\times h\eta(r - 4)e^{-hr} \\
 \eta^{-2}\partial\rho(r) = (2 - r)(r - 4)e^{-r} \\
 \eta^{-2}\partial^2\rho = [-(r - 4) + (2 - r) - (2 - r)(r - 4)]e^{-r}
\end{gather*}
Calculate the ratio $(\partial^2\rho)/(\partial\rho)$ and then the compute the wave pulse
$-s_+$ using (\ref{p2445n}), given by $-s_+ = -(1/2)\partial^2\rho/\partial\rho_m + u_+$, and (\ref{p9279}):
\begin{gather*}
  \fc{(\partial^2\rho)}{(\partial\rho)} = -(2 - r)^{-1} + (r - 4)^{-1} - 1 \\
    -s_+ =  \fc12\lt[1 - (r - 2)^{-1} - (r - 4)^{-1} - (r - 4)(r - 2)^{-1}\rt] \rbar
\end{gather*}
This equation and $(\ref{p9279})$ gives a formula for the signed Mach Ms, using $\Ms =
u_+/(-s_+)$.

\FIG{2s1}{The Mach speed Ma and the pressure $p$ with the density $\rho$ and
  the downhill momentum density $\rho u_-$ of the 2s state of the hydrogen atom and the flow-hill subset
  $[2,6]\subset[2,4,\infty)$. \zlabel{2s1}}
By examining a plot of the function Ms, as in Figures~(\ref{2s1}) and (\ref{2sl1}), it is
almost certain that Ms is non-negative. Also, there are two Mach 1 points, 2.59 and 5.41, and
these are on either side of the only antinode at 4.0.  Fig.~\ref{2s1} presents the same type of
Mach--pressure plot as (\ref{hosc1}) and (\ref{hosc3}), but for the hydrogen 2s state with the
flow-hill subset $[2,6]\subset[2,4,\infty)$.
The overall behavior exhibited by the functions are similar to the ones displayed in
Fig.~(\ref{hosc3}). However, as expected, the two points $r\in\mathbb{R}^3$ where $p(r) = 1,$
do not satisfy $\Ma(r) = 1$.

\FIG{2s2}{The downhill wave pulse $-s_-$ and velocity $u_-$ with the density $\rho$ and the
downhill momentum density $\rho u_-$ of the 2s state of the hydrogen atom and the flow-hill
subset $[2,6]\subset[2,4,\infty)$. \zlabel{2s2}}
Fig.~\ref{2s2} presents the same type of wave-pulse--velocity plot as (\ref{hosc2}) and
(\ref{hosc4}), but for the hydrogen 2s state with the flow-hill subset
$[2,6]\subset[2,4,\infty)$. The overall behavior exhibited by the functions are similar to the
ones displayed in Fig.~(\ref{hosc4}).

\FIG{2sl1}{The same functions as in both (\ref{2s1}) and (\ref{2s2}) but for the electron fall
  $[0,2]$. \zlabel{2sl1}}
Fig.~\ref{2sl1} presents the same functions as in both (\ref{2s1}) and (\ref{2s2}) but for the
electron fall $[0,2]$. All the functions, except for the pressure $p$, have finite limits at
the point-charge location $\mr_{\text{nuc}} = 0$. In particular, note that
$\Ma(\mr_{\text{nuc}}) = 8$. Also, the pressure has a similar shape as the one for the 1s flow,
Fig.~\ref{pressure-1s}. As in part {\bf 2} of the definition of ordinary Q1 flow, 
there exist a map $\rho(\mr)\rightarrow \,\Ma(\mr)$, for $\mr\in [0,2]$,
but the restriction of the Mach speed $\Ma$ to $\text{Range}(\rho|_{[0,2]})$ is a strictly
\emph{increasing}, not decreasing, function of~$\rho(\mr)\in \text{Range}(\rho|_{[0,2]})$.

By examining the pertinent functions, it is not difficult to demonstrate that
$|\rho u_\pm|\to 0$, $|s_\pm|\to 0$ and $\Ma \to \infty$, as $r \to \infty$.
Furthermore, it is readily verified, and it is also illustrated in the
three figures, that the hydrogen 2s flow is ordinary Q1 flow.

The behavior of the hydrogen 1s flow, with the single electron fall
$\sq\mr_{\text{nuc}},\mr_\infty\rd$ of a given streamline, can be compared to the 2s flow at
the limiting points $\mr_{\text{nuc}}$ and $\mr_\infty$.  The $\mr_\infty$ limits of $|\rho
u_\pm|$, $|s_\pm|$, and $|\Ma|$ for the hydrogen 1s state are the same as the 2s state, and the
limit of $|u_\pm|$ differ, but they are both finite.  Also all the functions, except for the
pressure $p$, have finite limits at the point-charge location $\mr_{\text{nuc}} = 0$.  However,
the hydrogenic flow is different from the others considered, because it his a single fall with
at point at infinity $\mr_\infty$ as an end point.

\section{Discussion \zlabel{p0822}}

\subsection{The compatability with quantum mechanics and the continuum assumption}

For later use, we note that we have used the time-dependent momentum balance (\ref{2392R}) and
continuity (\ref{cont2-time}) equations for the derivation of the speed of sound equation (\ref{p0024}).
Since a limit was taken near the end of the derivation, there are probably many other equations
which would give the same speed of sound equation. Hence, we have not extended the approach
into the time dependent realm. In order to do so, assuming it is possible, it seems reasonable
to expect the time dependent flow equation to be related in some way, or implied by, the
time-dependent Schr\"odinger equation.

There is a huge difference between the information contained in a Q1 flow state and a
corresponding Q1 probability state. For example, properties like the pressure of Q1 flow state
are represented by a time independent scalar field; while, an observable for a Q1 probability
state is represented by a single constant or a set of values with corresponding probabilities,
giving statistical information about measurement of the observable; there is no field. Still,
it is worth considering if a Q1 flow state can coexist with a corresponding Q1 probability
state without any contradictions.

Let $\phi$ be a normalized Q1 wavefunction.  Let $\chi$ be a normalized, nondegenerate
eigenvector with eigenvalue $\lambda$ of a Hermitian operator $\hat{O}$ with a discrete
spectrum representing an observable, such that the inner product $(\phi,\chi)$ of the $L^2$
Hilbert space satisfies $|(\phi,\chi)|^2\in (0,1)$. Hence, $\phi$ and $\chi$ are linderly
independent, but not orthogonal. According to an axiom of quantum mechanics, if a measurement
is made of the observable with operator $\hat{O}$ for the probability state represented by
$\phi$, then the probability of the measured value being equal to $\lambda$ is
$|(\chi,\phi)|^2$. Also, if the measured value is $\lambda$, then the quantum mechanical system
is in the probability state $\chi$ immediately after the measurement is made. In other words,
the measurement has transformed the state $\phi$ into $\chi$. Such a transformation for the
corresponding Q1 flow state involves a time-dependent process and non-steady flow. Hence, it is
outside the range of applicability of Q1 flow. The same conclusion is easily shown to hold for
observers represented by bounded self adjoint operators such that $\lambda$ corresponds to a
subspace, and to cases where the operator $\hat{O}$ is unbounded, but still self-adjoint, with
a continuous spectrum. Therefore the assumed existence of a Q1 flow state does not lead to any
contradictions with the corresponding existence of the Q1 probability state. Both states can
coexist, within the very limit range of applicability of Q1 flows.

In this work, certain one-body quantum systems have been endowed with a continuum with vector
and scalar fields. I do not have sufficient knowledge of experimental quantum-mechanics to know
if the properties represented by these fields can be measured, to determine if they ``exist''
by some reasonable definition. Therefore, in order to proceed, and make some progress on this
question, suppose these properties cannot be measured, nor can an experiment be performed that
shows that their existence contradicts experimental facts.

Under these circumstances, how is it to be determined at some point in the future if this
endowment is successful?  To approach an answer to this question we turn to Marion and
Thorton, where they discuss the appending of conservation postulates, including energy, on
systems that do not have conservative potentials:
\begin{quote}
We therefore extend the usual concept of energy to include ``electromagnetic energy'' to
satisfy our preconceived notion that energy must be conserved. This may seem an arbitrary and
drastic step to take, but nothing, it is said, succeeds as does success, and these conservation
``laws'' have been the most successful set of principles in physics. 
\end{quote}
Hence, the endowment will succeed if useful applications can be found.

\subsection{Q1 flow with the solution $s_\pm = u_\pm$ \zlabel{p0822a}}

Here two interpretations are given for a Q1 flow with the solution
$s_\pm = u_\pm$ from (\ref{8350n}).

{\bf 1)} Consider a a corresponding classical potential-free flow with Mach speed 1 and uniform
velocity $u\hat{\mz}$. If a wave pulse travels downstream and upstream, the speed of the wave
pulse in the static frame is 2u and 0u, respectively. (This same behavior is also observed for
a streamline $L$ of spherical flow with the wave pulse moving up or down the streamline $L$.)
Assuming this type of behavior also holds for Q1 flow, the solution $s_\pm = u_\pm$ for Q1 flow
is consistent with Mach 1, giving $u_\pm\hat{\mz} - s_\pm\hat{\mz} = \mathbf{0}$ for the
wave-pulse velocity in the static frame, even though, for some reason, the corresponding
solution $s_\pm = -u_\pm$, giving $u_\pm\hat{\mz} - s_\pm\hat{\mz} = 2u_\pm\hat{\mz}$ for
wave-pulse velocity in the static frame, is absent.

{\bf 2a)} Since, for $s_\pm = u_\pm$, the velocity of the wave pulse in the static coordinate
frame is $u_\pm\hat{\mz} - s_\pm\hat{\mz} = \mathbf{0}$, the solution $s_\pm = u_\pm$ is for
the case where the wave pulse frame and the control volume are at rest relative to the static
coordinate frame.  Hence, the solution is simply the one for a steady flow with velocity
$u_\pm$, and the ``speed-of-fluid'' equation is (\ref{u-quad}) with $\bar{u} =u$ and $\gamma =
0$, and there is no wave pulse.

{\bf 2b)} The possibility of time dependent flow cannot be ruled out because (\ref{u-quad}),
with $\bar{u} =u$ and $\gamma \ne 0$, also yields (\ref{p0024}) with $s_\pm = u_\pm$, where
$\rho_m$ is taken as the limit of a sequence of functions. (There is no reason to continue with
the derivation to (\ref{8350n}), since $s_\pm = u_\pm$ is set.)  Such a flow would be silent
with respect to the static frame, since there is no moving wave-pulse relative to the static
frame. Also, for such a flow, a disturbance is not carried downstream with the fluid velocity,
instead, just like steady flow, fluid particles are perturbed as they pass through a disturbed
region, but, since $\gamma \ne 0$ is permitted, the disturbance could be time dependent.
Hence, the existence of a Q1 system that cannot carry a wave pulse is conceivable.  The
situation is different for classical systems, since the wave pulse velocity of $(-s_\pm) =
-u_\pm$ is not always one of the solutions of the quadratic wave-pulse equation.

\subsection{The Q1 wave pulse direction and the sign of the signed Mach Ms \zlabel{p0822c}}

First note that it follows from (\ref{p0370}) that $\Ms > 0$, if and only if the the directions
of the wave-pulse velocity $(-s_\pm)$ is the same as the direction of the fluid velocity
$u_\pm$, i.e., $u_\pm$ and $(-s_\pm)$ have the same sign, where the direction of the fluid
velocity is considered specified. Also, the definition $\Ma(\mr) = \Ms(\mr)$ is only used in
regions where $\Ms(\mr) \ge 0$, since $\Ma(\mr) \ge 0$ is a requirement.

From (\ref{4944n}) and (\ref{p0370}) it follows that $\rho_m(\mr)[d(P\rho_m^{-1})/d\rho_m](\mr)
\ge 0$ if and only if $\Ms(\mr) \ge 0$. (The factor $\rho_m$ is included for convenience, since
$\rho_m$ is nonnegative.) Since the corresponding classical equations for the speed of sound
can be obtain, in part, by replacing $d(P\rho_m^{-1})$ by $\rho_m^{-1}dp$,
\begin{equation} \zlabel{p4420}
\rho_m\fc{d(P\rho_m^{-1})}{d\rho_m} > 0\quad \text{corresponds to } \quad \fc{dp}{d\rho_m} >0 
\end{equation}
and, from the stability conditions \cite{Kestin2,Callen} of thermodynamics, the rhs inequality
always holds for classical flows.

Since we have the correspondence (\ref{p4420}), and since in all the applications considered,
$\Ms(\mr) \ge 0$ for the domain of the flow, and, hence, $[d(P\rho^{-1})/d\rho_m](\mr) \ge 0$
for the domain of the flow, we make the following ``mathematical hypothesis:'' For Q1 flows
$[d(P\rho^{-1})/d\rho_m](\mr) \ge 0$, for the entire flow domain, and therefore $\Ms(\mr) \ge
0$. Hence. we can set $\Ma = \Ms$ in all the above equations.


\appendix

\section{Orbital Expresssions \zlabel{p4387}}

For every equality and formula involving the density $\rho$, there is a corresponding one
involving orbitals $\phi$, since $\rho = \phi^2$.  Here we derive some equalities  involving
orbitals.  \lbar Obtain orbital expressions for $p$ by raarranging (\ref{0382}) and using
(\ref{0003b}).
\begin{gather} \notag
  \\ \notag  -\fc14\nabla^2\rho =  -\fc12\phi\nabla^2\phi - \fc12\nabla\phi\cdot\nabla\phi \\
  \zlabel{4927}
p =  -\fc{\hbar^2}{2m}\phi\nabla^2\phi - \fc{\hbar^2}{2m}\nabla\phi\cdot\nabla\phi \rbar
\end{gather}

\lbar Use $\nabla\rho =2\phi\nabla\phi$ and (\ref{0003a}) to obtain an orbital expression for the
velocity $\bu_\pm$, the velocity component $u_\pm = \bu_\pm\cdot\hat{\mz}$, and the kinetic
energy density $mu^2/2$.
\begin{gather} 
\notag \bu_\pm = \pm\fc{\hbar}{2m}(\rho^{-1})\nabla\rho = \pm\fc{\hbar}{m}(\phi^{-2})\phi\nabla\phi \\
\zlabel{8927} \bu_\pm = \pm\fc{\hbar}{m}\fc{\nabla\phi}{\phi}
  \\ \zlabel{8925} u_\pm = \pm\fc{\hbar}{m}\fc{\partial\phi}{\phi}
  \\ \zlabel{8928}
  \fc12 mu^2 = \fc{\hbar^2}{2m}\phi^{-2}\nabla\phi\cdot\nabla\phi \rbar
\end{gather}

\lbar Obtain the orbital continuity equation from (\ref{8927}).
\begin{gather*}
  \nabla\cdot(\phi\bu) = \pm\fc{\hbar}{m}\nabla^2\phi\rbar
\end{gather*}

\bibliography{ref}


\pagebreak

\begin{figure}[!tbp] \input{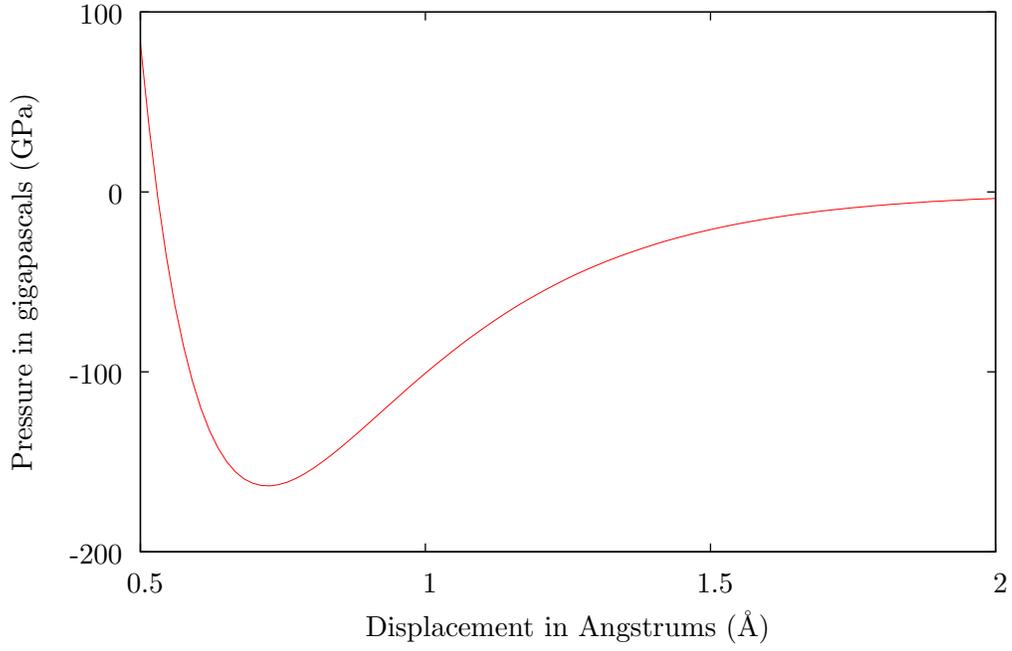} \caption{The pressure of hydrogen atom from Eq.~(\ref{p5220}). \zlabel{pressure-1s}} \end{figure} 
\begin{figure}[!tbp] \input{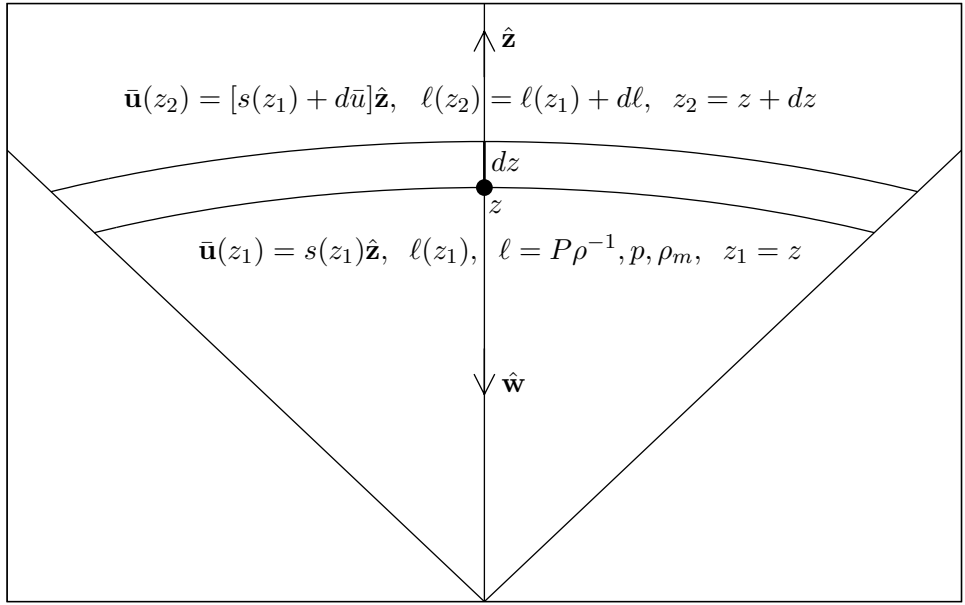} \caption{A representation of a spherical wave pulse and some of the functions needed
  for an ambient state with a Q1 flow and a corresponding classical potential-free
  flow.\zlabel{wavepulse}} \end{figure} 
\begin{figure}[!tbp] \input{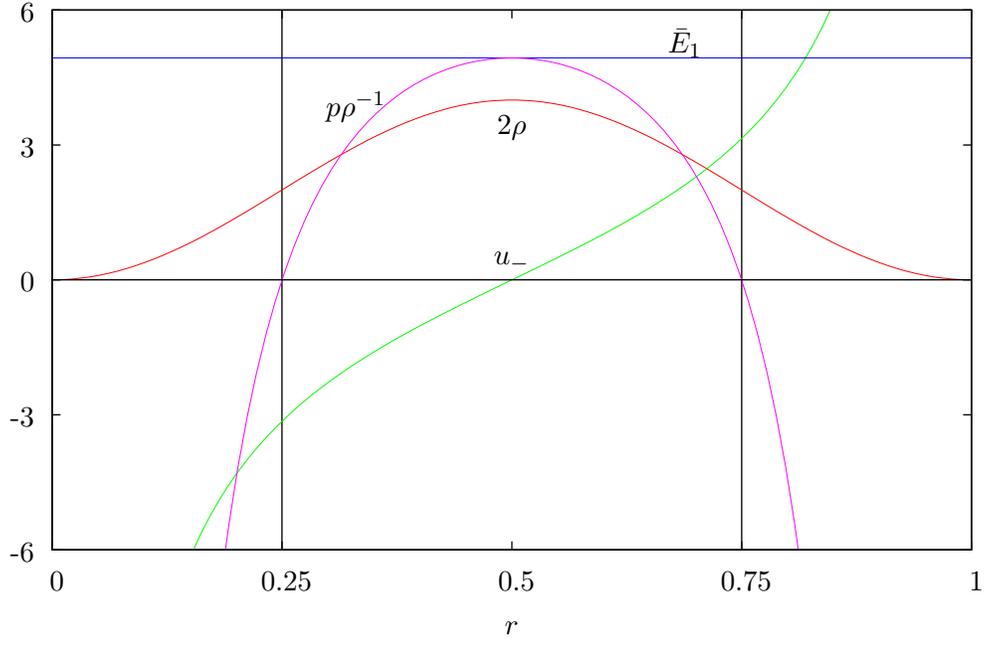} \caption{The specifc- compression $p\rho^{-1}$ and total $\bar{E}_1$ energies with the density $\rho$
  and downhill velocity $u_-$ of the ground-state of a fluid in a one-dimensional box of length $a_0$
  in atomic units. \zlabel{box1-energy}} \end{figure} 
\begin{figure}[!tbp] \input{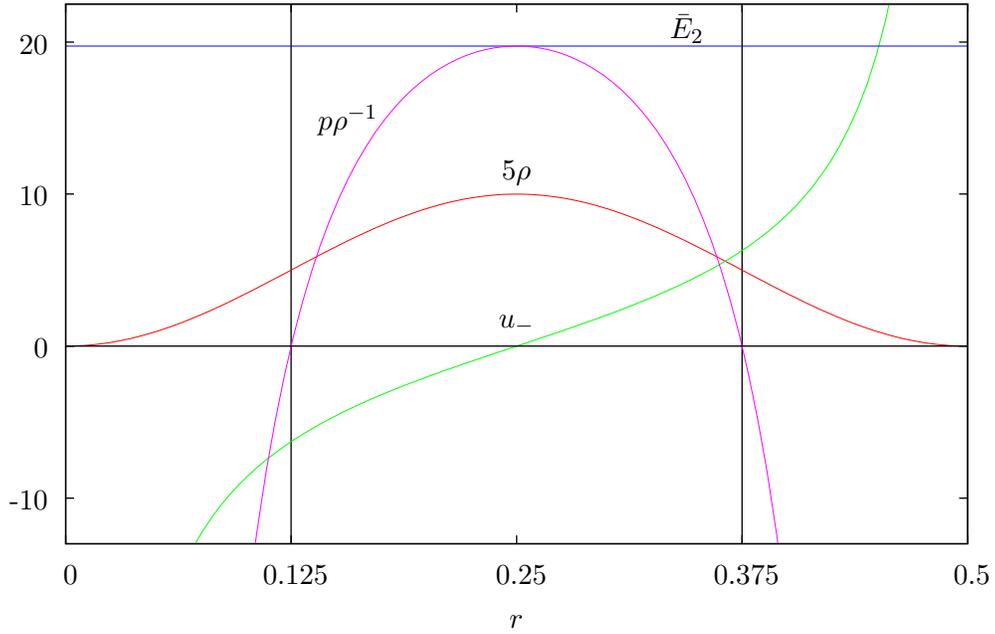} \caption{Same is Fig~(\ref{box1-energy}) but for the first excited state and
  for $r\in[0,0.5]$. \zlabel{box2-energy}} \end{figure} 
 \begin{figure}[!tbp] \input{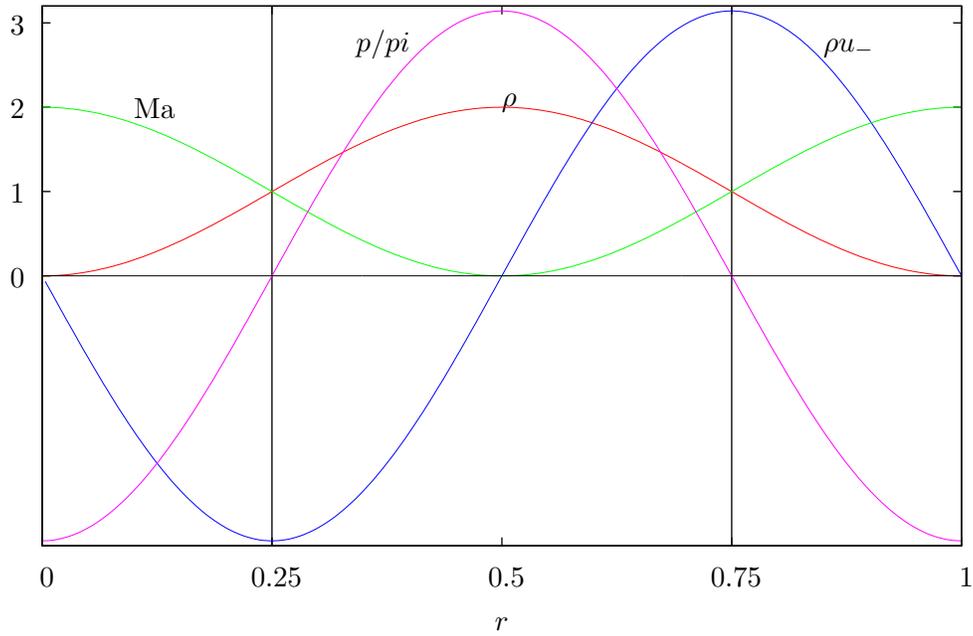} \caption{The Mach speed Ma and the pressure $p$ with the density $\rho$ and the
    downhill momentum density $\rho u_-$ of the ground-state of a fluid in a one-dimensional box of
    length $a_0$ in atomic units.\zlabel{box1}} \end{figure} 
\begin{figure}[!tbp] \input{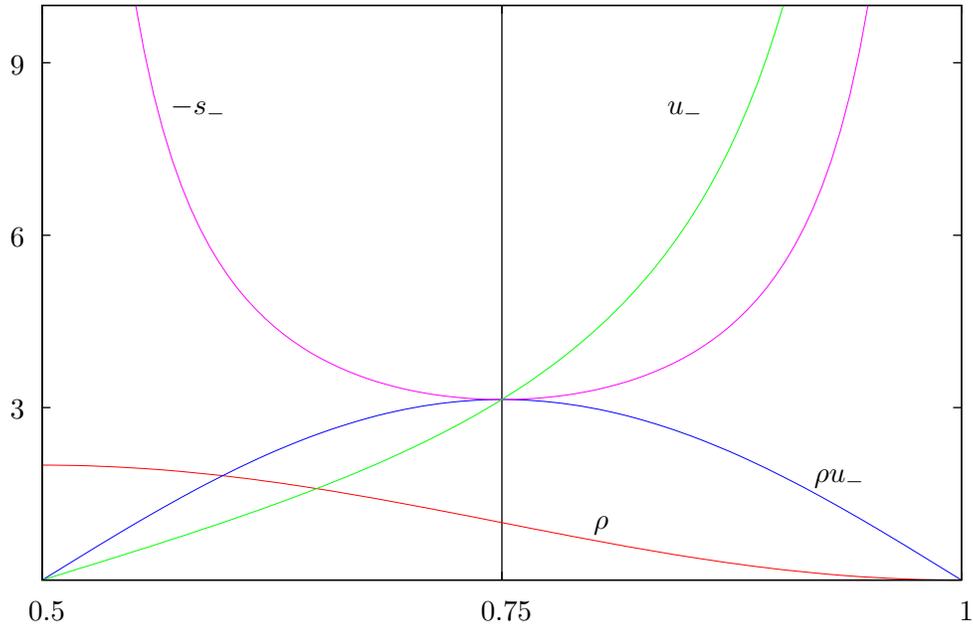} \caption{The downhill wave pulse $-s_-$ and velocity $u_-$ with the density $\rho$ and
  the downhill momentum density $\rho u_-$ of the ground-state of a fluid in a dimensional box of
  length $a_0$ in atomic units for $r\in[1/2,1]$. \zlabel{box2}} \end{figure} 
\begin{figure}[!tbp] \input{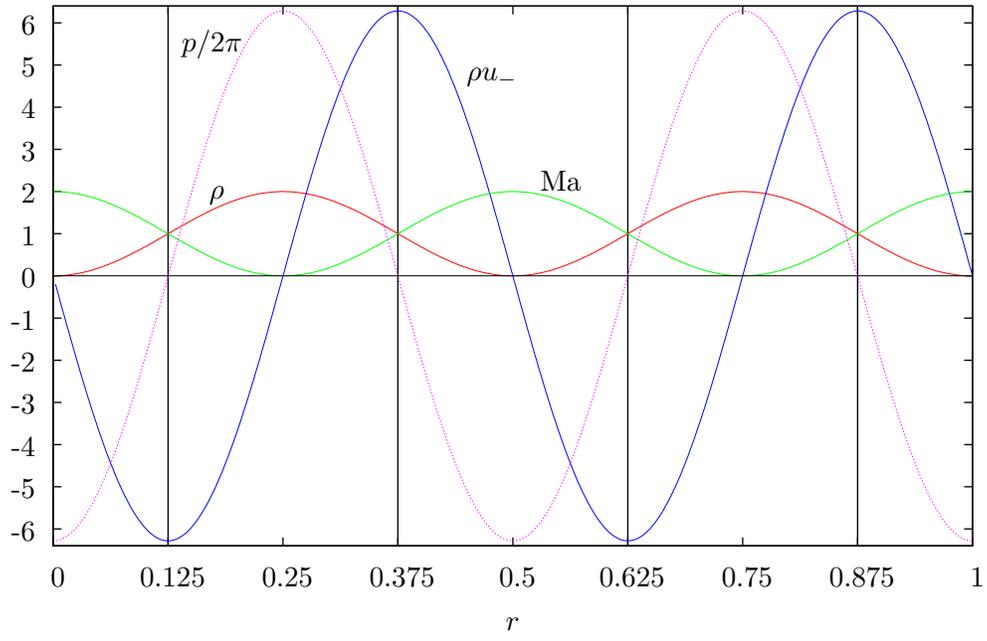} \caption{Same as (\ref{box1}), but for the first excited state
  with the two flow hill $(0,\,.25,\,.5)$ and $(.5,\,.75,\,1)$. \zlabel{box3}} \end{figure} 
\begin{figure}[!tbp] \input{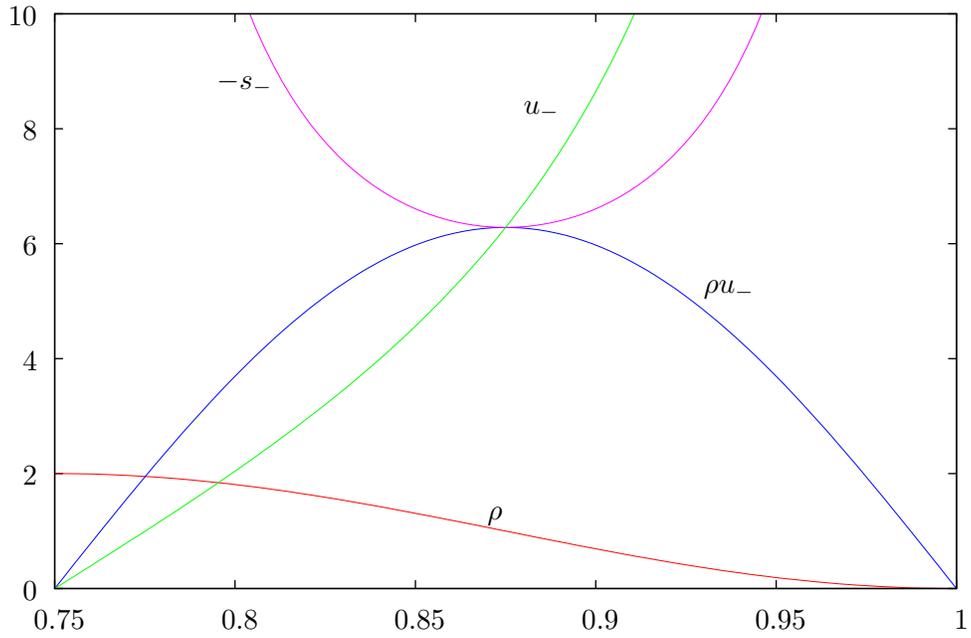} \caption{Same as (\ref{box2}), but for the first excited state and the
  primitive flow segment $[3/4,1]$. \zlabel{box4}} \end{figure} 
\begin{figure}[!tbp] \input{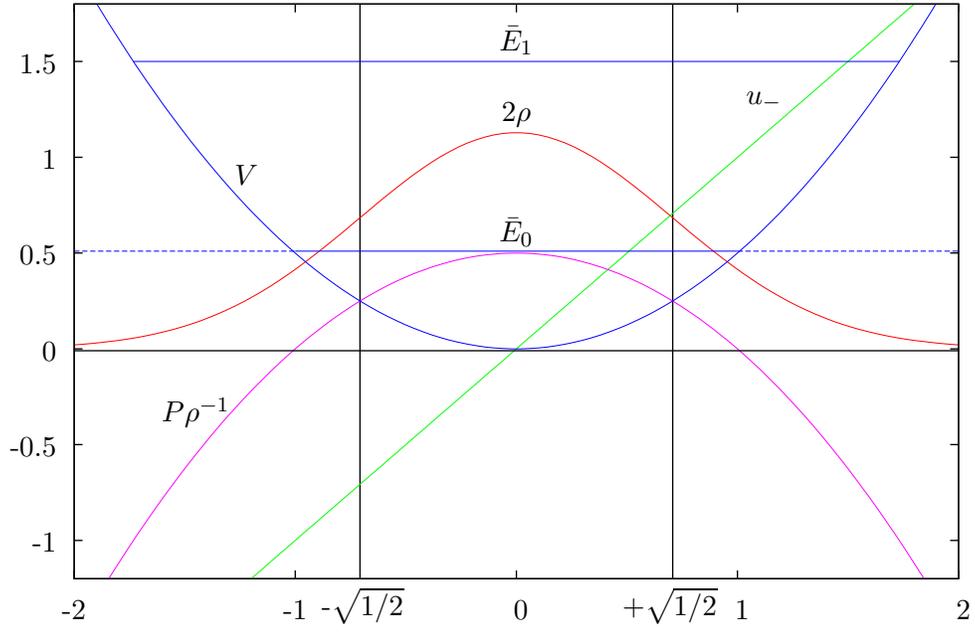} \caption{The specific- effective-compression $P\rho^{-1}$, potential $V(x) =
  x^2/2$. and total $\bar{E}_0$ \mbox{-energies} with the density $\rho$ and downhill velocity
  $u_-$ of the ground-state harmonic oscillator $(-\infty,0,+\infty)$ in atomic
  units. \zlabel{hosc1-energy}} \end{figure} 
\begin{figure}[!tbp] \input{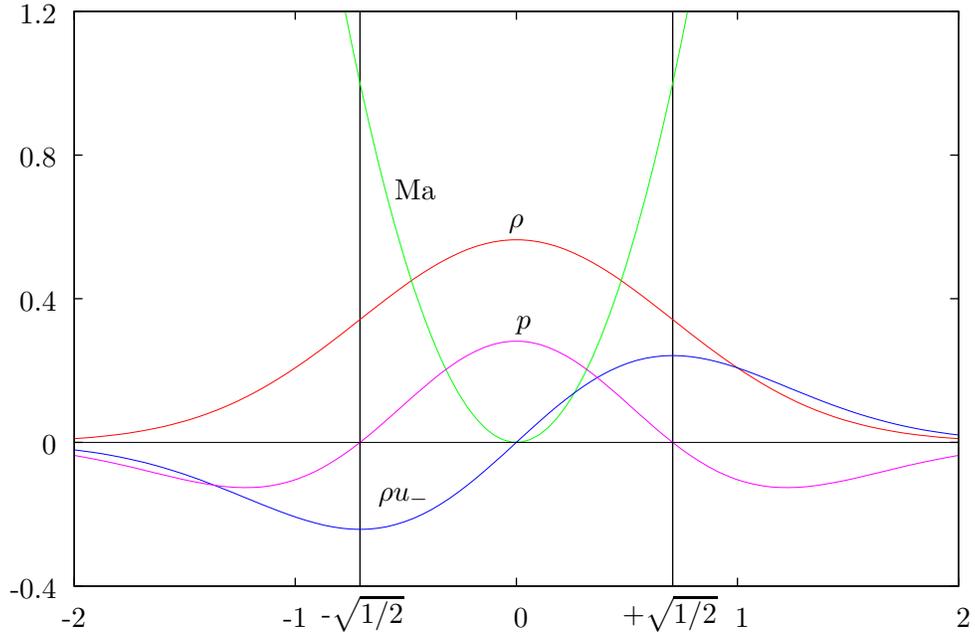} \caption{The Mach speed Ma and the pressure $p$ with the density $\rho$ and
  the downhill momentum density $\rho u_-$ of the ground-state harmonic oscillator $(-\infty,0,+\infty)$ in
  atomic units. \zlabel{hosc1}} \end{figure} 
\begin{figure}[!tbp] \input{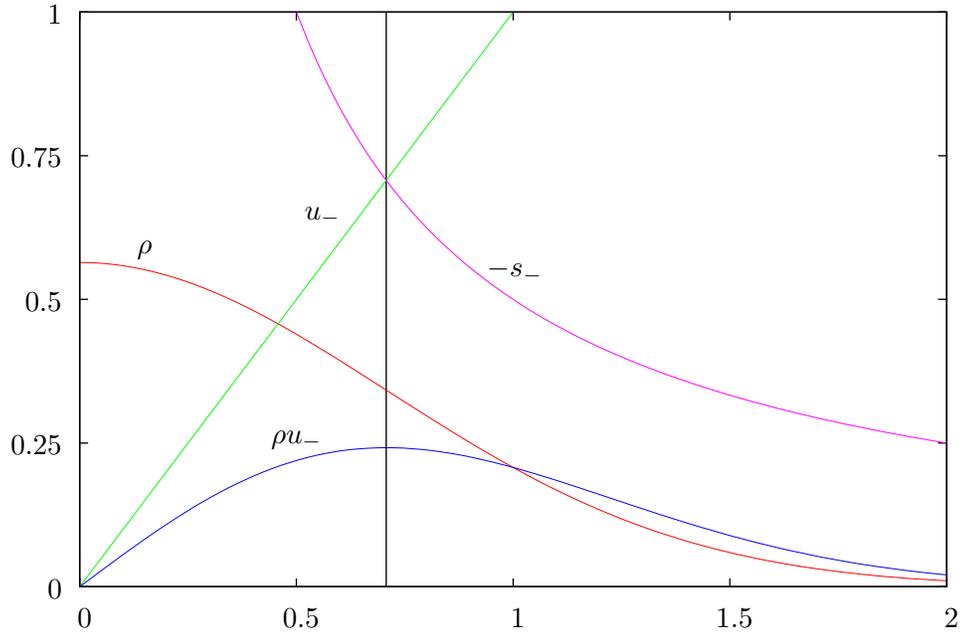} \caption{The downhill wave pulse $-s_-$ and velocity $u_-$ with the density $\rho$ and
  the downhill momentum density $\rho u_-$ of the ground-state harmonic oscillator in
  atomic units for the flow segment subset $[0,2]\subset[0,\infty)$. \zlabel{hosc2}} \end{figure} 
\begin{figure}[!tbp] \input{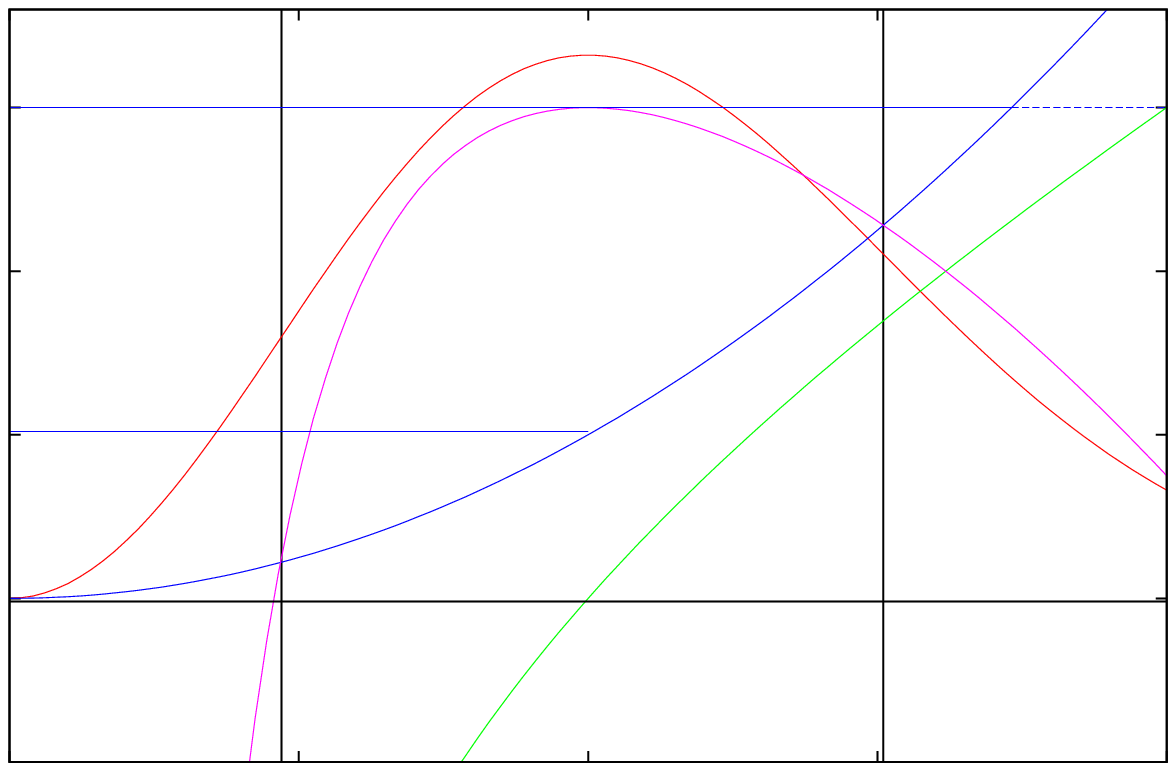} \caption{Same as (\ref{hosc1-energy}), but for the first excited state and the
  flow-hill subset $[0,2]\subset[0,1,\infty)$. \zlabel{hosc2-energy}} \end{figure} 
\begin{figure}[!tbp] \input{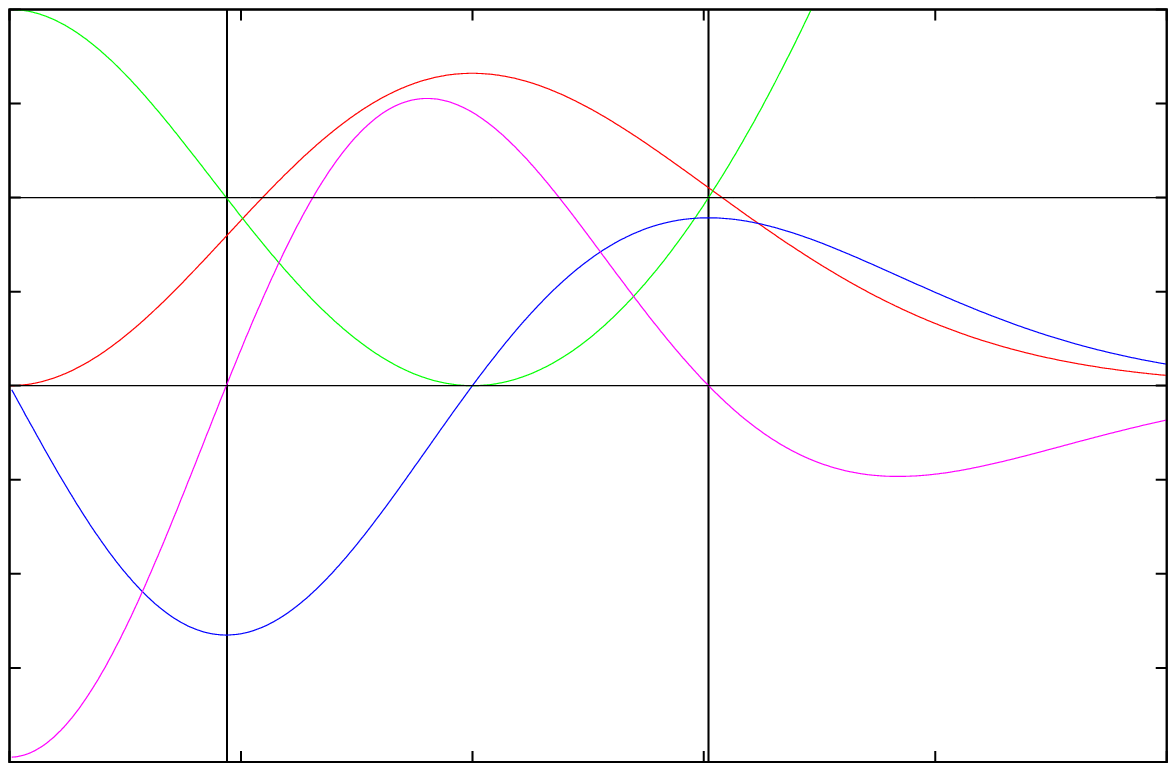} \caption{Same as (\ref{hosc1}), but for the first excited state and
  the flow-hill subset $[0,2.5]\subset[0,1,\infty)$. \zlabel{hosc3}} \end{figure} 
\begin{figure}[!tbp] \input{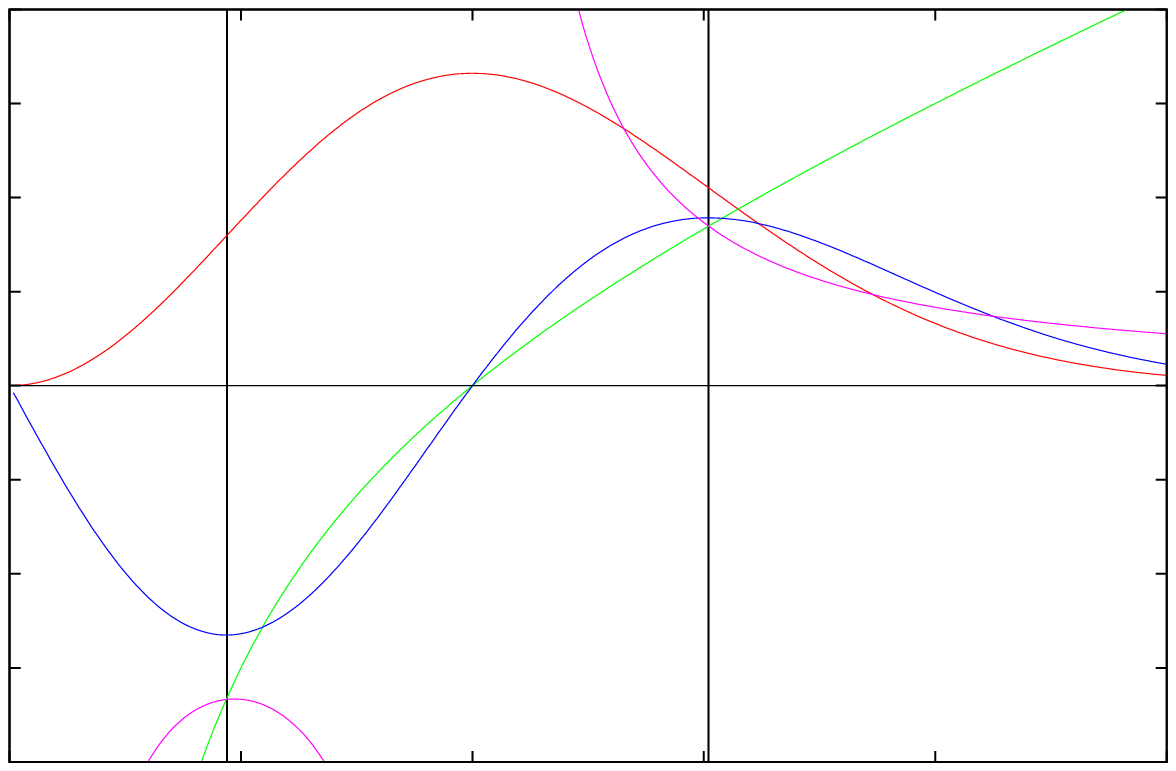} \caption{Same as (\ref{hosc2}), but for the first excited state and the
  flow-hill subset $[0,2.5]\subset[0,1,\infty)$. \zlabel{hosc4}} \end{figure} 
  \begin{figure}[!tbp] \input{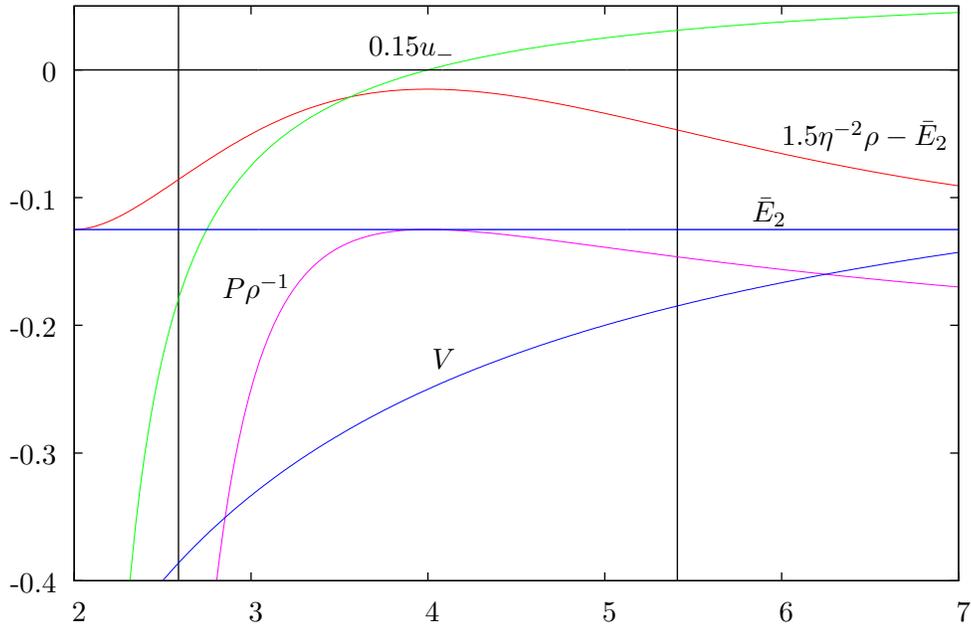} \caption{The specific- effective-compression $P\rho^{-1}$, potential $V(r) =
    -r^{-1}$, and total $\bar{E}_{2}$ \mbox{-energies} with the density $\rho$ and downhill
    velocity $u_-$ of the 2s state of the hydrogen atom and the flow-hill subset
    $[2,7]\subset[2,4,\infty)$, The normalizing factor is $\eta =
      1/(4\sqrt{2\pi})$. \zlabel{2s-energy}} \end{figure} 
  \begin{figure}[!tbp] \input{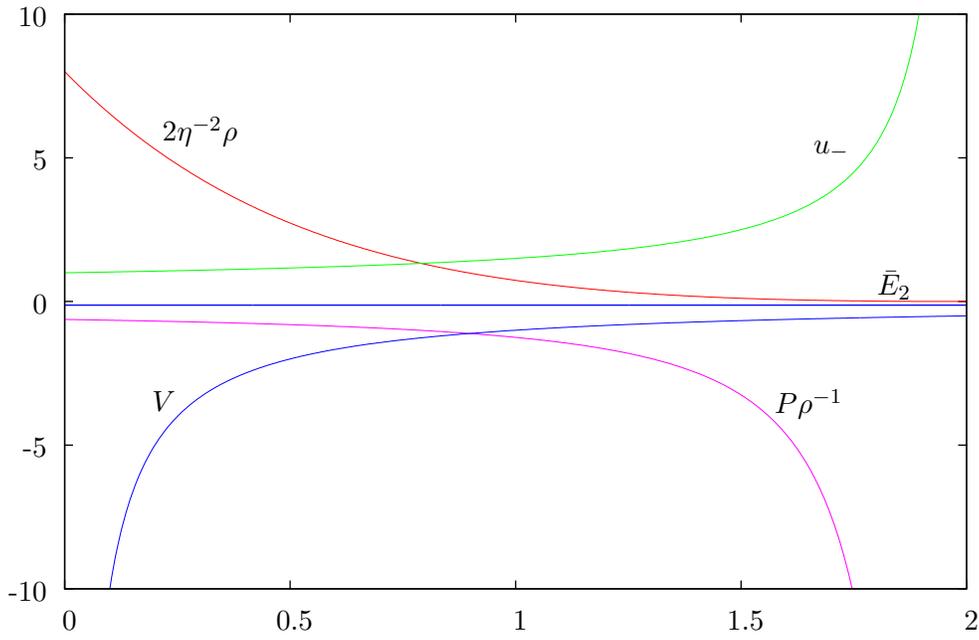} \caption{Same as (\ref{2s-energy}), but for the electron fall
    $[0,2]$. \zlabel{2sl-energy}} \end{figure} 
\begin{figure}[!tbp] \input{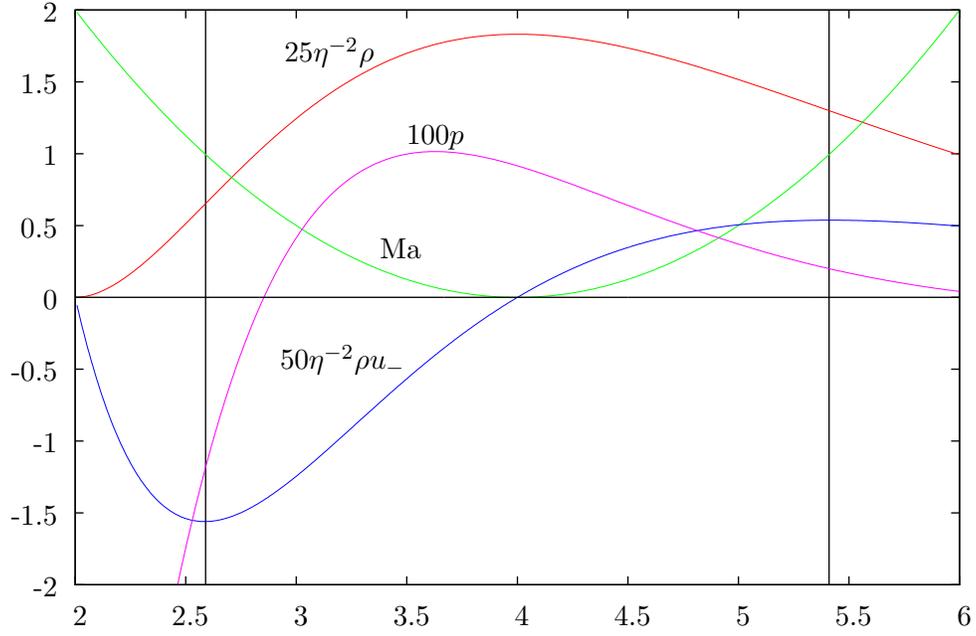} \caption{The Mach speed Ma and the pressure $p$ with the density $\rho$ and
  the downhill momentum density $\rho u_-$ of the 2s state of the hydrogen atom and the flow-hill subset
  $[2,6]\subset[2,4,\infty)$. \zlabel{2s1}} \end{figure} 
\begin{figure}[!tbp] \input{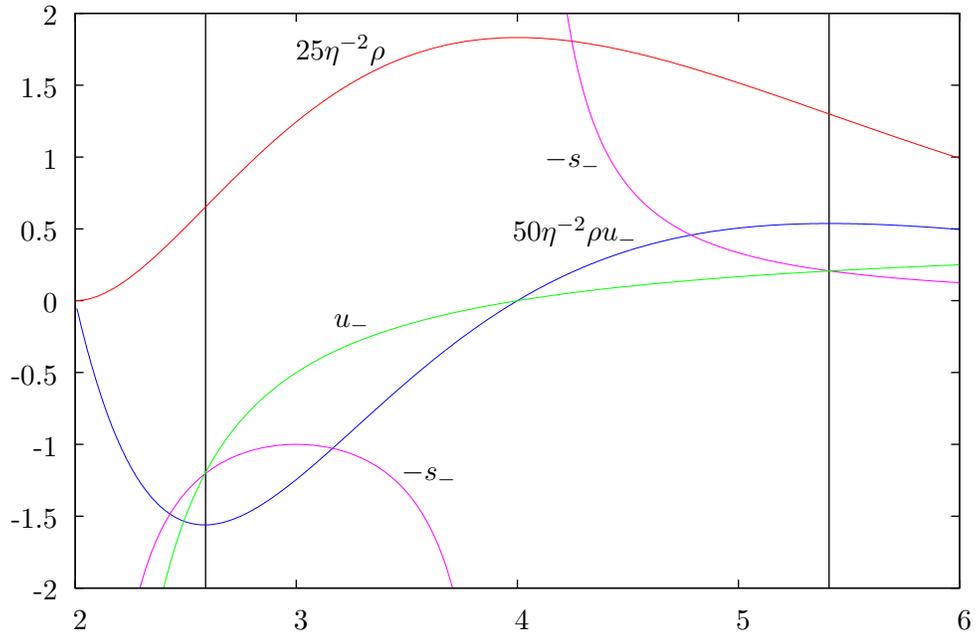} \caption{The downhill wave pulse $-s_-$ and velocity $u_-$ with the density $\rho$ and the
  downhill momentum density $\rho u_-$ of the 2s state of the hydrogen atom and the flow-hill
  subset $[2,6]\subset[2,4,\infty)$. \zlabel{2s2}} \end{figure} 

  \mbox{} \newpage \mbox{} 
 
\begin{figure}[!tbp] \input{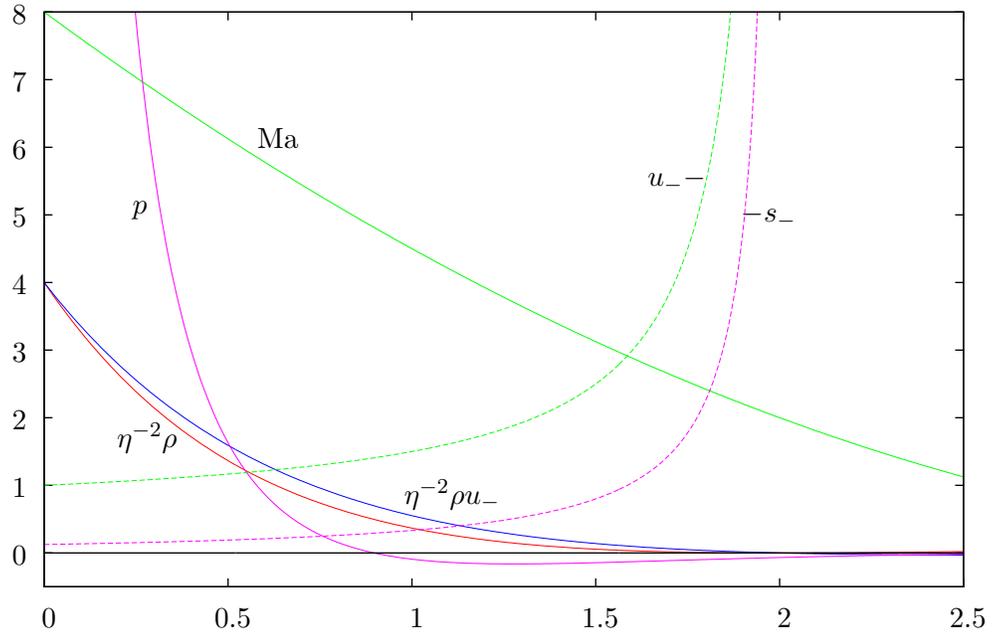} \caption{The same functions as in both (\ref{2s1}) and (\ref{2s2}) but for the electron fall
  $[0,2]$. \zlabel{2sl1}} \end{figure} 

\end{document}